\documentclass[journal=aamick,manuscript=article]{achemso}

\pdfoutput=1

\usepackage[utf8]{inputenc}
\usepackage{amsmath}
\usepackage{mathtools}
\usepackage{physics}
\usepackage{upgreek}
\usepackage{color}
\usepackage{siunitx} 

\usepackage{graphicx}
\usepackage{color}
\usepackage{float}

\usepackage[font=small,labelfont=bf,margin=5pt, format=plain]{caption}

\usepackage{subcaption} 
\usepackage{placeins} 
\usepackage{tikz}
\usetikzlibrary{calc}
\usepackage{pgfplots}
\pgfplotsset{compat=1.16}
\usepackage{gensymb}
\usepackage{comment}

\usepackage{xr}
\makeatletter
\newcommand*{\addFileDependency}[1]{
  \typeout{(#1)}
  \@addtofilelist{#1}
  \IfFileExists{#1}{}{\typeout{No file #1.}}
}
\makeatother
\newcommand*{\myexternaldocument}[1]{%
    \externaldocument{#1}%
    \addFileDependency{#1.tex}%
    \addFileDependency{#1.aux}%
}
\myexternaldocument{SI_text}

\title{Optical properties of MoSe$_2$ monolayer implanted with ultra-low energy Cr ions}

\author{Minh N. Bui}
\email{m.bui@fz-juelich.de}
\affiliation{Peter Gr\"{u}nberg Institute 9 (PGI-9), Forschungszentrum J\"{u}lich, 52425 J\"{u}lich, Germany}
\alsoaffiliation{Department of Physics, RWTH Aachen University, 
52074 Aachen, Germany}

\author{Stefan Rost}
\affiliation{Peter Gr\"{u}nberg Institute 1 (PGI-1) and Institute for Advanced Simulation 1 (IAS-1), Forschungszentrum J\"{u}lich and JARA, 52425 J\"{u}lich, Germany}
\alsoaffiliation{Department of Physics, RWTH Aachen University, 
52074 Aachen, Germany}

\author{Manuel Auge}
\affiliation{II. Institute of Physics, University of G\"{o}ttingen, 37077 G\"{o}ttingen, Germany}

\author{Lanqing Zhou}
\affiliation{Peter Gr\"{u}nberg Institute 9 (PGI-9), Forschungszentrum J\"{u}lich, 52425 J\"{u}lich, Germany}
\alsoaffiliation{Department of Physics, RWTH Aachen University, 
52074 Aachen, Germany}

\author{Christoph Friedrich}
\affiliation{Peter Gr\"{u}nberg Institute 1 (PGI-1) and Institute for Advanced Simulation 1 (IAS-1), Forschungszentrum J\"{u}lich and JARA, 52425 J\"{u}lich, Germany}

\author{Stefan Bl\"{u}gel}
\affiliation{Peter Gr\"{u}nberg Institute 1 (PGI-1) and Institute for Advanced Simulation 1 (IAS-1), Forschungszentrum J\"{u}lich and JARA, 52425 J\"{u}lich, Germany}
\alsoaffiliation{Department of Physics, RWTH Aachen University, 
52074 Aachen, Germany}

\author{Silvan Kretschmer}
\affiliation{Institute of Ion Beam Physics and Materials Research, Helmholtz‐Zentrum Dresden‐Rossendorf, 01328 Dresden, Germany}

\author{Arkady V. Krasheninnikov}
\affiliation{Institute of Ion Beam Physics and Materials Research, Helmholtz‐Zentrum Dresden‐Rossendorf, 01328 Dresden, Germany}
\alsoaffiliation{Department of Applied Physics, Aalto University School of Science, P.O.Box 11100, 00076 Aalto, Finland}

\author{Kenji Watanabe}
\affiliation{Research Center for Functional Materials, National Institute for Materials Science, 1-1 Namiki, Tsukuba 305-0044, Japan}

\author{Takashi Taniguchi}
\affiliation{International Center for Materials Nanoarchitectonics, National Institute for Materials Science,  1-1 Namiki, Tsukuba 305-0044, Japan}

\author{Hans C. Hofsäss}
\affiliation{II. Institute of Physics, University of G\"{o}ttingen, 37077 G\"{o}ttingen, Germany}

\author{Detlev Gr\"{u}tzmacher}
\author{Beata E.~Kardyna\l}
\email{b.kardynal@fz-juelich.de}
\affiliation{Peter Gr\"{u}nberg Institute 9 (PGI-9), Forschungszentrum J\"{u}lich, 52425 J\"{u}lich, Germany}
\alsoaffiliation{Department of Physics, RWTH Aachen University, 
52074 Aachen, Germany}


\begin{document}
\newpage

\begin{abstract} 
    
     The paper explores the optical properties of an exfoliated MoSe$_2$ monolayer implanted with Cr$^+$ ions, accelerated to 25\,eV.
    Photoluminescence of the implanted MoSe$_2$ reveals an emission line from Cr-related defects that is present only under weak electron doping. Unlike band-to-band transition, the Cr-introduced emission is characterised by non-zero activation energy, long lifetimes, and weak response to the magnetic field. To rationalise the experimental results and get insights into the atomic structure of the defects, we modelled the Cr-ion irradiation process
    using ab-initio molecular dynamics simulations followed by the electronic structure calculations of the system with defects. 
    The experimental and theoretical results suggest that the recombination of electrons on the acceptors, which could be introduced by the Cr implantation-induced defects, with the valence band holes is the most likely origin of the low energy emission.
    Our results demonstrate the potential of low-energy ion implantation as a tool to tailor the properties of
    2D materials by doping.
    \\

\end{abstract}

\section{Introduction}

The properties of semiconductors, especially atomically thin monolayer (ML) semiconductors, depend strongly on the types and densities of defects in their crystal lattices. The most technologically relevant defects are dopants, i.e., foreign atoms in substitutional positions in the crystal lattice.
Shallow dopants introduce free electrons or holes into the conduction or valence band and change thus semiconductor conductivity. As such they facilitate the fabrication of p-n junctions, which underpins most active optoelectronic devices.

Doping with transition metal atoms has been shown to introduce ferromagnetic order in p-doped semiconductors \cite{Dietl2014}. Impurity atoms can also trap electrons or holes or bind excitons. Radiative recombination involving such states can be detected as sub-bandgap photoluminescence (PL). Single foreign atoms binding excitons have been explored for single photon sources \cite{Ikezawa2012}. Alternatively, if the dopant atom has a functionality of a spin qubit, the bound excitons provide an optical readout of its state \cite{Vasileios22,  Yamamoto09, Vandersypen17}. The binding of excitons to the dopant atoms depends not only on electron and hole masses but also the dielectric constant of the semiconductors. Because of that, excitonic effects in bulk semiconductors are only observed at cryogenic temperatures. Foreign atoms can also act as colour centres in semiconductors and insulators. Spin qubits based on the colour centres have been realised in diamond \cite{Sekiguchi2021, Metsch2019, Ruf2021, Pezzagna2021} or SiC \cite{Wolfowicz2020, Lohrmann2015}.

In two-dimensional (2D) semiconducting transition metal dichalcogenides (TMDs), which feature weak electrostatic screening, substitutional atoms tend to introduce deep levels in the bandgaps \cite{Ma2017}. While excitons have considerable binding energies, they are predicted to be very weakly bound to individual doping atoms \cite{Mostaani17}. Optical transitions involving defect states have been observed \cite{Rivera2021}, with defects identified as vacancies \cite{Borghardt2020, Klein2019, Mitterreiter2021, Rivera2021}. The transition responsible for the PL was found to occur between the hybridised defect states and 2D lattice electronic states.

Among several methods of doping bulk semiconductors, ion implantation offers the highest flexibility in choosing implanted elements. Ion energies of tens of keV are used for implantation since functional layers can be even a hundred nanometers below the surface. High-energy ion implantation has been used \cite{Xu2017, He2019, Prucnal2021, Wang2021, Jadwiszczak2020} to modify 2D materials, but its efficiency is low in this case, as most atoms go through the 2D target \cite{Krasheninnikov2020}. Moreover, the ions penetrating through the ML can cause undesirable effects, e.g., trapped charges in the substrate. Implantation into 2D materials has the highest implantation efficiency with ion energies in the range of tens of eV. At these energies, the implantation efficiency and threshold energy depend on the ions’ mass and also chemical properties \cite{Kretschmer2022}. Ultra-low energy ion implantation \cite{Auge2022, Junge2022} has recently been demonstrated to be efficient in doping graphene using 40 eV Mn ions \cite{Lin2021, Lin2022}, or in Se ion implantation into MoS$_2$ with the ion energy of 20\,eV \cite{Bui2022, Bangert2017}. The ratio of the replaced S atoms with Se in the top sublattice was sufficient to form a Janus compound MoS$_\text{2-2x}$Se$_\text{2x}$ as indicated by Raman spectroscopy and the transmission electron microscopy imaging.

Here, we study the optical properties of MoSe$_2$ ML implanted with 25 eV $^{52}$Cr$^+$ ions. Sub-bandgap defect-induced PL emission was observed only at the low n-doping level and with saturation behaviour, characteristic of defects with low density.
Ab-initio molecular dynamics (MD) simulations of the implantation process were performed, and possible configurations of the Cr atoms in the MoSe$_2$ ML lattice were outlined to understand the atomic structure of the implanted MLs. The optical properties of the MoSe$_2$ ML with such defects were calculated using density functional theory (DFT). The most probable defect configurations were identified by combining experimental data and theoretical calculations.

\section{Results}

\subsection{Sample preparation and ion implantation}

\begin{figure}[h]
\begin{subfigure}[c]{.42\textwidth}
    \caption{}
    \centering
    \includegraphics[width=\textwidth]{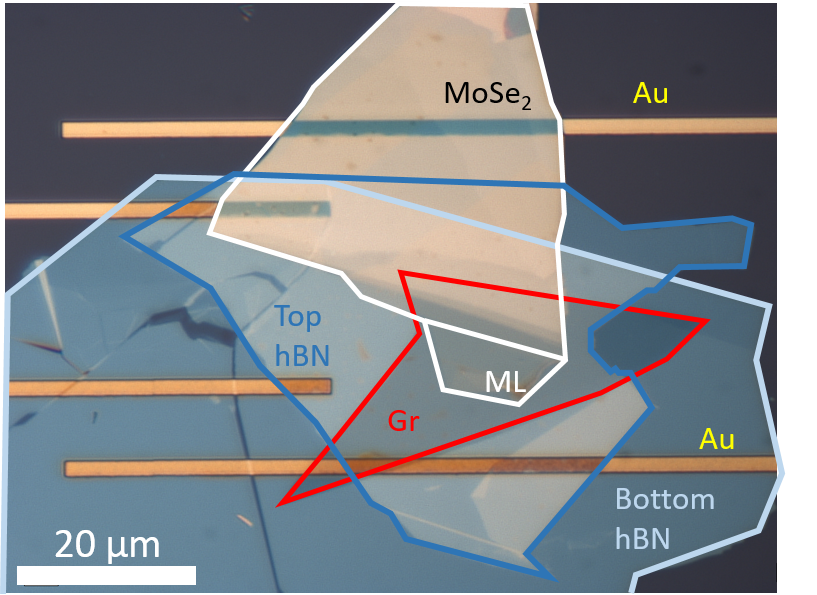}
    \label{fig:sample micrograph}
    \end{subfigure}
\begin{subfigure}[c]{.34\textwidth}
    \caption{}
    \centering
    \includegraphics[width=\textwidth]{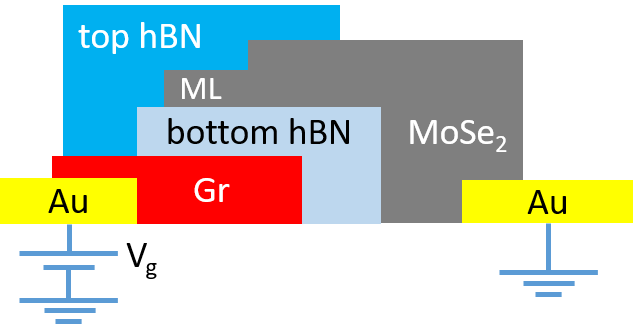}
    \label{fig:sample diagram}
    \end{subfigure}
    \vspace{-10pt}
\caption{\textbf{Cr-implanted MoSe$_2$ ML with hBN encapsulation and graphite backgate.}
(a) Micrograph of the finished  device. The ML part of the exfoliated MoSe$_2$ flake is encapsulated between two thin hBN flakes. The few-layer graphite back gate and the thick part of MoSe$_2$ flake make contacts with the two Ti/Au lines to the right. (b) Schematic diagram of the device cross-section. Back gate voltage $V_g$ can be applied to the graphite back gate via the Au contact, while the MoSe$_2$ flake is grounded via the other Au contact.}
\label{fig:sample picture}
\end{figure}

A sample for ion implantation was prepared by mechanical exfoliation of the MoSe$_2$, graphene, and hBN flakes and their sequential transfer onto the Si/SiO$_2$ substrate with pre-patterned Ti/Au contacts. The use of the dry-viscoelastic transfer technique \cite{Castellanos_Gomez2014} ensured that the surface of the ML was sufficiently clean for the implantation.
The MoSe$_2$ ML has to be grounded during the implantation. An electric contact to the ML was provided by placing the multilayer part of the exfoliated MoSe$_2$ flake on a Ti/Au metal contact. The ML was placed atop a graphite gate connected to another Ti/Au contact. The ML was separated from the gate with an hBN flake. Once completed, the device was implanted with $^{52}$Cr$^+$ ions at 25 eV and a fluence of 3\,$\times$\,10$^{12}$ cm$^{-2}$ (equivalent to 0.003 Cr per ML MoSe$_2$ unit cell, using the MoSe$_2$ ML in-plane lattice constant 3.32 \AA~ \cite{Kang2013}). The implantation was performed with the device heated to 220 $\degree$C. Following the implantation and initial characterisation, another hBN flake was deposited on the ML MoSe$_2$, and the device was annealed at 200 $\degree$C. The complete device is shown in figure \ref{fig:sample picture}. More details are available in the Methods section.

\subsection{Optical spectroscopy data}

\begin{figure}[h]
\begin{subfigure}[c]{0.45\textwidth}
    \caption{}
    \vspace{-15pt}
    \hspace{-15pt}
    \centering
\includegraphics[trim={30 0 30 0},clip,width=\textwidth]{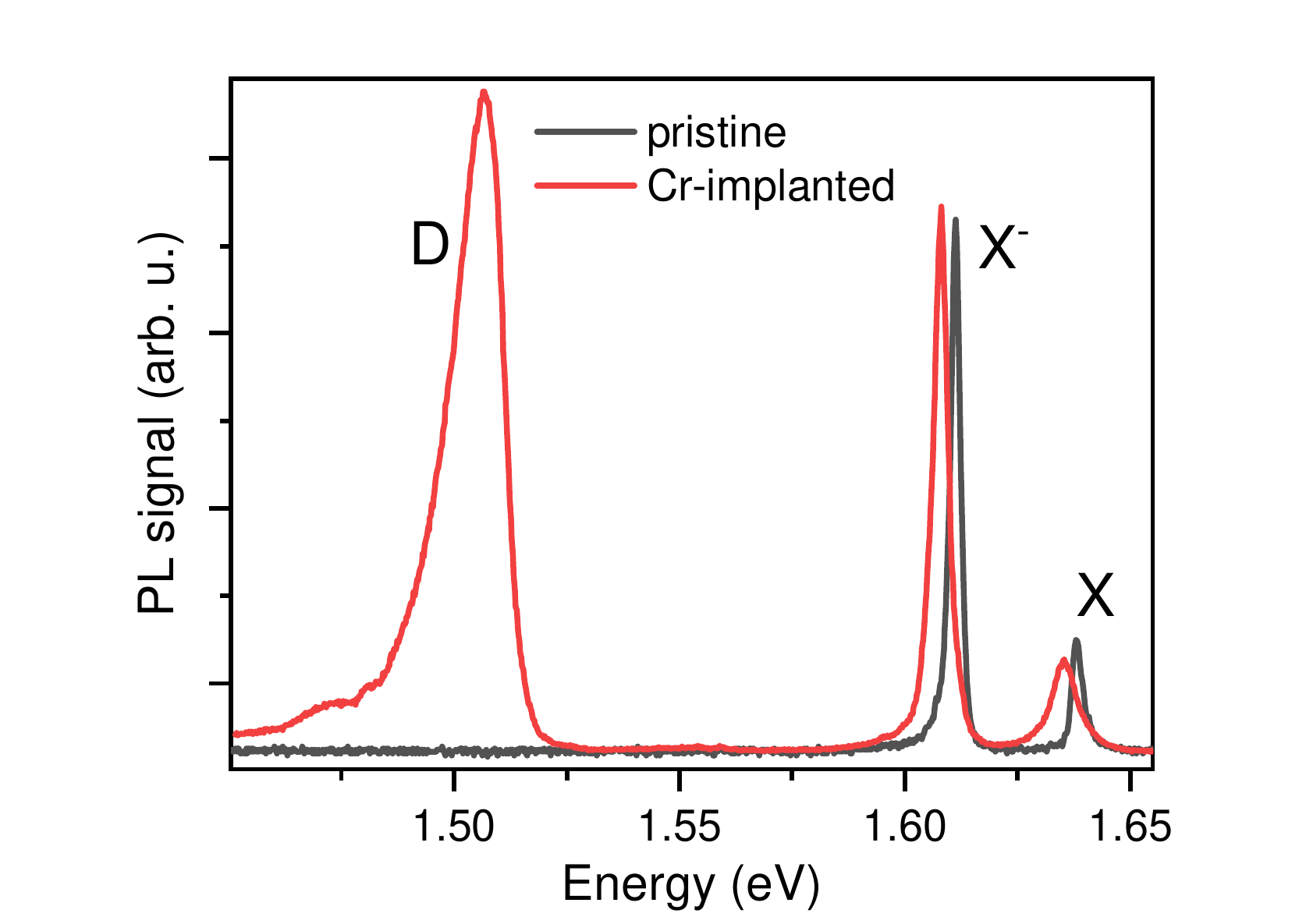}
\label{fig:pristine vs implanted}
\end{subfigure}
\begin{subfigure}[c]{.45\textwidth}
    \caption{}
    \centering
    \includegraphics[width=\textwidth]{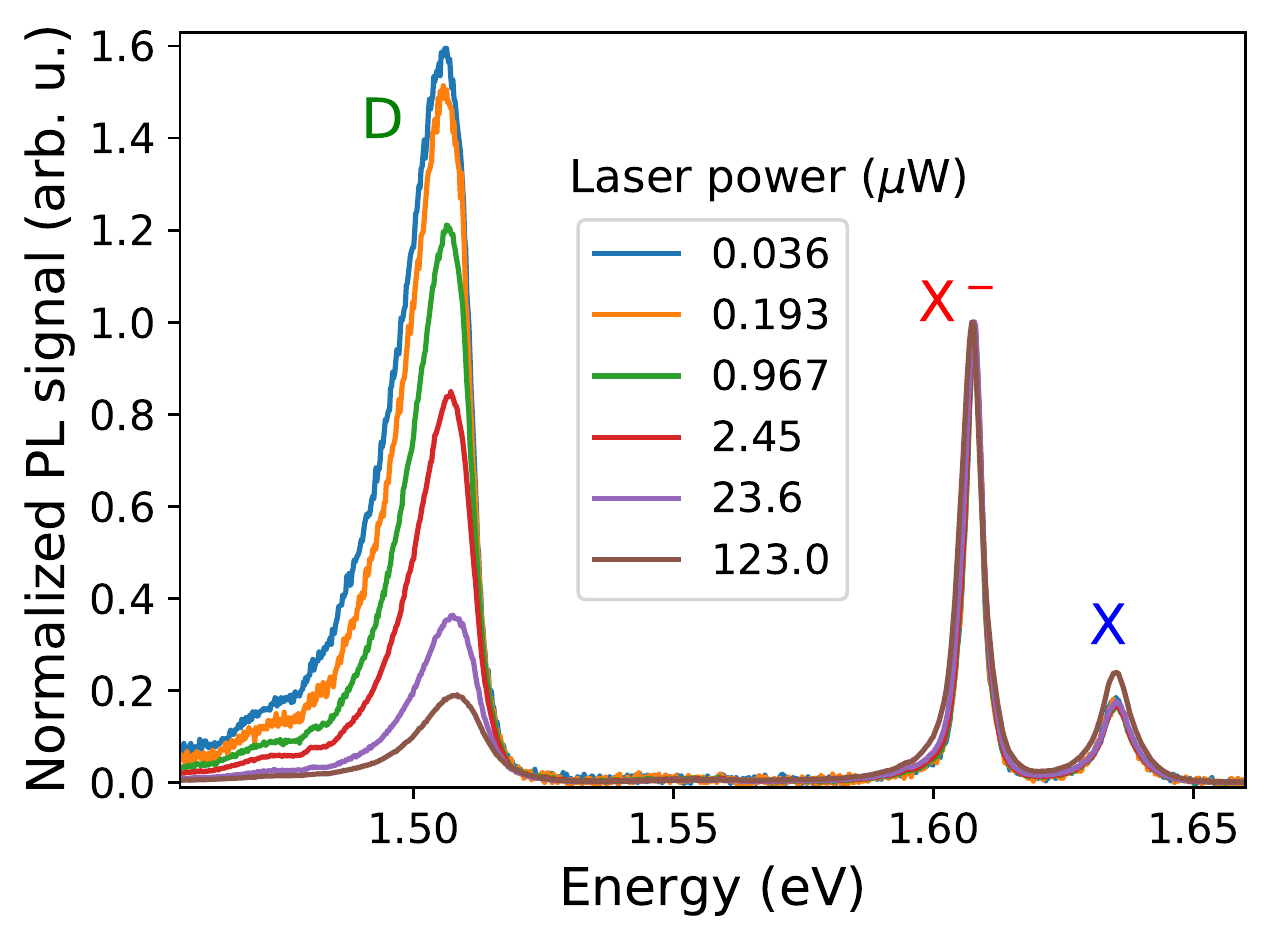}
    \label{fig:power spectra}
\end{subfigure}
\begin{subfigure}[c]{.46\textwidth}
\vspace{-15pt}
    \caption{}
    \centering
    \includegraphics[width=\textwidth]{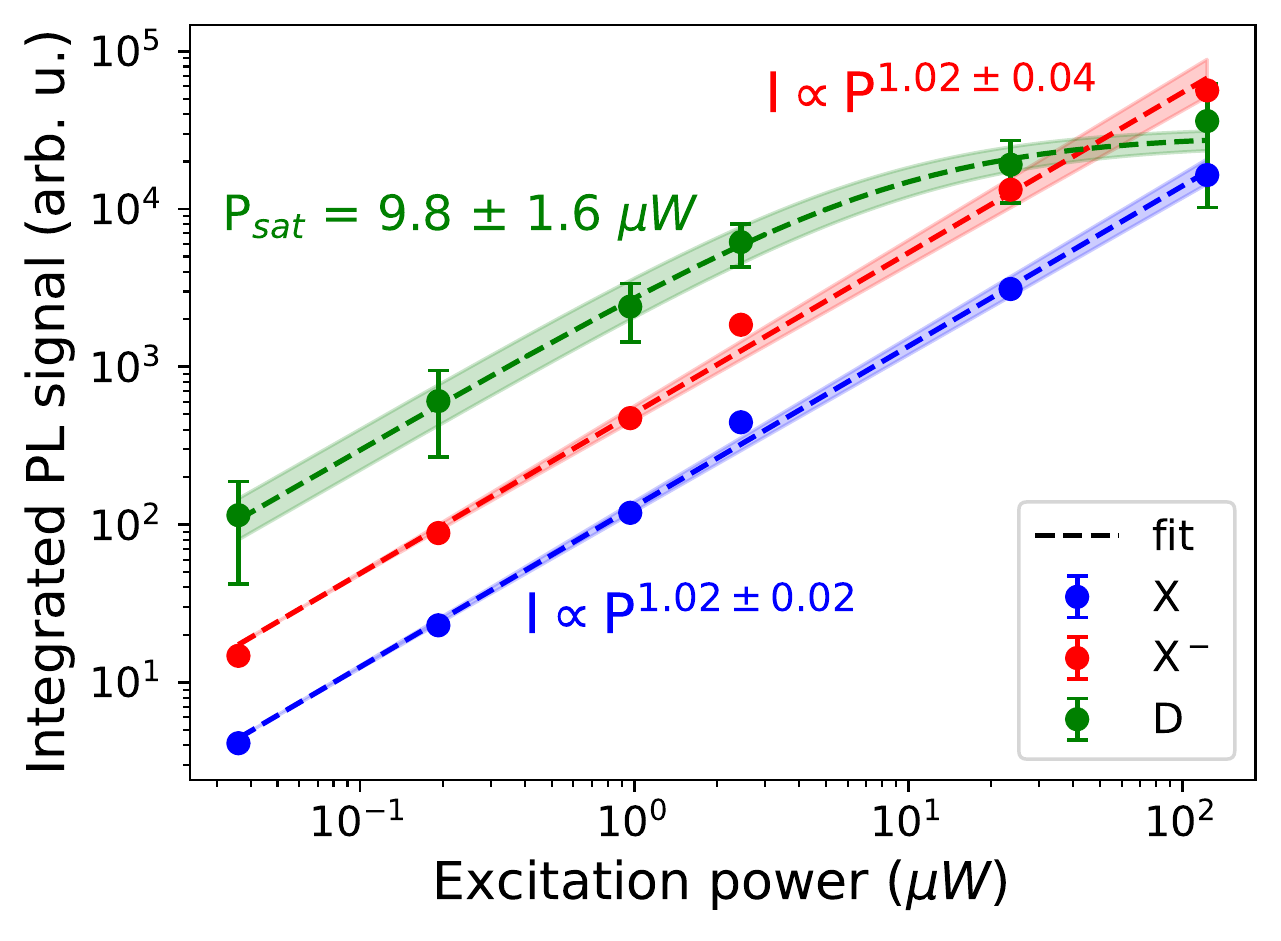}
    \label{fig:power fit}
    \end{subfigure}
    \hspace{10pt}
\begin{subfigure}[c]{0.47\textwidth}
\vspace{-15pt}
    \caption{}
    \vspace{-10pt}
    \hspace{-5pt}
    \centering
    \includegraphics[trim={5 0 20 0},clip,width=\textwidth]{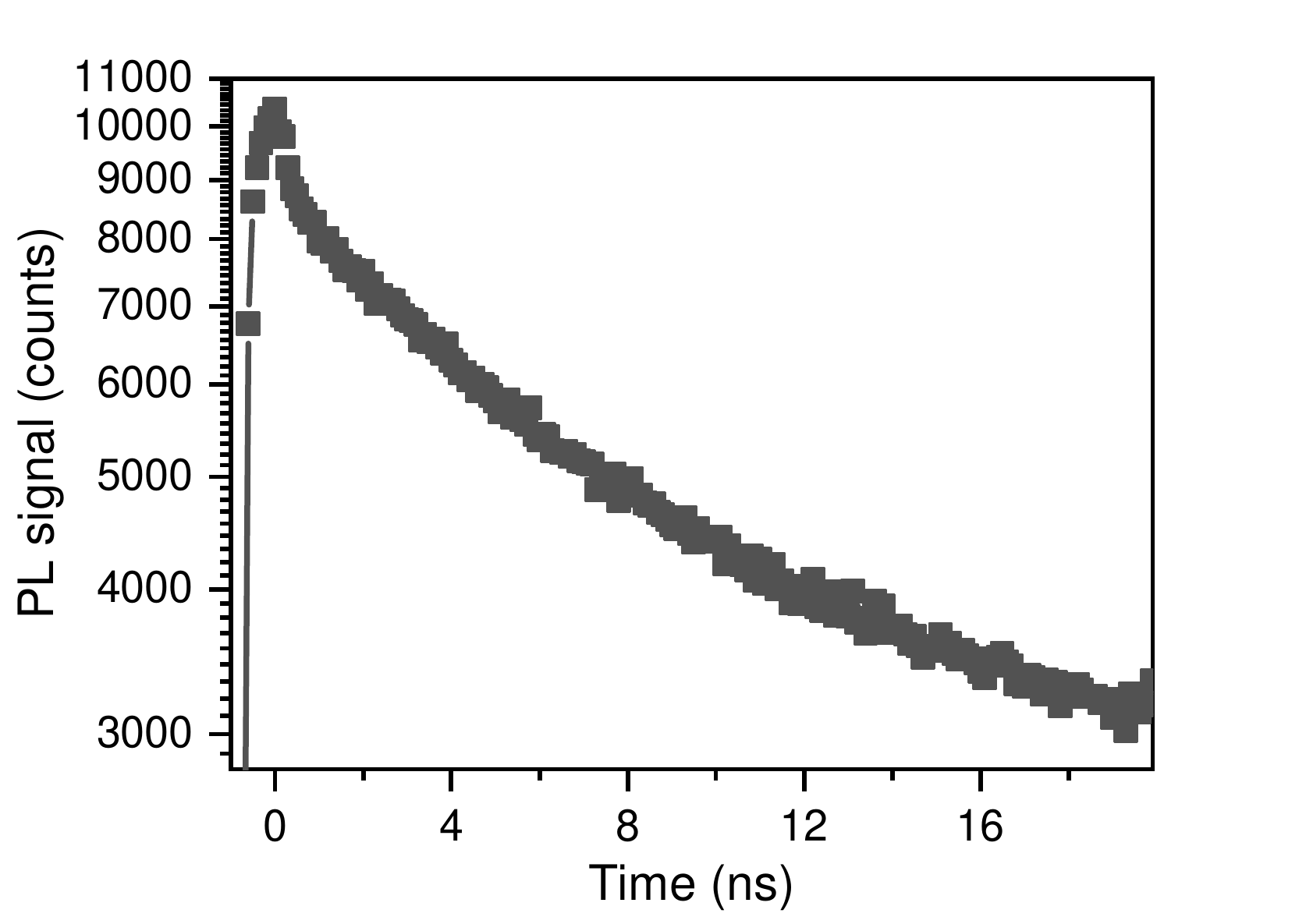}
    \label{fig:TRPL}
\end{subfigure}
\vspace{-10pt}
\caption{\textbf{PL of Cr-implanted MoSe$_2$ ML at 10 K.}
(a) PL spectra of Cr-implanted MoSe$_2$ ML (red curve) at low n-doping ($V_g$ = 0.8 V), plotted with that of pristine MoSe$_2$ ML (black). In addition to the X$^-$ and X from MoSe$_2$ ML, the Cr-implanted sample also shows the broad D peak at around 1.51 eV. (b) PL spectra of Cr-implanted ML under laser power ranging from 36 nW to 123 $\upmu$W. Spectra are normalised to X$^-$. Here the sample is slightly n-doped at $V_g$ = 0.8 V. (c) Power dependence of PL. Best fit lines (dashed), with their standard deviations (shaded region around the lines), are plotted together with extracted intensity from PL spectra (dots). Unless explicitly shown, the error bars are smaller than the size of the data points. X$^-$ and X are fitted with power law $I \propto P^{\alpha}$, and D is fitted with the saturation curve described by equation (\ref{eqn:sat}). (d) Time-resolved PL of Cr-implanted MoSe$_2$. 1/$e$ time is around 14\,ns. 
}
\label{fig:PL 10 K}
\end{figure}

Figure \ref{fig:pristine vs implanted} shows PL spectra of weakly electron-doped pristine and Cr-implanted MoSe$_2$ MLs, measured at 10\, K with the same laser power of 1 $\upmu$W. Both spectra show similar features around the bandgap transitions, with an emission line from neutral excitons (X) and negative trions (X$^-$). The red shift of these transitions in the Cr-implanted sample (figure \ref{fig:pristine vs implanted}) is most likely due to a difference in the dielectric environment or strain between the samples. The level of implantation is too low to expect changes in the bandgap \cite{Ho2019}.
Significant homogeneous broadening of the X line was determined by Voigt function fitting. It resulted in the Lorentzian width of nearly 7 meV for the implanted sample, compared to about 2 meV for the pristine sample, suggesting a much shorter lifetime of the former. The most significant difference between the samples is a broad emission at around 1.51 eV, which we label D.

\FloatBarrier

The relative intensity of the D peak compared to X$^-$ and X depends on the excitation power $P$ (figure \ref{fig:power spectra}). It is the most intense line at low laser excitation powers ($P < 1\,\upmu$W) but saturates as the laser power increases, while X$^-$ and X continue to grow linearly. The saturating behaviour of D intensity, shown in figure \ref{fig:power fit}), can be expressed phenomenologically as:
\begin{equation}
    I \propto \frac{P}{P+P_{sat}}
    \label{eqn:sat}
\end{equation}
with saturation power $P_\text{sat}$ $\approx$ 10 $\upmu$W. The saturating behaviour is expected when the exciton generation rate exceeds the recombination rate of the states responsible for D. The saturation threshold depends on the density of states and the lifetime of the recombining carriers \cite{Klein2019, Rivera2021}. The low threshold for D is consistent with the low implantation level. The lifetime of carriers was measured from time resolved PL. The decay of the population of the excited states  contributing to the D-peak after a pulsed excitation can be seen in figure \ref{fig:TRPL}. As can be expected from measuring an ensemble of emitters, the decay is not a single exponential. The very fast initial decay of population by about 10\%, which is faster than the time resolution of the experiment, is followed by a slower decay with
the 1/$e$ decay time of around 14 ns. These decay times are 2-3 orders of magnitude longer than the lifetime of free excitons in MoSe$_2$ MLs \cite{Wang2015, Fang2019, Selig2016, Robert2016} and one order of magnitude longer than that from the isolated, confined excitons \cite{Yu2021}, pointing to a low oscillator strength of the emitters.

\begin{figure}[h]
\begin{subfigure}[c]{.52\textwidth}
    \caption{}
    \vspace{-10pt}
    \hspace{-10pt}
    \centering
    \includegraphics[width=\textwidth]{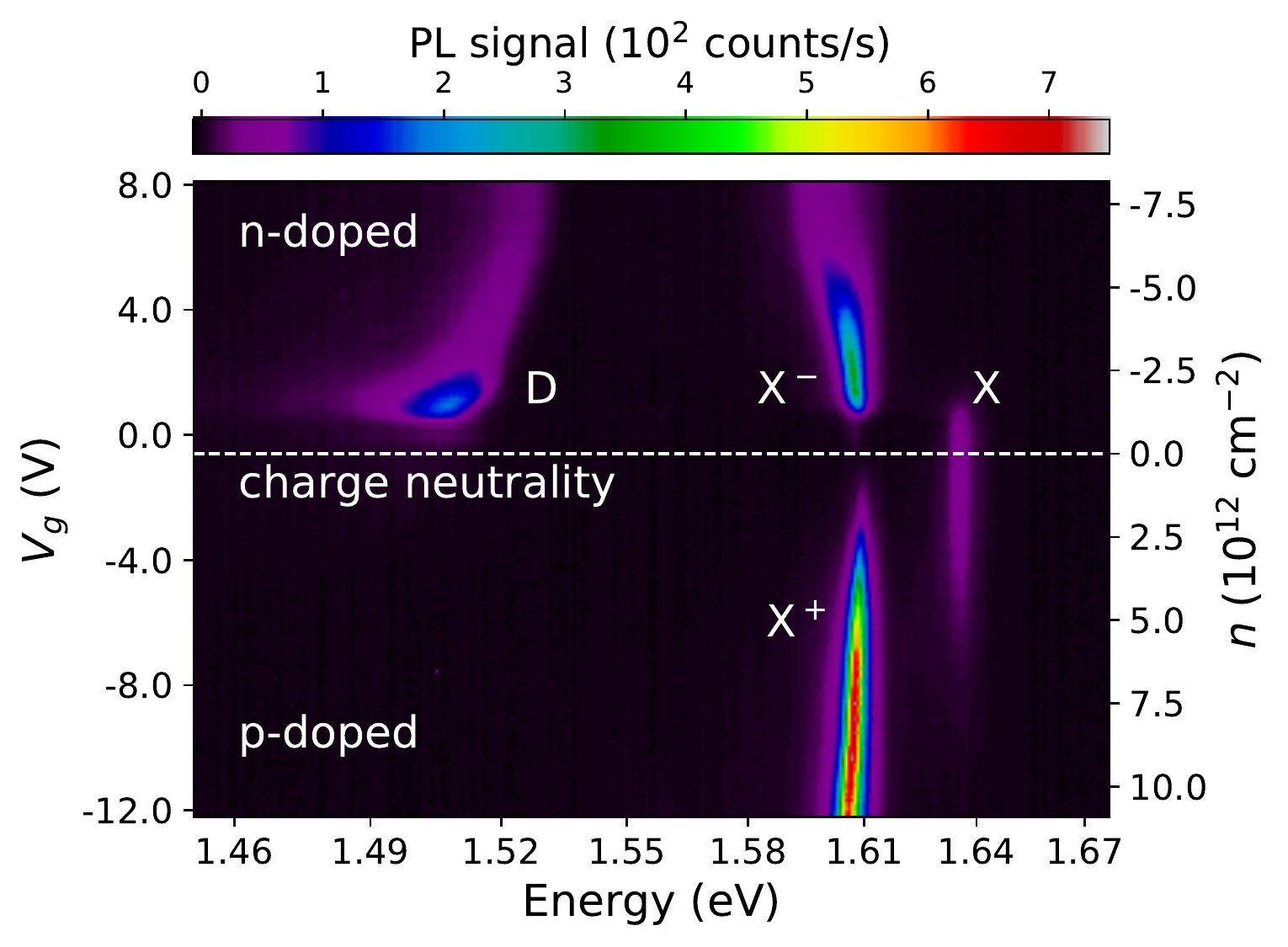}
    \label{fig:gate dependence}
\end{subfigure}
\begin{subfigure}[c]{0.47\textwidth}
    \caption{}
    \vspace{-5pt}
    \hspace{-5pt}
    \centering
    \includegraphics[width=\textwidth]{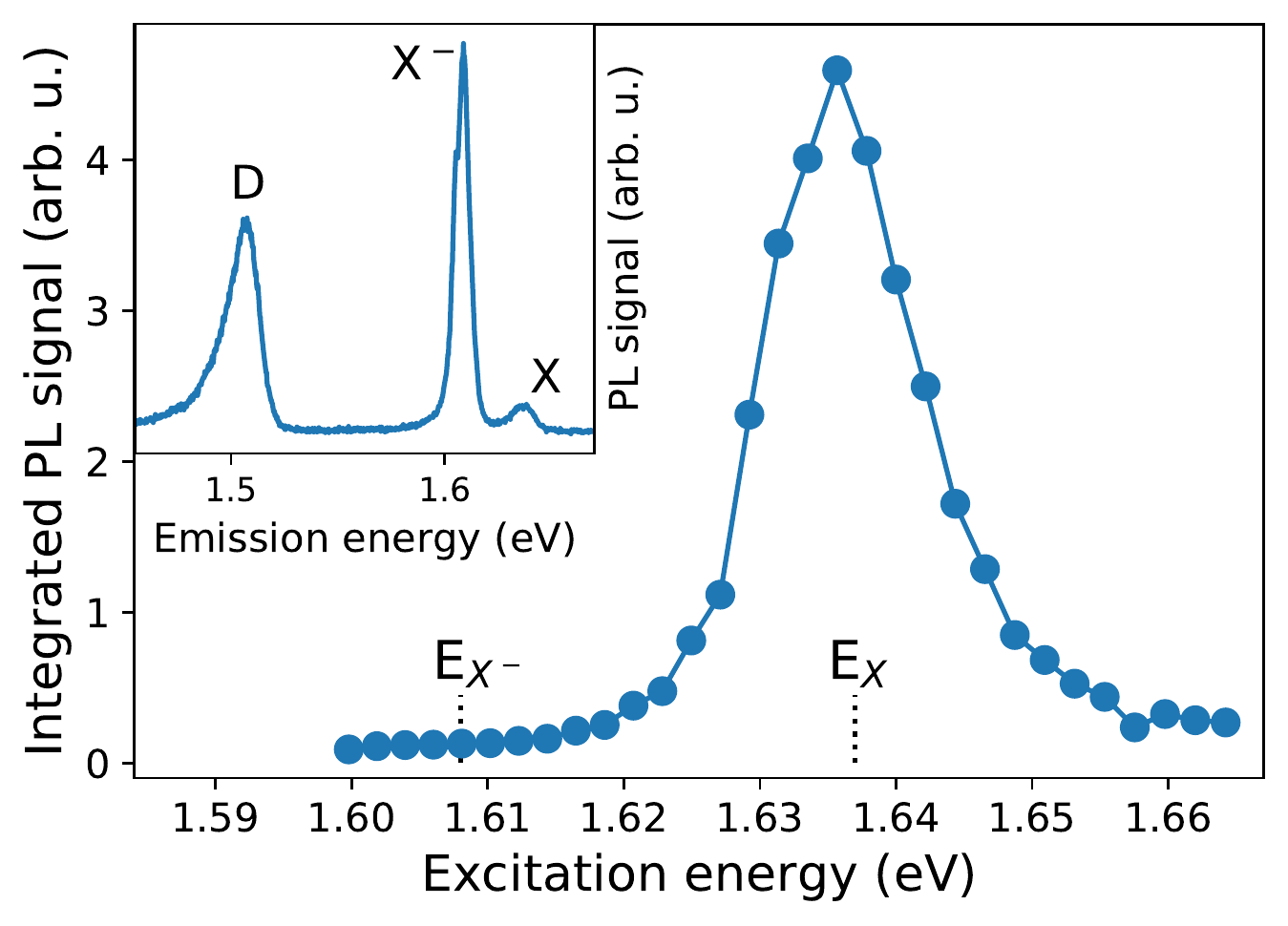}
    \label{fig:PLE}
\end{subfigure}
\caption{\textbf{Doping level and excitation energy dependencies of PL.}
(a) Gate dependent PL, where the back gate voltage $V_g$ was varied from -12 to 8 V to tune the doping level in the ML from p- via neutral to n-doping. The carrier concentration $n$ is calculated using the simple parallel plate capacitor model (more details in supplementary note). (b) PLE of D peak, taken at $V_g$ = 0.7 V. The D peak intensity was integrated around its PL emission energy between 1.48 and 1.52 eV. Inset: PL spectrum under 688\,nm (1.80\,eV) excitation under the gate voltage as applied for the PLE measurements.}
\end{figure}

The doping-dependent PL from neutral and charged excitons shown in figure \ref{fig:gate dependence} is typical of MoSe$_2$ MLs. When increasing the gate voltage, the D emission only emerges after the signal from the positive trion, X$^+$, entirely disappears. It reaches the maximum intensity around 1\,V, the gate voltage of the transition between X and X$^-$ dominated spectra. The D line weakens strongly as the X$^-$ line intensifies with a further gate voltage increase. This behaviour suggests a competition between the exciton capture at the defect state and the formation of an X$^-$. This scenario is supported by PL excitation (PLE) spectroscopy measurement (figure \ref{fig:PLE}), which shows that the intensity of D emission is maximum when the excitation wavelength is resonant with the energy of X around 1.637 eV. D emission was not excited with laser resonant with X$^-$ (around 1.608 eV) even though the PL spectrum, measured at the same doping level, clearly shows that the ML is doped with electrons (X$^-$ emission is the strongest PL signal). 

\begin{figure}[h]
\begin{subfigure}[c]{0.50\textwidth}
    \caption{}
    \vspace{-10pt}
    \hspace{-10pt}
    \centering
    \includegraphics[trim={10 0 10 0},clip,width=\textwidth]{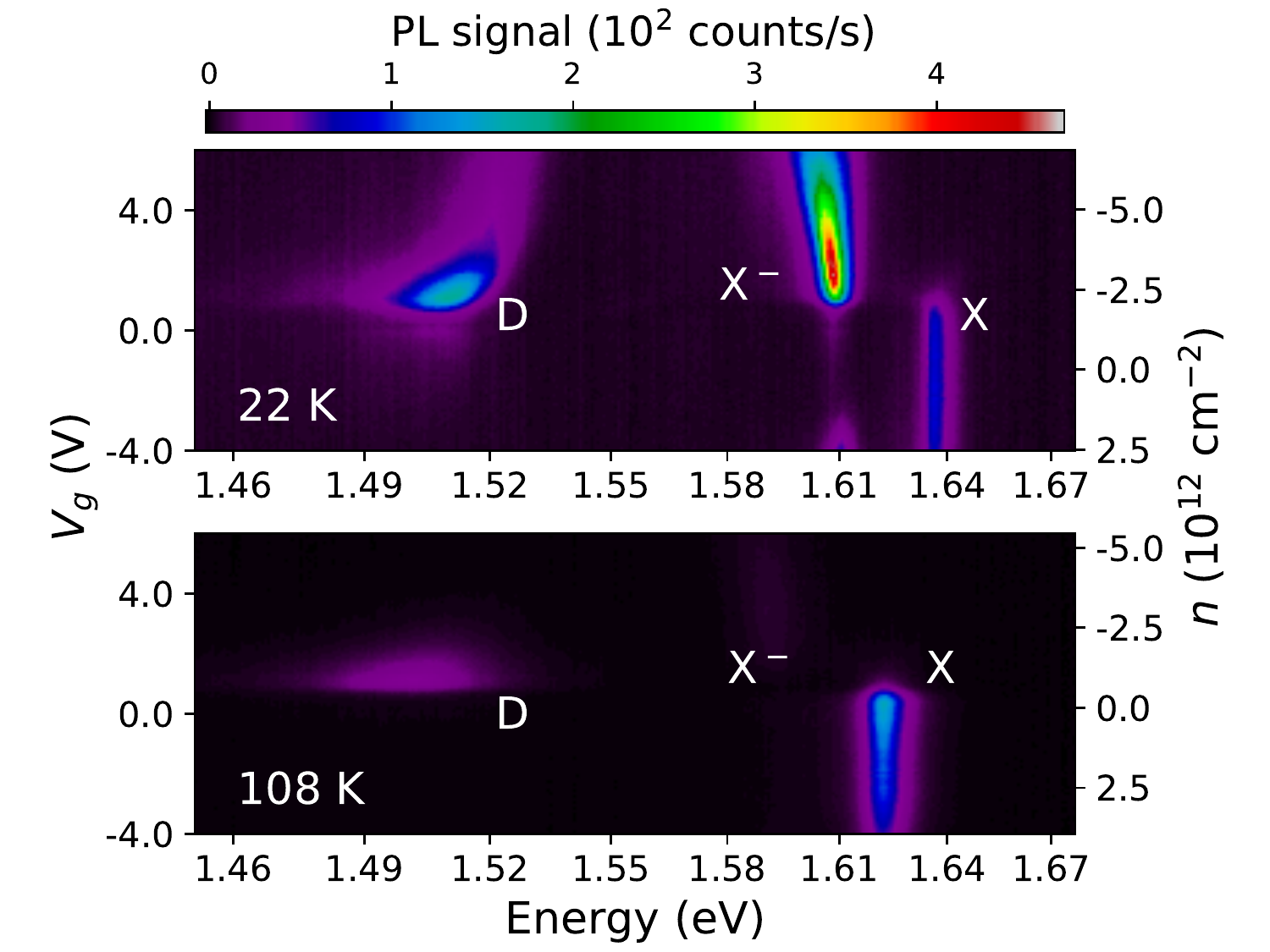}
    \label{fig:gate scan 22 & 108 K}
\end{subfigure}
\begin{subfigure}[c]{0.2475\textwidth}
    \caption{}
    \vspace{-0pt}
    \hspace{-10pt}
    \centering
    \includegraphics[trim={10 0 10 0},clip,width=\textwidth]{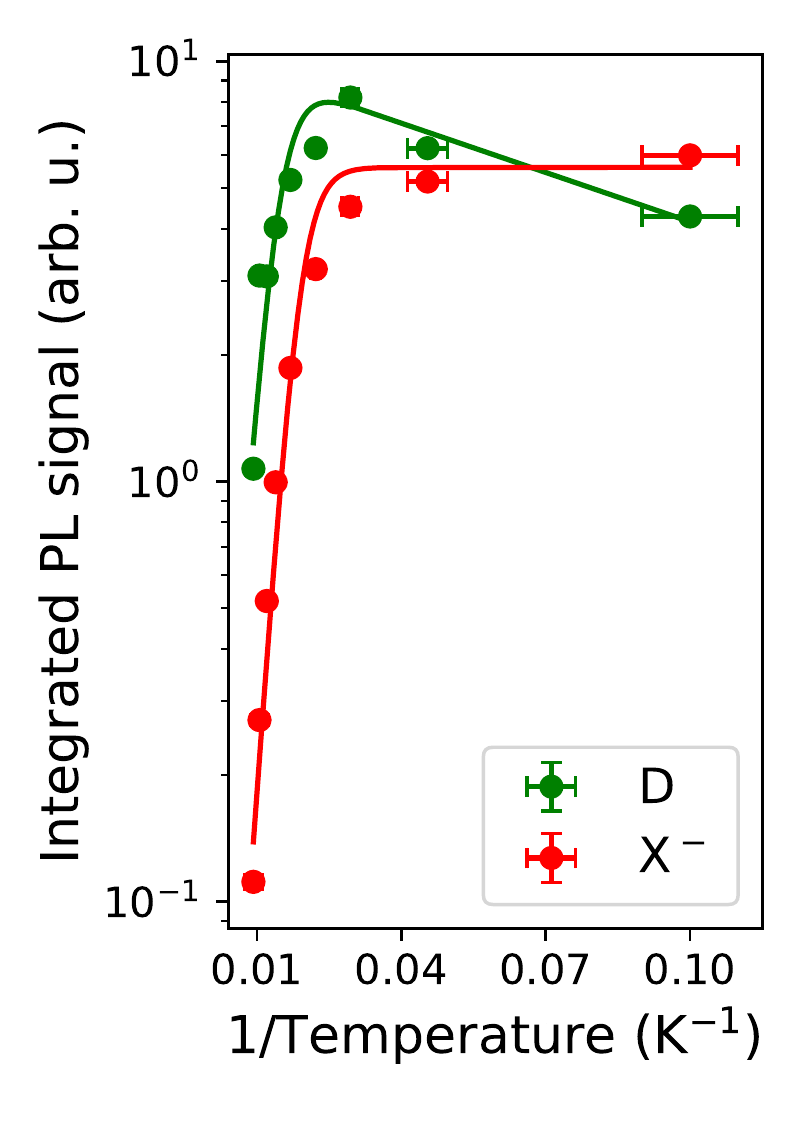}
    \label{fig:Arrhenius}
\end{subfigure}
\begin{subfigure}[c]{0.235\textwidth}
    \caption{}
    \vspace{-0pt}
    \hspace{-10pt}
    \centering
    \includegraphics[trim={5 0 5 0},clip,width=\textwidth]{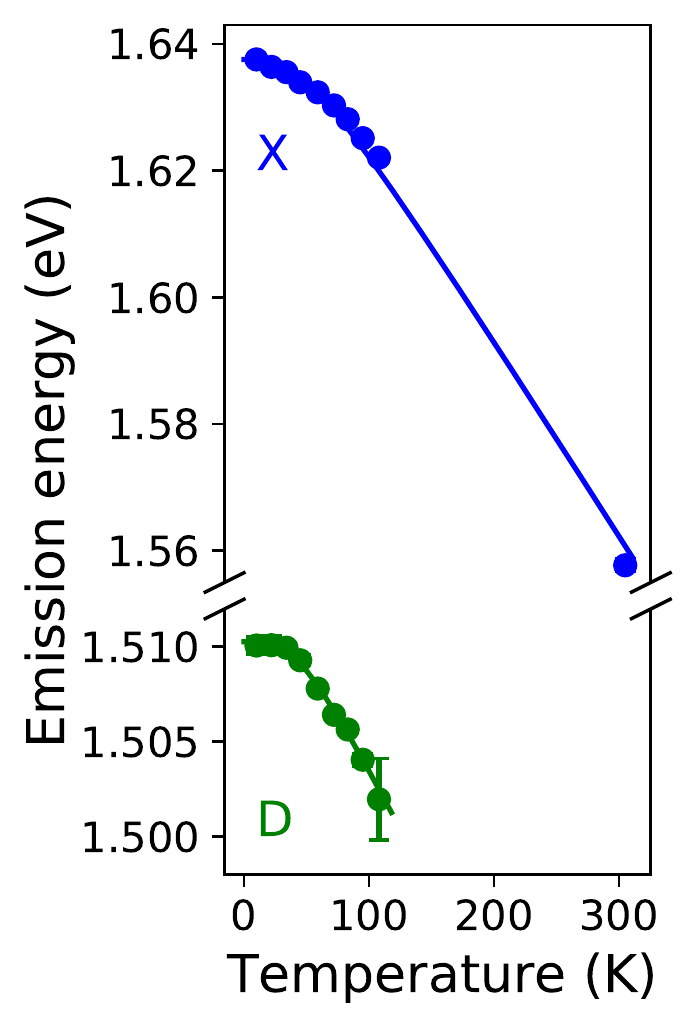}
    \label{fig:bandgap}
\end{subfigure}
\caption{\textbf{Temperature dependence of PL emission.}
(a) Gate voltage-dependent PL at 22 K and 108 K. (b) Arrhenius plot of X$^-$ and D (symbols). At each temperature, the two peaks' integrated intensities were acquired at the doping levels, where each emission is the brightest. The intensities were then normalised to that of X (at the voltage where X is most intense). Lines: the best-fit line was according to equations (\ref{eqn:quench1}) and (\ref{eqn:quench2}). (c) The temperature-dependent bandgap of X and D (symbols). Line: the best-fit line was according to equation (\ref{eqn:bandgap})
}
\label{fig:Temp dependence}
\end{figure}

Figure \ref{fig:gate scan 22 & 108 K} compares gate-dependent PL at  22\,K and 108\,K.  While X$^-$ emission is the most intense line in the spectra at the lower temperature, it is very weak at the higher temperature. X becomes the strongest line, but D diminished less than X$^-$ and remains up to room temperature (supplementary figure \ref{fig:RT PL}). 
We trace the change of PL signal counts from X$^-$ and D, both normalised to X signal counts, on the Arrhenius plot shown in figure \ref{fig:Arrhenius}. X$^-$ dissociates into higher energy X and an electron at higher temperature. Its intensity can be fitted with the standard Arrhenius formula \cite{Reshchikov2018, Reshchikov2021, Huang2016}:
\begin{equation}
    I(T) = \frac{I(0)}{1+A \exp(\frac{-E_a}{k_BT})}
    \label{eqn:quench1}
\end{equation}
where $I(0)$ is the PL intensity at temperature 0\,K, $A$ is a proportionality constant, $E_a$ is the activation energy for the dissociation of X$^-$, and $k_B$ is the Boltzmann constant. Fitting the formula to the data gives $E_a$ $\approx$ 32\,$\pm$\,5\,meV, which is expected for a binding energy of the trion \cite{Mostaani17, Szyniszewski2017}. D emission intensity first increased with temperature up to around 34\, K before diminishing. To account for this initial increase in intensity, we assume that trapping of carriers that recombine requires overcoming activation energy. We use a modified multi-level model for the temperature dependence of D intensity \cite{Shibata1998, Watanabe2006, Lin2012, Tangi2017, Huang2016}:
\begin{equation}
    I(T) = I(0)\frac{1+A_1\exp(\frac{-E_{a1}}{k_BT})}{1+A_2\exp(\frac{-E_{a2}}{k_BT})}
    \label{eqn:quench2}
\end{equation}
where $A_1$, $A_2$ are proportionality constants, and $E_{a1}$ and $E_{a2}$ are activation energies for trapping and detrapping of carriers. Fitting of the D peak yields $E_{a1}$ $\approx$ 1\,$\pm$\,6\,meV, and $E_{a2}$ $\approx$ 30\,$\pm$\,9\,meV. 

Temperature affects not only the intensity but also the D emission energy. The temperature-dependent energy shift of X and D lines can be described by the modified Varshni relation \cite{ODonnell1991, Li2017, Klein2019} as
\begin{equation}
    E_g(T) = E_g(0) - S \langle \hbar \omega \rangle [\mathrm{coth} \frac{\langle \hbar \omega \rangle }{2k_BT} - 1]
    \label{eqn:bandgap}
\end{equation}

where $E_g(0)$ is the emission energy at 0\,K, $S$ is the electron-phonon coupling, and $\langle \hbar \omega \rangle$ is the average phonon energy. Fitting gives $\langle \hbar \omega \rangle$ = 11.1\,$\pm$\,1.3 meV and $S$ = 1.82\,$\pm$\,0.14 for X, similar to the reported values ($\langle \hbar \omega \rangle$ $\approx$ 12 - 20\,meV, $S$ $\approx$ 2 \cite{Tongay2012, Ross2013, Li2017, Choi2017, Kioseoglou2016}). The fitted $\langle \hbar \omega \rangle$ and $S$ for D emission are smaller, at 10.7\,$\pm$\,1.5 meV and 0.79\,$\pm$\,0.09, respectively. 
Smaller $S$ constant compared to excitonic lines has been reported for vacancy-induced PL emissions from TMD MLs \cite{Klein2019, Mitterreiter2021, Parto2021} and explained as a result of the defect being decoupled from the conduction band, which varies with the temperature. Similar scenario is also likely to be the case in our sample.

\begin{figure}[h]
\begin{subfigure}[c]{.35\textwidth}
 \caption{}
 \vspace{-10pt}
 \hspace{-10pt}
    \centering
    \includegraphics[width=\textwidth]{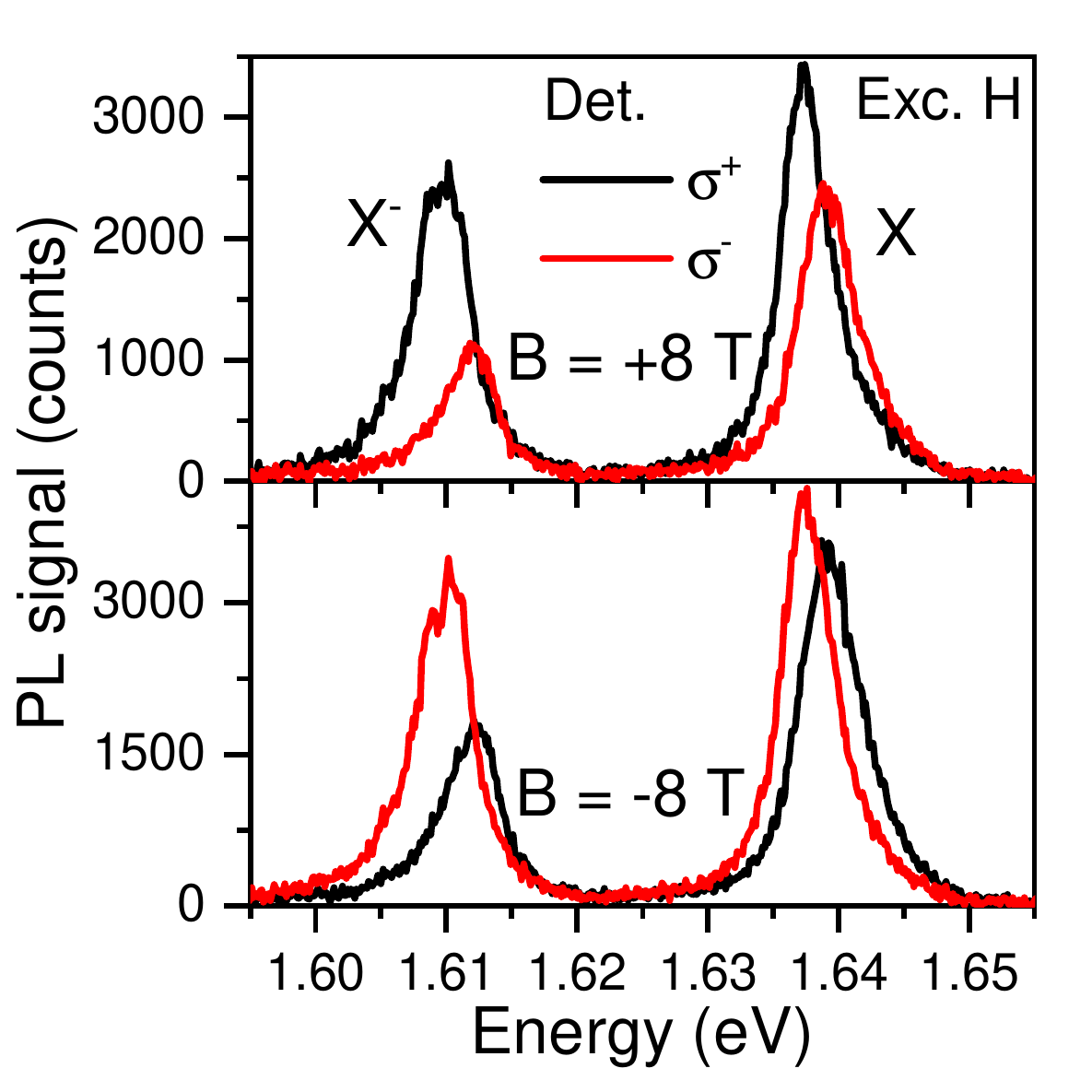}
    \label{fig:Zeeman X}
    \end{subfigure}
\begin{subfigure}[c]{.35\textwidth}
 \caption{}
 \vspace{-10pt}
 \hspace{-10pt}
    \centering
    \includegraphics[width=\textwidth]{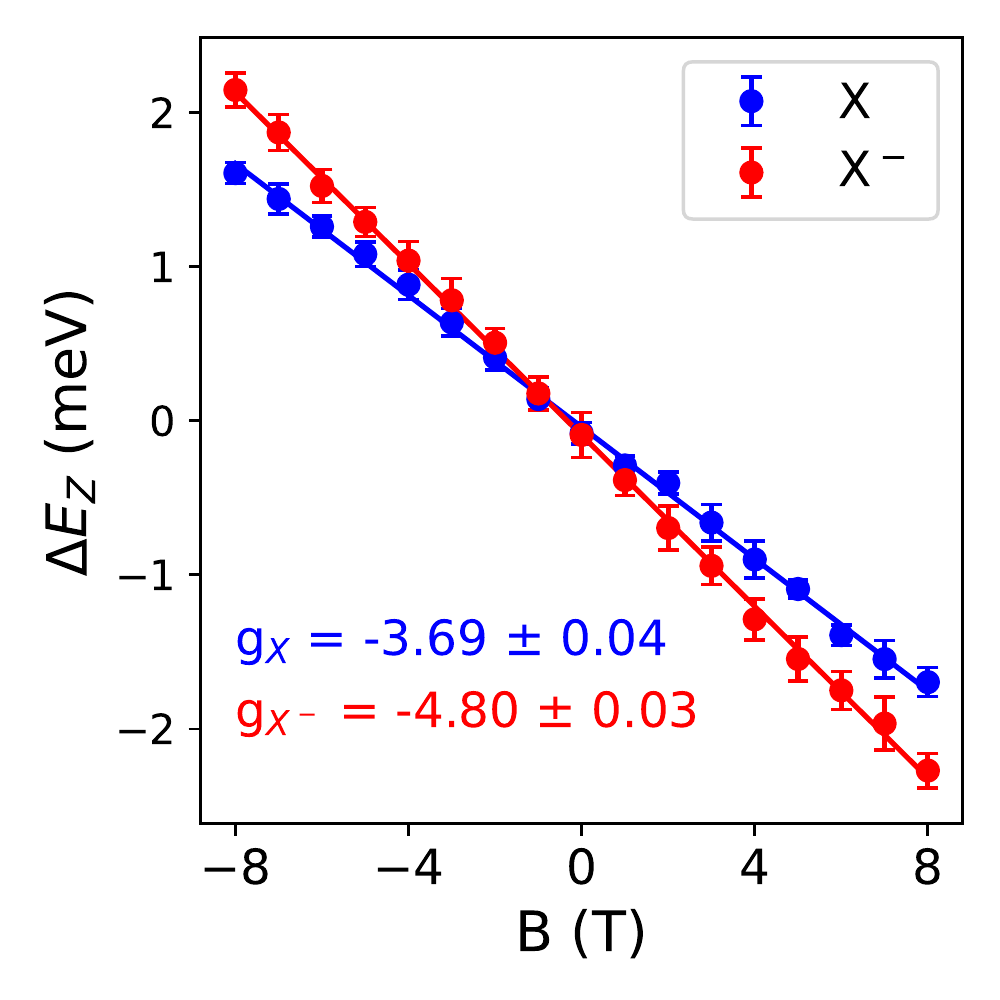}
    \label{fig:Zeeman X splitting}
    \end{subfigure}
\begin{subfigure}[c]{.35\textwidth}
 \caption{}
 \vspace{-5pt}
 \hspace{-10pt}
    \centering
    \includegraphics[width=\textwidth]{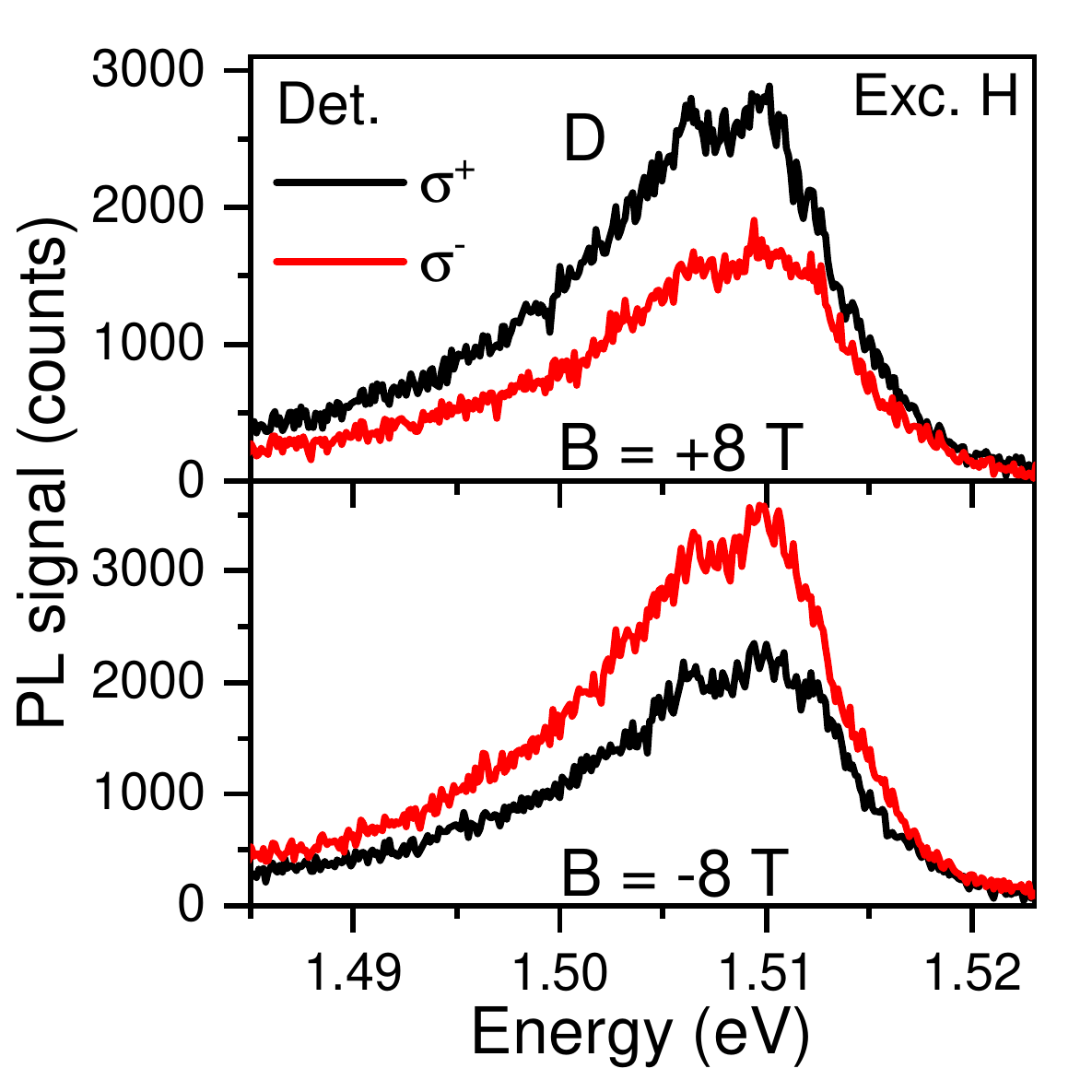}
    \label{fig:Zeeman D}
    \end{subfigure}
\begin{subfigure}[c]{.35\textwidth}
 \caption{}
 \vspace{-10pt}
 \hspace{-10pt}
    \centering
    \includegraphics[width=\textwidth]{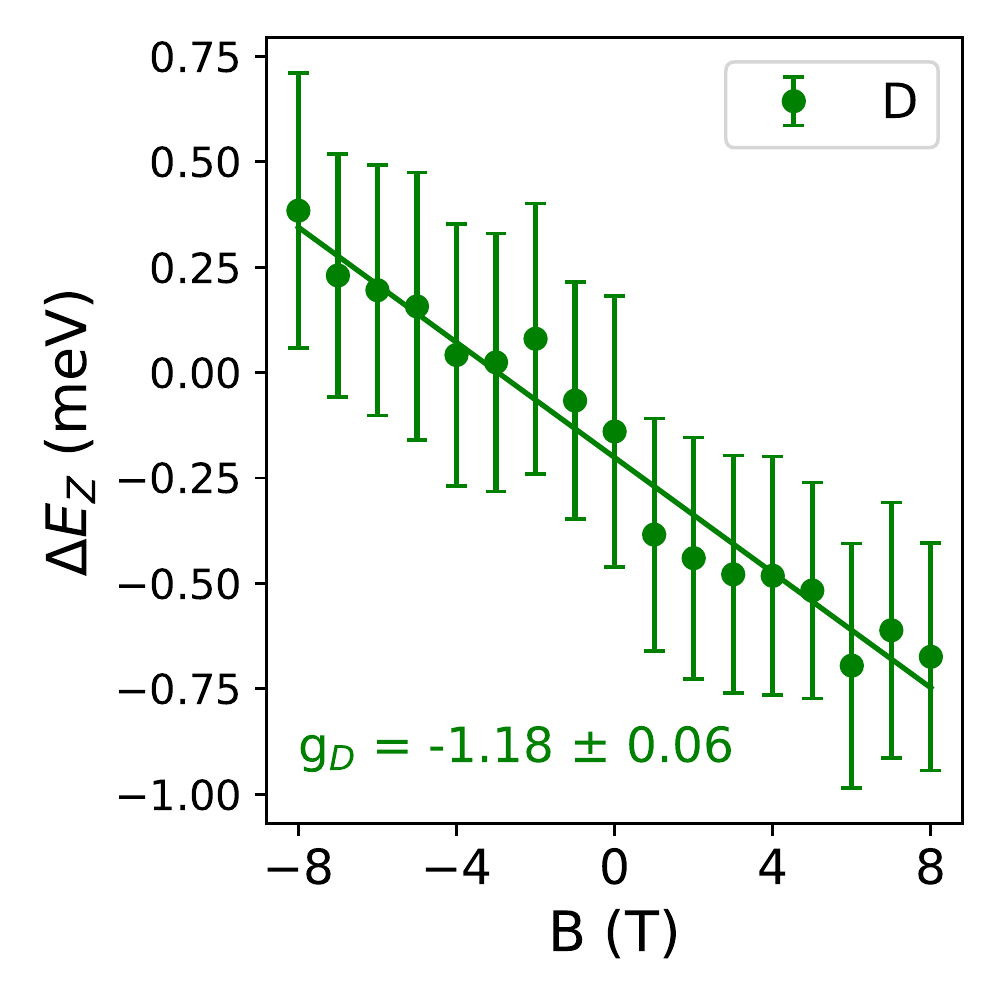}
    \label{fig:Zeeman D splitting}
    \end{subfigure}
\caption{\textbf{Magneto-PL measurement of Cr-implanted MoSe$_2$ ML at 1.8 K.} PL spectra acquired with out-of-plane magnetic field $B$ (varying between -8 and 8 T) applied to the sample and excited with an H-polarised laser. The detection is set to measure either $\upsigma^+$ (black) or $\upsigma^-$ (red) polarisation states. The figure shows the polarisation-resolved PL spectra of (a) X and X$^-$, (c) D, with their Zeeman splitting $\Delta E_Z$ shown in (b) and (d). The splitting was calculated from the peak positions (extracted from fitting Voigt functions to the PL spectra), with the error bars representing the propagated standard deviation of the fit procedure. The Zeeman splitting of the three emissions reveals expected $g$-factors around -4 for X and X$^-$, only around -1.18 for D.}
\label{fig:Zeeman}
\end{figure}

To gain further insight, we measured the PL emission from the sample under out-of-plane magnetic field $B$ varying from -8 to 8 T. Figures \ref{fig:Zeeman X} show the splitting of X and X$^-$ spectra in two circular polarisation detection states under applied $B$-field. The valley splitting, caused by the Zeeman effect (\cite{Srivastava2015, Wang2015b, Koperski2019}) and defined as
\begin{equation}
    \Delta E_Z = E_{\sigma^+} - E_{\sigma^-} = g \mu_B B 
    \label{eqn:Zeeman}
\end{equation}
(where $E_{\sigma^+}$ and $E_{\sigma^-}$ are the emission energy in the detected circular polarisation basis $\sigma^+$ and $\sigma^-$ respectively, $g$ is the Land\'e $g$-factor, $\mu_B$ is the Bohr magneton) changes linearly with the applied magnetic field (figures \ref{fig:Zeeman X splitting}). The 
$g$-factors derived from the data are  -3.69\,$\pm$\,0.04 and -4.80\,$\pm$\,0.03, for X and X$^-$, respectively. The $g$-factor value for X is close to the ones from previous experimental work \cite{Koperski2019, Li2014, MacNeill2015, Back2017, Wang2015b, Goryca2019, Cadiz2017} which are between -3.8 and -4.3  and well within the expected range from -3.22 to -3.82 predicted by recent ab initio calculations \cite{Deilmann2020,Wozniak2020}. The $g$-factor for X$^-$ is slightly higher than for X but similar to the values observed for samples under higher doping level \cite{Li2014, MacNeill2015, Back2017}. On the other hand, D emission shows little change with the magnetic field (figure \ref{fig:Zeeman D}). Comparing the energy of photons from D peak in both circular polarisation gives $g$-factor of about -1.18\,$\pm$\,0.06. The peak position was determined by fitting the data with three Voigt functions and then taking the maximum of the fitted line. The uncertainty here is high, partly owing to the D peak's large width.

\FloatBarrier

\subsection{First-principles molecular dynamics simulation of Cr ion implantation into MoSe$_2$ ML}

\begin{figure}[h]
    \centering
    \includegraphics[width=0.6\textwidth]{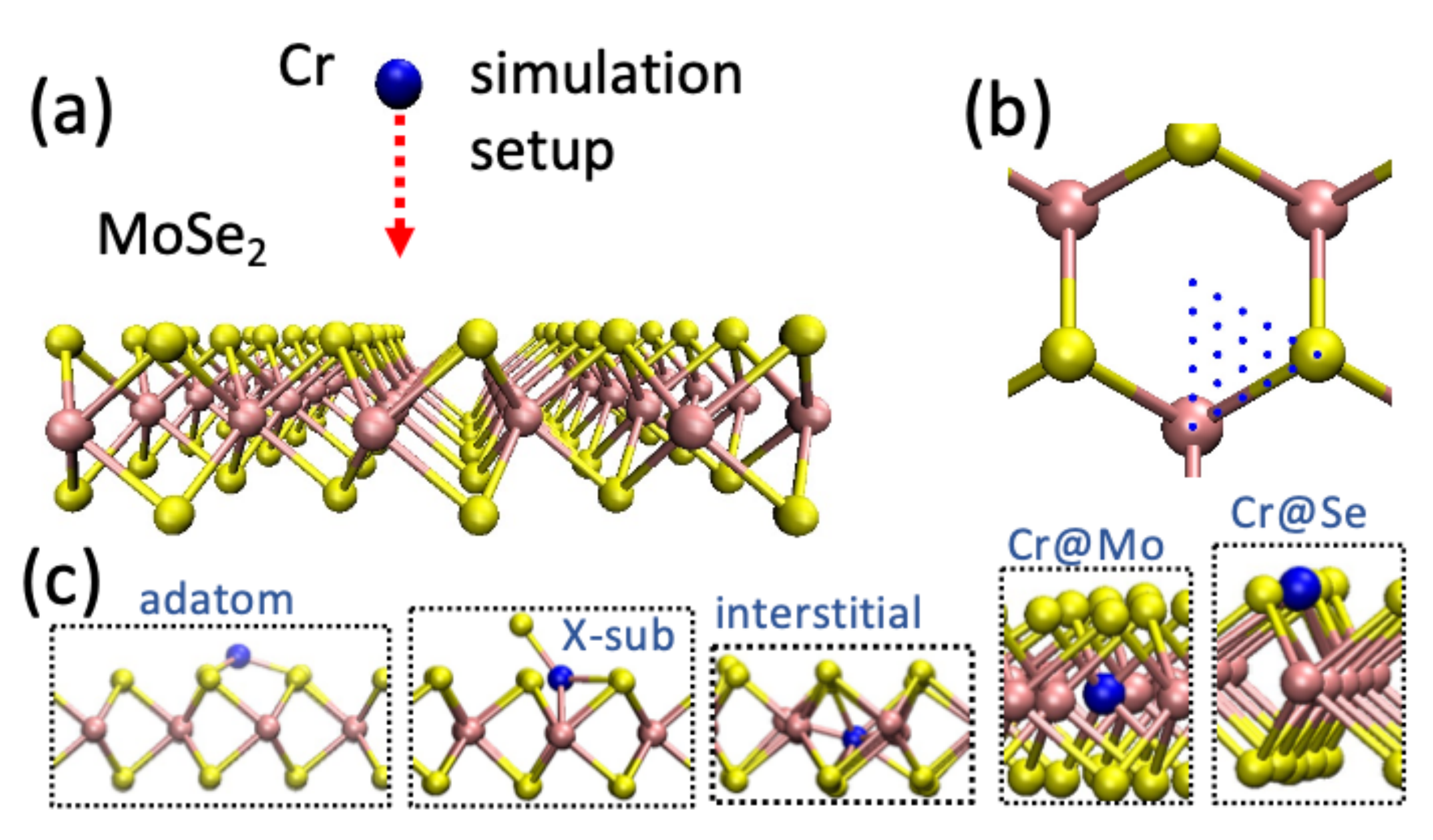}
\caption{\textbf{First-principles MD simulation of ion implantation process into ML MoSe$_2$.} (a) The setup for simulations of ion impacts. (b) Impact sites used in the simulation. (c) Atomic structures of the defects likely to appear upon impacts of energetic Cr ions.}
\label{fig:simulation}
\end{figure}

To get insights into the defect formation process and types of defects which can 
appear upon impacts of energetic Cr ions, we carried out DFT MD simulations, as described below. 
The atomic structure of a free-standing MoSe$_2$ rectangular slab containing 90 atoms
was fully optimised, then a Cr atom was placed 6 \AA~ above its surface, figure \ref{fig:simulation}(a)
and kinetic energy of 25 eV was assigned to the atom. Normal incidence was simulated; that is, the initial velocity vector of the projectile was oriented perpendicular to the surface of the ML.
The projectile was assumed to be a neutral atom, as at such low energies and low charge states, its neutralisation must
occur well before it reaches the surface. We note that DFT MD on the Born-Oppenheimer surface cannot describe the evolution of charge transfer anyway, and the Ehrenfest 
dynamics \cite{Gruber-2016,Ojanpera-2014} should be used. 21 impact points were selected in the
irreducible area of the primitive cell of MoSe$_2$, figure \ref{fig:simulation}(b), 
and the outcomes of the simulations were averaged with the corresponding weights.  The MD runs
continued until the kinetic energy brought up by the projectile was distributed
over the whole supercell (normally after a few picoseconds), then the system's temperature
was quenched to zero, and the atomic structure was analysed. Although the effects of substrate
on defect generation in a 2D system can be significant for ions with much higher (keV range) energies
\cite{Kalbac-2013,Kretschmer-2018,Standop13}, the role of the substrate should be minimal for 
impacts of 25 eV Cr ions onto MoSe$_2$, 
so that a free-standing slab was simulated. Spin-polarised calculations were carried out.
Although computationally more efficient non-spin-polarised method with a correction for isolated atom polarisation
energies can be used to simulate irradiation effects \cite{Kretschmer2022}, the account for spin effects is
particularly important for Cr, as it is magnetic, which affects the energetics of defect configurations.
 
Figure \ref{fig:simulation}(c) shows the most common atomic configurations which
appear after Cr atom impacts. These are Cr adatoms, X-sub configuration (Cr at Se sites with a Se adatom \cite{Karthikeyan19}), interstitials
(the Cr atom between Mo atoms) and substitutional defects in Mo and Se sites, which are Cr@Mo and Cr@Se respectively.
Table \ref{tab:MDsim} lists the probabilities for the defects to appear.
Ion irradiation also gives rise to the sputtering of Se atoms, that is the formation of
Se vacancies (V$_\text{Se}$), but these events were not so common.

\begin{table}[h]
\caption{\textbf{Results of DFT MD simulations of 25 eV Cr ion irradiation on single layer MoSe$_2$.} 
The probabilities $p$ of likely defect configurations to appear along with the formation energies $E_f$ of
these configurations are listed. }
\centering
\begin{tabular}{lrr}
                   & $p$ & $E_f$ [eV]  \\
adatom             & 0.16 & -0.85         \\
X-sub              & 0.41 & -0.76        \\
interstitial       & 0.08 & -0.25      \\
Cr@Mo              & 0.21 &  3.03          \\
Cr@Se              & 0.04 &  2.57         \\
V$_\text{Se}$      & 0.01 &  5.41           \\
passed through     & 0.09 & 0.00            \\                  
\end{tabular}
\label{tab:MDsim}
\end{table}

According to the DFT MD simulations, the most probable defects which
appear upon 25 eV Cr ion irradiation are Cr adatoms, Cr@Mo, and X-sub defects.
The Cr atoms which pass through the MoSe$_2$ sheet will likely form
adatoms attached to the bottom of MoSe$_2$. 
Self-annealing of defects at finite temperatures at which irradiation was carried out
in the experiment
can affect their concentrations in the implanted samples. To
get insight into the possible evolution of defects, we assessed the
defect formation energies $E_f$, as done previously \cite{Karthikeyan19}.
For adatoms, interstitials and X-sub defects, $E_f$ was calculated as the
energy difference between the system with Cr atom and the pristine system plus isolated
Cr atom. For the Cr@Mo, Cr@Se and V$_\text{Se}$ configurations,
the energies of isolated Mo and Se atoms were also taken as a reference.
We note that the listed defect formation energies for the Cr@Mo, Cr@Se, and V$_\text{Se}$ cannot be used to assess the equilibrium concentrations of these defects, as the chemical potentials were chosen to match isolated, that is sputtered, atoms. This can be done, though, if the chemical potentials of the displaced Se and Mo atoms are chosen in such a way that they reflect the actual experimental conditions that the potential can be anywhere between the values corresponding to the Se or Mo-rich limits. This would result in lower formation energies, as the sputtered atoms would be incorporated in the lattice. It can also be assumed that the displaced Se atoms form Se clusters at the surface, which would give rise the lowering of Cr@Se defect energies. 

As evident from Table \ref{tab:MDsim}, $E_f$ for adatoms is lower than for the interstitials, 
so that at finite temperatures, the interstitials will most likely be 'pushed away' from the Mo plane and form adatoms.
We note that this result was obtained for a relatively small 90-atom supercell, and 
in the larger system, the difference between these energies is smaller, as reported earlier \cite{Karthikeyan19}. Nevertheless, even for equal formation energies at zero temperature, with account for the entropic term in the Gibbs energy, the probabilities for the adatoms should be higher due to a  larger configurational space.
Some X-sub defects may also be converted to Cr@Se configurations, especially in the Mo-rich limit, when
Se vacancies are present, but the energetics of this process naturally depends on the experimental conditions, that is, the choice of Se chemical potential.
The Cr@Se defects can also appear due to the adsorption of Cr atoms on Se vacancies,
as this is energetically favourable due to the saturation of dangling bonds. 
Thus one can expect that the most prolific defects in the samples are Cr adatoms 
(or Cr clusters on top of MoSe$_2$), X-sub, as well as Cr@Mo and Cr@Se substitutional
configurations.

\subsection{DFT calculations of optical properties}

\begin{figure}[h]
  \centering
    \includegraphics[width=0.55\textwidth]{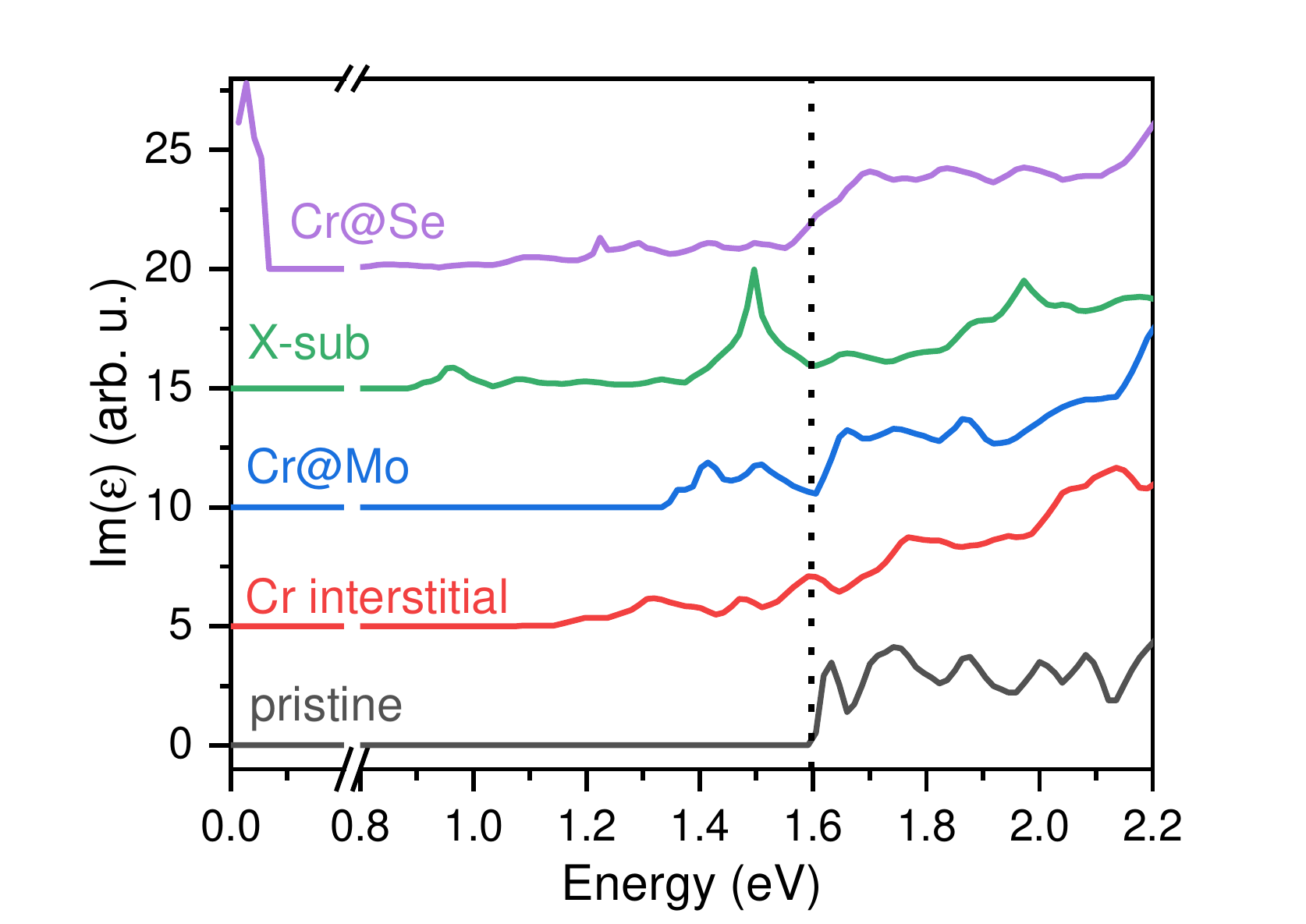}
  \caption{\textbf{Comparison of calculated absorption functions}
  for pristine MoSe$_2$ ML (black line) and MoSe$_2$ ML with Cr in various positions of the crystal structure: Cr at interstitial position (red), Cr at Mo position (Cr@Mo - blue), Cr at Se position with additional Se adatom (X-sub - green) and Cr at Se position (Cr@Se - purple). Spectra are offset vertically for clarity. The vertical dotted line at about 1.6 eV marks the calculated bandgap of pristine MoSe$_2$ ML.
}
  \label{fig:mose2_cr_compare_imag_multiple_pristine}
 \end{figure}

To investigate if Cr defects introduce states in the bandgap of the MoSe$_2$ ML that are optically active we have simulated optical absorption spectra for MoSe$_2$ with Cr defects in various positions. The simulations are based on a 5x5 supercell. Each supercell hosts one Cr defect.
The unfolded band structures are shown in the supplementary figure \ref{fig:DFT_band}. All defects 
give rise to states in the bandgap.
The absorption spectrum $\Im[\varepsilon(\omega)]$ with the energy dependent macroscopic dielectric function $\varepsilon(\omega)$ has been calculated within the random-phase approximation in the limit $\mathbf{k}\rightarrow\mathbf{0}$. Optical matrix elements and local-field effects are taken into account.
It should be pointed out that self-energy corrections (such as $GW$) or electron-hole interactions (as described by the Bethe-Salpeter equation) are neglected. Self-energy corrections and electron-hole interactions are known to have a partially compensating effect on the bandgap \cite{Liu2015}: While the former tends to increase the bandgap, the electron-hole interactions make the optical bandgap smaller. Due to this compensating effect, the present theoretical results can be seen as approximate spectra specifically showing the impact of the defects. However, quantitative differences between theory and experiment should be expected due to the neglect of many-body effects and also due to the difference in the dielectric environment.
The theoretical spectra are shown in figure \ref{fig:mose2_cr_compare_imag_multiple_pristine}. Absorption below the bandgap (around 1.6\,eV) is present for all defects and originates from transitions involving the defect states.

There is an optical transition for X-sub at 1.5\,eV, which is in the same energy range of D emission from the PL spectra, between the valence band and an acceptor state of X-sub. This state results from the coupling of the conduction band at the K point with the Cr defect state. 
The energy of this transition is similar to that of the MoSe$_2$ ML with a vacancy \cite{Rost2023, Iberi2016, Samani2016, Shafqat2017}. 

The weak transition involving a deep acceptor state at 0.9\,eV is outside the spectral range of our experiments. We note that the coupling between the conduction band and the defect state shifts the conduction band minimum from K towards the $\Gamma$ point
(supplementary figure \ref{fig:DFT_band}). However, the resulting suppression of PL would not be visible in the experiment due to the low density of defects and only the local opening of the bandgap.

Well-defined spin-degenerate acceptor levels are also introduced by Cr substituting the Mo atom in the lattice (Cr@Mo). Optical transitions from this defect state into the valence band states can be seen in figure \ref{fig:mose2_cr_compare_imag_multiple_pristine} in the range between 1.4 and 1.5\,eV, which is also in a similar energy range to D peak PL emission. The band-to-band transition is shifted to higher energy compared to the pristine MoSe$_2$ ML because of the coupling between the conduction band and the defect state. However, similar to the X-sub configuration discussed above, it is unlikely to observe this blueshift in the PL spectra due to the low defect density. 

The coupling of the defect and conduction band results in a gradual increase of the above-bandgap absorption for Cr substitution into the Se site (Cr@Se). This defect type also  introduces a donor state at the Fermi level and two single-spin, deep defect levels. The signal from the donor state merges with the band-to-band absorption. Otherwise, the Cr defect at the Se site hardly affects the MoSe$_2$ band structure. Several weak optical transitions are present at a large range of energies (down to 500 meV below the bandgap).

Interstitial Cr introduces several deep defect levels in the bandgap, and again the highest state couples to the conduction band shifting the conduction band minimum to the $\Lambda$ point. The absorption spectrum does not contain discrete absorption lines but a gradually increasing absorption from 1.2\,eV.

\section{Discussion}

Radiative recombination of an electron (e$^-$) bound to a defect state with the valence band hole (h$^+$) can explain the measured PL. Considering that our DFT calculations do not show donor states at high enough energy, the electron here is likely to occupy an acceptor. In this scenario, an exciton bound to a negatively charged acceptor (A$^-$X) dissociates into  A$^-$h$^+$ and a free electron in the conduction band. Following radiative recombination, A$^-$h$^+$ becomes neutral acceptor A$^0$. Theoretical modelling of A$^-$X indicated a binding energy of only a few meV compared with A$^0$\,+\,e$^-$ state \cite{Mostaani17}, which is of the same order of magnitude as the activation energy of the D-line determined from the Arrhenius plot.  Among the potential defects identified by MD calculations, Cr@Mo, X-sub, and Cr@Se have non-zero matrix elements for optical transitions between acceptor states and valence band. Other configurations, e.g. interstitial Cr or Se vacancies, are unlikely to be present. Besides, neither would explain the data (see figure \ref{fig:mose2_cr_compare_imag_multiple_pristine} and supplementary information note 4). 

The measured 1/$e$ recombination time is longer than the lifetimes reported for band-to-band and localised states recombination in MoSe$_2$. Low oscillator strength of the transition can result from the spatial separation of electrons and holes, as for Cr@Se or X-sub. However, since this lifetime is longer than that of the spin dark states in WSe$_2$, which is only a few ns \cite{Tang19}, this transition could also be from a spin forbidden state. Such a state would correspond to the charge configuration of A$^0$ for Cr@Se defect in the absence of exchange interactions between electrons in the conduction band.

The $g$-factor of D emission is negative but much smaller than the ones for X or X$^-$. With large $g$-factors for electrons in the valence band, it implies either a large $g$-factor for an electron on the acceptor level near the conduction band (e.g. in Cr@Mo or X-sub configuration) and valley-selective transitions or reduced $g$-factor of holes in the valence band. The latter could be caused by the hybridisation of the valence band with the defect level as in the Cr@Se configuration. Further insight would require higher magnetic field measurements and theoretical input.

\section{Conclusion}

In conclusion, we demonstrated the ultra-low energy ion implantation of Cr ions (at 25\,eV) into a MoSe$_2$ ML. The implantation was performed with ions at 25\,eV and the ion fluence of 3\,$\times$\,10$^{12}$ cm$^{-2}$, the resultant material retains high optical quality as evidenced by clear excitonic PL. Implanted Cr ions introduce an additional low energy PL signal at around 1.51 eV visible at the onset of n-doping. Molecular dynamics calculations identified defects that can be generated by implantation. We found that Cr atoms can substitute for both Mo and Se atoms. In the latter case, the Cr atom is slightly more likely to bind an additional Se atom than not. The defects' stability, including interstitial Cr, depends on the post-implantation treatment and the final configuration of Se and Mo, which are not in the lattice. DFT calculations revealed that all the probable defects introduce one or more defect states in the MoSe$_2$ bandgap with non-zero matrix elements for optical transitions. It is impossible to identify with certainty which defect is the origin of D-line, Cr at Se site with Se adatom (X-sub), and perhaps Cr at Mo (Cr@Mo) seem to fit best with the measured data. Further experiments, for example, implantation of Cr only into the Se sub-lattice or implantation through the hBN protective layer to avoid environmental changes, could be considered to distinguish between the cases. 

More generally, this study shows that while implantation of heavier elements into metal sub-lattice of TMD MLs is possible without a visible loss of the material quality, the implantation process is complex, and simulations of the possible outcomes are necessary to identify material systems of the desired properties. In the search for single photon emitting sites, it is also worth noting that upon implantation with a very low fluence, it should be possible to address individual Cr atoms at different lattice sites.

\section{Methods}

\subsection{Atomistic simulation}

We used DFT-MD as implemented in the
VASP code \cite{Kresse1993, Kresse1996}. 
The Perdew-Burke-Ernzerhof (PBE) exchange and correlation functional was employed \cite{Perdew1996_PRL}.
The evolution of the system
was modelled using the microcanonical ensemble. A cutoff value of
300\,eV was chosen for DFT MD, and sampling over the Brillouin zone was done using a $3\times3\times1$ 
k-point mesh.
The time step was chosen to be 0.1\,fs, which provided energy conservation better than 0.1\,eV.

\subsection{Band structure and absorption spectra calculation}
Density functional theory (DFT) simulations were performed in supercells of 5$\times$5 primitive unit cells. Each constructed with lattice constants of $a=3.28\,\si{\angstrom}$ and $c=12.918\,\si{\angstrom}$ of the hexagonal lattice. An internal structure parameter of $z=0.125$ was used. The defect systems were spatially relaxed using FLEUR~\cite{flapw} until the residual atomic forces had fallen below 5~$\times$~10$^{-2}$\,eV/$\si{\angstrom}$. The subsequent calculation of the macroscopic dielectric function in SPEX~\cite{Friedrich2010, Friedrich2021} is based on the random-phase approximation~\cite{Adler1962,Wiser63} and includes local-field effects. Calculations of 2D materials with 3D periodic boundary conditions are computationally expensive because the decoupling of neighbouring layers in the $z$ direction requires large supercells in this direction. In the case of 2D systems with defects, the computational cost grows considerably, particularly in the case of low defect concentrations, because suppressing the unwanted defect-defect coupling requires large supercells in the $x$ and $y$ directions. To facilitate the calculations of the dielectric function, we had to reduce the reciprocal cutoff radius from $4.1$ to $3.6$~Bohr$^{-1}$ in the case of the X-sub defect system. However, this should not affect the form of the respective spectrum shown in figure~\ref{fig:mose2_cr_compare_imag_multiple_pristine}.
\\

The band structures presented in the Supplementary Information 
are made up of 320 $\mathbf{k}$ points along the unfolded high-symmetry path $\Gamma-\text{M}-\text{K}-\Gamma$. Here, "unfolded" 
means that the high-symmetry points refer to the ones of the defect-free MoSe$_2$ ML. The necessary unfolding of the band structures of the defect systems has been carried out with a new implementation~\cite{Rost2023} in the FLEUR code adapting the technique described in Ref.~\cite{Rubel2014} to the LAPW basis~\cite{Anderson1975}. In this technique, a spectral weight is assigned to each state plotted in the band structure. The weight 
$w_n(\mathbf{k})$ for the $n$-th state at $\mathbf{k}$ of the unfolded path is given by
\begin{align}
    w_n(\mathbf{k})=\sum_{\tilde{\mathbf{G}}, \mathbf{G}'} C^{*}_{\mathbf{k}'n}({\tilde{\mathbf{G}}}) \cdot C_{\mathbf{k}'n}(\mathbf{G}') \cdot S_{{\tilde{\mathbf{G}}}\mathbf{G}'}(\mathbf{k}')
\end{align}
where $\mathbf{k}'=\mathbf{k}+\mathbf{G}''$ with a suitable reciprocal lattice vector $\mathbf{G}''$ that folds $\mathbf{k}$ back into the (smaller) Brillouin zone of the defect system. The $\mathbf{G}'$ sum runs over the set of all reciprocal lattice vectors (of the defect system) at $\mathbf{k}'$, and the $\tilde{\mathbf{G}}$ sum runs over the set of reciprocal lattice vectors (of the pristine system) at $\mathbf{k}$. The latter is a subset of the former.
The wave functions are represented in the LAPW basis $\{\chi_{\mathbf{kG}}(\mathbf{r})\}$ with coefficients $C_{\mathbf{k}n}(\mathbf{G})$ and overlap matrix $S_{\mathbf{G}\mathbf{G}'}(\mathbf{k})=\bra{\chi_{\mathbf{k}\mathbf{G}}}\ket{\chi_{\mathbf{k}\mathbf{G}'}}$ ~\cite{Rost2023}.

\subsection{Sample preparation}
Si with 90 nm thick dry-thermally grown SiO$_2$ chips with 60 nm thick Ti/Au contacts (pre-patterned by electron beam lithography) were used as the substrate. Before flake transfer, the chips were cleaned in acetone and isopropanol (IPA) under bath sonication, blown dry with N$_2$ and treated with oxygen plasma (300 W, 200 sccm for 10 minutes). Few-layer graphite, MoSe$_2$ (from 2D Semiconductors) MLs, and hBN (from Takashi Taniguchi and Kenji Watanabe) multilayers were mechanically exfoliated from bulk crystal using polydimethylsiloxane (PDMS) stamps (Gel-pak DGL X4 films) and transferred onto the substrate using dry viscoelastic transfer process \cite{Castellanos_Gomez2014}. The process was performed in a N$_2$ filled glovebox. After transferring the graphite (5.5 nm thick) and bottom-hBN (20 nm thick) flakes, the sample was annealed in H$_2$/Ar (1:10 ratio) atmosphere at 300 $\degree$C for 3 hours to improve the top surface for the subsequent MoSe$_2$ ML transfer. After transferring the top-hBN (15 nm thick), the sample was annealed in low vacuum (5\,$\times$\,10$^{-3}$ mbar) at 200 $\degree$C for 2 hours to improve interfaces in the vdW stack. Electrical contacts, provided by the Ti/Au lines, were made to the MoSe$_2$ flake and the graphite back gate. After each transfer, the heterostructure surface was checked with atomic force microscopy to ensure a sufficiently flat area in the stack and to obtain the flakes' thickness. MoSe$_2$ ML's quality was confirmed with Raman and PL spectroscopy at room temperature \cite{Tonndorf2013}.

\subsection{Ion implantation}

Bronze tips were used to fix the sample on a holder, making contact with the sample's Au pads and, thus, the ML. To remove volatile contamination from the sample, the sample chamber was then evacuated to 10$^{-9}$ mbar for several hours. The sample was heated to 150 $\degree$C for 10 minutes to remove residual volatile adsorbates, then to 220 $\degree$C during the implantation.
A foil is used as the feedstock to provide $^{52}$Cr$^+$ ions. After extraction, the ions are decelerated from 30 keV to 25 eV directly in front of the sample. Since the deceleration voltage is set relative to the potential of the source anode, this energy represents the upper limit, with a tail towards lower energies. The fluence of the ions was set to 3\,$\times$\,10$^{12}$ cm$^{-2}$. The fluence was verified by test implantations using Rutherford backscatter spectrometry (more information in Supplementary information \ref{section:RBS}). A detailed description of the source and the implantation system can be found in the references \cite{Junge2022, Auge2022}.

\subsection{Optical measurements}
PL spectroscopy was performed at 10 K (unless otherwise specified) in a He-cooled cold-finger cryostat (Cryoindustries) with a heating element (allowing a sample temperature range from 10 to 300 K). For PL measurements, the laser beam - 688 nm (1.80 eV) from a Ti:Sa laser - is passed through a 680\,$\pm$\,5 nm band-pass filter before being focused by an aspheric lens (NA = 0.47) into a spot of 1.6 $\upmu$m in diameter on the sample. Unless otherwise specified, the laser power on the sample was at 1 $\upmu$W for PL experiments. PL signal is collected by the same lens and passed through a 700-nm low pass filter before being focused by an achromatic doublet (NA = 0.24) through the entrance slit of a Czerny-Turner spectrometer, dispersed by a 600 l/mm grating onto a CCD camera. For gate dependence and temperature dependence PL, the laser power was kept at 1 $\upmu$W. For PLE, excitation power ranged from 2 to 5 $\upmu$W, and PL intensity is normalised to the power density for final data. 

For time-resolved PL, the excitation was done using a pulsed laser at 660 nm (1.88 eV) with 200ps pulse length, 2.5 MHz repetition rate and 2.8 $\upmu$W average power. The PL signal is directed through an 800 nm (1.55\,eV) low-pass filter on the detection path before entering an avalanche photodiode with 30\,ps time resolution. The histogram of the time difference between the laser pulses and PL emission was acquired with a time tagger.

The sample was mounted on an x-y-z Attocube stage in a He flow cryostat (attoDRY2100) for magneto-optics measurement at temperature T = 1.8 K. A magnetic field up to $\pm$8 T was applied perpendicularly to the sample (Faraday configuration). The excitation laser beam (688 nm, i.e. 1.80 eV at 4 $\upmu$W) was passed through a 680 nm bandpass filter and a linear polariser in an H-configuration. The laser was focused by an aspheric lens (NA = 0.47) into a spot of $\approx$ 1.6 $\upmu$m in diameter on the sample. The emitted PL was collected by the same objective. It was passed through a combination of $\uplambda$/4, $\uplambda$/2 waveplates, and a linear polariser set to pass  $\upsigma^{\pm}$ polarised light. It was then propagated via a single mode optical fiber towards the entrance slit of a Czerny-Turner spectrometer, where it was dispersed by a 600 l/mm grating onto a CCD camera. A long pass filter (with 700 nm band edge) was inserted between the fiber output and the spectrometer entrance to remove any remaining laser light.

\section*{Data availability}
The data supporting the findings of this study are available within the paper and its supplementary information files. Data are also available from the corresponding author upon reasonable request.

\section*{Code availability}
All scripts used to generate the results in this study are available from the corresponding author upon reasonable request.

\section*{Acknowledgement}

This project is supported by the “Integration of Molecular Components in Functional Macroscopic Systems” initiative of Volkswagen Foundation.
We would like to thank the staff at the Helmholtz Nano Facility \cite{HNF} of Forschungszentrum J\"{u}lich for helping with substrate fabrication, and Felix Junge (II. Institute of Physics, University of Göttingen, Göttingen, Germany) for organizing the RBS data.
We acknowledge the computing time granted through JARA-HPC on the supercomputer JURECA at Forschungszentrum J\"{u}lich. 
A.V.K. acknowledges funding from the German Research Foundation (DFG), project KR 4866/8-1 and the collaborative research center “Chemistry of Synthetic 2D Materials” SFB-1415-417590517.
Generous grants of computer time from the Technical University of Dresden computing cluster (TAURUS) and the High Performance Computing Center (HLRS) in Stuttgart, Germany, are gratefully appreciated.
K.W. and T.T. acknowledge support from the JSPS KAKENHI (Grant Numbers 19H05790 and 20H00354).

\section*{Author contribution}
B.E.K., H.C.H., and M.N.B conceived and designed the experiments. S.K. and A.V.K. performed MD simulations. S.R., C.F. and S.B. performed DFT calculations for band structures and absorption spectra. T.T. and K.W. grew hBN crystals. L.Z. processed the Si/SiO$_2$ substrate with patterned markers and metal contacts. M.N.B. prepared the samples on Si/SiO$_2$ substrate. M.A. and H.C.H. performed ion implantation. M.N.B. acquired and analysed PL and Raman data. All authors discussed the results and contributed to the writing of this manuscript.

\section*{Competing interests}
The authors declare that there are no competing interests.

\bibliography{Main_text.bib}

\providecommand{\latin}[1]{#1}
\makeatletter
\providecommand{\doi}
  {\begingroup\let\do\@makeother\dospecials
  \catcode`\{=1 \catcode`\}=2 \doi@aux}
\providecommand{\doi@aux}[1]{\endgroup\texttt{#1}}
\makeatother
\providecommand*\mcitethebibliography{\thebibliography}
\csname @ifundefined\endcsname{endmcitethebibliography}
  {\let\endmcitethebibliography\endthebibliography}{}
\begin{mcitethebibliography}{88}
\providecommand*\natexlab[1]{#1}
\providecommand*\mciteSetBstSublistMode[1]{}
\providecommand*\mciteSetBstMaxWidthForm[2]{}
\providecommand*\mciteBstWouldAddEndPuncttrue
  {\def\EndOfBibitem{\unskip.}}
\providecommand*\mciteBstWouldAddEndPunctfalse
  {\let\EndOfBibitem\relax}
\providecommand*\mciteSetBstMidEndSepPunct[3]{}
\providecommand*\mciteSetBstSublistLabelBeginEnd[3]{}
\providecommand*\EndOfBibitem{}
\mciteSetBstSublistMode{f}
\mciteSetBstMaxWidthForm{subitem}{(\alph{mcitesubitemcount})}
\mciteSetBstSublistLabelBeginEnd
  {\mcitemaxwidthsubitemform\space}
  {\relax}
  {\relax}

\bibitem[Dietl and Ohno(2014)Dietl, and Ohno]{Dietl2014}
Dietl,~T.; Ohno,~H. Dilute ferromagnetic semiconductors: Physics and spintronic
  structures. \emph{Rev. Mod. Phys.} \textbf{2014}, \emph{86}, 187--251\relax
\mciteBstWouldAddEndPuncttrue
\mciteSetBstMidEndSepPunct{\mcitedefaultmidpunct}
{\mcitedefaultendpunct}{\mcitedefaultseppunct}\relax
\EndOfBibitem
\bibitem[Ikezawa \latin{et~al.}(2012)Ikezawa, Sakuma, Zhang, Sone, Mori,
  Hamano, Watanabe, Sakoda, and Masumoto]{Ikezawa2012}
Ikezawa,~M.; Sakuma,~Y.; Zhang,~L.; Sone,~Y.; Mori,~T.; Hamano,~T.;
  Watanabe,~M.; Sakoda,~K.; Masumoto,~Y. Single-photon generation from a
  nitrogen impurity center in GaAs. \emph{Applied Physics Letters}
  \textbf{2012}, \emph{100}, 042106\relax
\mciteBstWouldAddEndPuncttrue
\mciteSetBstMidEndSepPunct{\mcitedefaultmidpunct}
{\mcitedefaultendpunct}{\mcitedefaultseppunct}\relax
\EndOfBibitem
\bibitem[Niaouris \latin{et~al.}(2022)Niaouris, Durnev, Linpeng, Viitaniemi,
  Zimmermann, Vishnuradhan, Kozuka, Kawasaki, and Fu]{Vasileios22}
Niaouris,~V.; Durnev,~M.~V.; Linpeng,~X.; Viitaniemi,~M. L.~K.; Zimmermann,~C.;
  Vishnuradhan,~A.; Kozuka,~Y.; Kawasaki,~M.; Fu,~K.-M.~C. Ensemble spin
  relaxation of shallow donor qubits in ZnO. \emph{Phys. Rev. B} \textbf{2022},
  \emph{105}, 195202\relax
\mciteBstWouldAddEndPuncttrue
\mciteSetBstMidEndSepPunct{\mcitedefaultmidpunct}
{\mcitedefaultendpunct}{\mcitedefaultseppunct}\relax
\EndOfBibitem
\bibitem[Yamamoto \latin{et~al.}(2009)Yamamoto, Ladd, Press, Clark, Sanaka,
  Santori, Fattal, Fu, Höfling, Reitzenstein, and Forchel]{Yamamoto09}
Yamamoto,~Y.; Ladd,~T.~D.; Press,~D.; Clark,~S.; Sanaka,~K.; Santori,~C.;
  Fattal,~D.; Fu,~K.~M.; Höfling,~S.; Reitzenstein,~S.; Forchel,~A. Optically
  controlled semiconductor spin qubits for quantum information processing.
  \emph{Physica Scripta} \textbf{2009}, \emph{2009}, 014010\relax
\mciteBstWouldAddEndPuncttrue
\mciteSetBstMidEndSepPunct{\mcitedefaultmidpunct}
{\mcitedefaultendpunct}{\mcitedefaultseppunct}\relax
\EndOfBibitem
\bibitem[Vandersypen \latin{et~al.}(2017)Vandersypen, Bluhm, Clarke, Dzurak,
  Ishihara, Morello, Reilly, Schreiber, and Veldhorst]{Vandersypen17}
Vandersypen,~L. M.~K.; Bluhm,~H.; Clarke,~J.~S.; Dzurak,~A.~S.; Ishihara,~R.;
  Morello,~A.; Reilly,~D.~J.; Schreiber,~L.~R.; Veldhorst,~M. Interfacing spin
  qubits in quantum dots and donors—hot, dense, and coherent. \emph{npj
  Quantum Information} \textbf{2017}, \emph{3}, 34\relax
\mciteBstWouldAddEndPuncttrue
\mciteSetBstMidEndSepPunct{\mcitedefaultmidpunct}
{\mcitedefaultendpunct}{\mcitedefaultseppunct}\relax
\EndOfBibitem
\bibitem[Sekiguchi \latin{et~al.}(2021)Sekiguchi, Tsurumoto, Koga, Reyes, and
  Kosaka]{Sekiguchi2021}
Sekiguchi,~Y.,~Yuheiand~Yasui; Tsurumoto,~K.; Koga,~Y.; Reyes,~R.; Kosaka,~H.
  Geometric entanglement of a photon and spin qubits in diamond.
  \emph{Communications Physics} \textbf{2021}, \emph{4}, 264\relax
\mciteBstWouldAddEndPuncttrue
\mciteSetBstMidEndSepPunct{\mcitedefaultmidpunct}
{\mcitedefaultendpunct}{\mcitedefaultseppunct}\relax
\EndOfBibitem
\bibitem[Metsch \latin{et~al.}(2019)Metsch, Senkalla, Tratzmiller, Scheuer,
  Kern, Achard, Tallaire, Plenio, Siyushev, and Jelezko]{Metsch2019}
Metsch,~M.~H.; Senkalla,~K.; Tratzmiller,~B.; Scheuer,~J.; Kern,~M.;
  Achard,~J.; Tallaire,~A.; Plenio,~M.~B.; Siyushev,~P.; Jelezko,~F.
  Initialization and Readout of Nuclear Spins via a Negatively Charged
  Silicon-Vacancy Center in Diamond. \emph{Phys. Rev. Lett.} \textbf{2019},
  \emph{122}, 190503\relax
\mciteBstWouldAddEndPuncttrue
\mciteSetBstMidEndSepPunct{\mcitedefaultmidpunct}
{\mcitedefaultendpunct}{\mcitedefaultseppunct}\relax
\EndOfBibitem
\bibitem[Ruf \latin{et~al.}(2021)Ruf, Wan, Choi, Englund, and Hanson]{Ruf2021}
Ruf,~M.; Wan,~N.~H.; Choi,~H.; Englund,~D.; Hanson,~R. Quantum networks based
  on color centers in diamond. \emph{Journal of Applied Physics} \textbf{2021},
  \emph{130}, 070901\relax
\mciteBstWouldAddEndPuncttrue
\mciteSetBstMidEndSepPunct{\mcitedefaultmidpunct}
{\mcitedefaultendpunct}{\mcitedefaultseppunct}\relax
\EndOfBibitem
\bibitem[Pezzagna and Meijer(2021)Pezzagna, and Meijer]{Pezzagna2021}
Pezzagna,~S.; Meijer,~J. Quantum computer based on color centers in diamond.
  \emph{Applied Physics Reviews} \textbf{2021}, \emph{8}, 011308\relax
\mciteBstWouldAddEndPuncttrue
\mciteSetBstMidEndSepPunct{\mcitedefaultmidpunct}
{\mcitedefaultendpunct}{\mcitedefaultseppunct}\relax
\EndOfBibitem
\bibitem[Wolfowicz \latin{et~al.}(2020)Wolfowicz, Anderson, Diler, Poluektov,
  Heremans, and Awschalom]{Wolfowicz2020}
Wolfowicz,~G.; Anderson,~C.~P.; Diler,~B.; Poluektov,~O.~G.; Heremans,~F.~J.;
  Awschalom,~D.~D. Vanadium spin qubits as telecom quantum emitters in silicon
  carbide. \emph{Science Advances} \textbf{2020}, \emph{6}, eaaz1192\relax
\mciteBstWouldAddEndPuncttrue
\mciteSetBstMidEndSepPunct{\mcitedefaultmidpunct}
{\mcitedefaultendpunct}{\mcitedefaultseppunct}\relax
\EndOfBibitem
\bibitem[Lohrmann \latin{et~al.}(2015)Lohrmann, Iwamoto, Bodrog, Castelletto,
  Ohshima, Karle, Gali, Prawer, McCallum, and Johnson]{Lohrmann2015}
Lohrmann,~A.; Iwamoto,~N.; Bodrog,~Z.; Castelletto,~S.; Ohshima,~T.;
  Karle,~T.~J.; Gali,~A.; Prawer,~S.; McCallum,~J.~C.; Johnson,~B.~C.
  Single-photon emitting diode in silicon carbide. \emph{Nature Communications}
  \textbf{2015}, \emph{6}, 7783\relax
\mciteBstWouldAddEndPuncttrue
\mciteSetBstMidEndSepPunct{\mcitedefaultmidpunct}
{\mcitedefaultendpunct}{\mcitedefaultseppunct}\relax
\EndOfBibitem
\bibitem[Ma \latin{et~al.}(2017)Ma, Yu, and Zhang]{Ma2017}
Ma,~J.; Yu,~Z.~G.; Zhang,~Y.-W. Tuning deep dopants to shallow ones in 2D
  semiconductors by substrate screening: The case of
  ${\mathrm{X}}_{\mathrm{S}}$ (X = Cl, Br, I) in ${\mathrm{MoS}}_{2}$.
  \emph{Phys. Rev. B} \textbf{2017}, \emph{95}, 165447\relax
\mciteBstWouldAddEndPuncttrue
\mciteSetBstMidEndSepPunct{\mcitedefaultmidpunct}
{\mcitedefaultendpunct}{\mcitedefaultseppunct}\relax
\EndOfBibitem
\bibitem[Mostaani \latin{et~al.}(2017)Mostaani, Szyniszewski, Price, Maezono,
  Danovich, Hunt, Drummond, and Fal'ko]{Mostaani17}
Mostaani,~E.; Szyniszewski,~M.; Price,~C.~H.; Maezono,~R.; Danovich,~M.;
  Hunt,~R.~J.; Drummond,~N.~D.; Fal'ko,~V.~I. Diffusion quantum Monte Carlo
  study of excitonic complexes in two-dimensional transition-metal
  dichalcogenides. \emph{Phys. Rev. B} \textbf{2017}, \emph{96}, 075431\relax
\mciteBstWouldAddEndPuncttrue
\mciteSetBstMidEndSepPunct{\mcitedefaultmidpunct}
{\mcitedefaultendpunct}{\mcitedefaultseppunct}\relax
\EndOfBibitem
\bibitem[Rivera \latin{et~al.}(2021)Rivera, He, Kim, Liu, Rubio-Verdú, Moon,
  Mennel, Rhodes, Yu, Taniguchi, Watanabe, Yan, Mandrus, Dery, Pasupathy,
  Englund, Hone, Yao, and Xu]{Rivera2021}
Rivera,~P. \latin{et~al.}  Intrinsic donor-bound excitons in ultraclean
  monolayer semiconductors. \emph{Nature Communications} \textbf{2021},
  \emph{12}, 871\relax
\mciteBstWouldAddEndPuncttrue
\mciteSetBstMidEndSepPunct{\mcitedefaultmidpunct}
{\mcitedefaultendpunct}{\mcitedefaultseppunct}\relax
\EndOfBibitem
\bibitem[Borghardt \latin{et~al.}(2020)Borghardt, Tu, Taniguchi, Watanabe, and
  Kardyna\l{}]{Borghardt2020}
Borghardt,~S.; Tu,~J.-S.; Taniguchi,~T.; Watanabe,~K.; Kardyna\l{},~B.~E.
  Interplay of excitonic complexes in $p$-doped ${\mathrm{WSe}}_{2}$
  monolayers. \emph{Phys. Rev. B} \textbf{2020}, \emph{101}, 161402\relax
\mciteBstWouldAddEndPuncttrue
\mciteSetBstMidEndSepPunct{\mcitedefaultmidpunct}
{\mcitedefaultendpunct}{\mcitedefaultseppunct}\relax
\EndOfBibitem
\bibitem[Klein \latin{et~al.}(2019)Klein, Lorke, Florian, Sigger, Sigl, Rey,
  Wierzbowski, Cerne, Müller, Mitterreiter, Zimmermann, Taniguchi, Watanabe,
  Wurstbauer, Kaniber, Knap, Schmidt, Finley, and Holleitner]{Klein2019}
Klein,~J. \latin{et~al.}  Site-selectively generated photon emitters in
  monolayer ${\mathrm{MoS}}_{2}$ via local helium ion irradiation. \emph{Nat.
  Commun.} \textbf{2019}, \emph{10}, 2755\relax
\mciteBstWouldAddEndPuncttrue
\mciteSetBstMidEndSepPunct{\mcitedefaultmidpunct}
{\mcitedefaultendpunct}{\mcitedefaultseppunct}\relax
\EndOfBibitem
\bibitem[Mitterreiter \latin{et~al.}(2021)Mitterreiter, Schuler, Micevic,
  Hernangómez-Pérez, Barthelmi, Cochrane, Kiemle, Sigger, Klein, Wong,
  Watanabe, Taniguchi, Lorke, Jahnke, Finley, Schwartzberg, Qiu,
  Refaely-Abramson, Holleitner, Weber-Bargioni, and Kastl]{Mitterreiter2021}
Mitterreiter,~E. \latin{et~al.}  The role of chalcogen vacancies for atomic
  defect emission in ${\mathrm{MoS}}_{2}$. \emph{Nat. Commun.} \textbf{2021},
  \emph{12}, 3822\relax
\mciteBstWouldAddEndPuncttrue
\mciteSetBstMidEndSepPunct{\mcitedefaultmidpunct}
{\mcitedefaultendpunct}{\mcitedefaultseppunct}\relax
\EndOfBibitem
\bibitem[Xu \latin{et~al.}(2017)Xu, Zhao, Lin, Long, Wang, Chan, and
  Chai]{Xu2017}
Xu,~K.; Zhao,~Y.; Lin,~Z.; Long,~Y.; Wang,~Y.; Chan,~M.; Chai,~Y. Doping of
  two-dimensional MoS$_2$ by high energy ion implantation. \emph{Semiconductor
  Science and Technology} \textbf{2017}, \emph{32}, 124002\relax
\mciteBstWouldAddEndPuncttrue
\mciteSetBstMidEndSepPunct{\mcitedefaultmidpunct}
{\mcitedefaultendpunct}{\mcitedefaultseppunct}\relax
\EndOfBibitem
\bibitem[He \latin{et~al.}(2019)He, Huang, Liou, Woon, and Su]{He2019}
He,~S.-M.; Huang,~C.-C.; Liou,~J.-W.; Woon,~W.-Y.; Su,~C.-Y. Spectroscopic and
  Electrical Characterizations of Low-Damage Phosphorous-Doped Graphene via Ion
  Implantation. \emph{ACS Applied Materials \& Interfaces} \textbf{2019},
  \emph{11}, 47289--47298, PMID: 31746197\relax
\mciteBstWouldAddEndPuncttrue
\mciteSetBstMidEndSepPunct{\mcitedefaultmidpunct}
{\mcitedefaultendpunct}{\mcitedefaultseppunct}\relax
\EndOfBibitem
\bibitem[Prucnal \latin{et~al.}(2021)Prucnal, Hashemi, Ghorbani-Asl, Hübner,
  Duan, Wei, Sharma, Zahn, Ziegenrücker, Kentsch, Krasheninnikov, Helm, and
  Zhou]{Prucnal2021}
Prucnal,~S.; Hashemi,~A.; Ghorbani-Asl,~M.; Hübner,~R.; Duan,~J.; Wei,~Y.;
  Sharma,~D.; Zahn,~D. R.~T.; Ziegenrücker,~R.; Kentsch,~U.;
  Krasheninnikov,~A.~V.; Helm,~M.; Zhou,~S. Chlorine doping of MoSe$_2$ flakes
  by ion implantation. \emph{Nanoscale} \textbf{2021}, \emph{13},
  5834--5846\relax
\mciteBstWouldAddEndPuncttrue
\mciteSetBstMidEndSepPunct{\mcitedefaultmidpunct}
{\mcitedefaultendpunct}{\mcitedefaultseppunct}\relax
\EndOfBibitem
\bibitem[Wang \latin{et~al.}(2021)Wang, Ho, Ho, Lu, Hsieh, Huang, Chiu, Chen,
  Chang, White, Tang, and Woon]{Wang2021}
Wang,~Y.-H.; Ho,~H.-M.; Ho,~X.-L.; Lu,~L.-S.; Hsieh,~S.-H.; Huang,~S.-D.;
  Chiu,~H.-C.; Chen,~C.-H.; Chang,~W.-H.; White,~J.~D.; Tang,~Y.-H.;
  Woon,~W.-Y. Photoluminescence Enhancement in WS$_2$ Nanosheets Passivated
  with Oxygen Ions: Implications for Selective Area Doping. \emph{ACS Applied
  Nano Materials} \textbf{2021}, \emph{4}, 11693--11699\relax
\mciteBstWouldAddEndPuncttrue
\mciteSetBstMidEndSepPunct{\mcitedefaultmidpunct}
{\mcitedefaultendpunct}{\mcitedefaultseppunct}\relax
\EndOfBibitem
\bibitem[Jadwiszczak \latin{et~al.}(2020)Jadwiszczak, Maguire, Cullen,
  Duesberg, and Zhang]{Jadwiszczak2020}
Jadwiszczak,~J.; Maguire,~P.; Cullen,~C.~P.; Duesberg,~G.~S.; Zhang,~H. Doping
  Graphene with Substitutional Mn. \emph{Beilstein J. Nanotechnol.}
  \textbf{2020}, \emph{11}, 1329–1335\relax
\mciteBstWouldAddEndPuncttrue
\mciteSetBstMidEndSepPunct{\mcitedefaultmidpunct}
{\mcitedefaultendpunct}{\mcitedefaultseppunct}\relax
\EndOfBibitem
\bibitem[Krasheninnikov(2020)]{Krasheninnikov2020}
Krasheninnikov,~A.~V. {Are two-dimensional materials radiation tolerant?}
  \emph{Nanoscale Horizons} \textbf{2020}, \emph{5}, 1447--1452\relax
\mciteBstWouldAddEndPuncttrue
\mciteSetBstMidEndSepPunct{\mcitedefaultmidpunct}
{\mcitedefaultendpunct}{\mcitedefaultseppunct}\relax
\EndOfBibitem
\bibitem[Kretschmer \latin{et~al.}(2022)Kretschmer, Ghaderzadeh, Facsko, and
  Krasheninnikov]{Kretschmer2022}
Kretschmer,~S.; Ghaderzadeh,~S.; Facsko,~S.; Krasheninnikov,~A.~V. Threshold
  Ion Energies for Creating Defects in 2D Materials from First-Principles
  Calculations: Chemical Interactions Are Important. \emph{The Journal of
  Physical Chemistry Letters} \textbf{2022}, \emph{13}, 514--519, PMID:
  35005978\relax
\mciteBstWouldAddEndPuncttrue
\mciteSetBstMidEndSepPunct{\mcitedefaultmidpunct}
{\mcitedefaultendpunct}{\mcitedefaultseppunct}\relax
\EndOfBibitem
\bibitem[Auge \latin{et~al.}(2022)Auge, Junge, and Hofsäss]{Auge2022}
Auge,~M.; Junge,~F.; Hofsäss,~H. Laterally controlled ultra-low energy ion
  implantation using electrostatic masking. \emph{Nuclear Instruments and
  Methods in Physics Research Section B: Beam Interactions with Materials and
  Atoms} \textbf{2022}, \emph{512}, 96--101\relax
\mciteBstWouldAddEndPuncttrue
\mciteSetBstMidEndSepPunct{\mcitedefaultmidpunct}
{\mcitedefaultendpunct}{\mcitedefaultseppunct}\relax
\EndOfBibitem
\bibitem[Junge \latin{et~al.}(2022)Junge, Auge, and Hofsäss]{Junge2022}
Junge,~F.; Auge,~M.; Hofsäss,~H. Sputter hot filament hollow cathode ion
  source and its application to ultra-low energy ion implantation in 2D
  materials. \emph{Nuclear Instruments and Methods in Physics Research Section
  B: Beam Interactions with Materials and Atoms} \textbf{2022}, \emph{510},
  63--68\relax
\mciteBstWouldAddEndPuncttrue
\mciteSetBstMidEndSepPunct{\mcitedefaultmidpunct}
{\mcitedefaultendpunct}{\mcitedefaultseppunct}\relax
\EndOfBibitem
\bibitem[Lin \latin{et~al.}(2021)Lin, Villarreal, Achilli, Bana, Nair, Tejeda,
  Verguts, De~Gendt, Auge, Hofsäss, De~Feyter, Di~Santo, Petaccia, Brems,
  Fratesi, and Pereira]{Lin2021}
Lin,~P.-C. \latin{et~al.}  Doping Graphene with Substitutional Mn. \emph{ACS
  Nano} \textbf{2021}, \emph{15}, 5449--5458, PMID: 33596385\relax
\mciteBstWouldAddEndPuncttrue
\mciteSetBstMidEndSepPunct{\mcitedefaultmidpunct}
{\mcitedefaultendpunct}{\mcitedefaultseppunct}\relax
\EndOfBibitem
\bibitem[Lin \latin{et~al.}(2022)Lin, Villarreal, Bana, Zarkua, Hendriks, Tsai,
  Auge, Junge, Hofsäss, Tosi, Lacovig, Lizzit, Zhao, Di~Santo, Petaccia,
  De~Feyter, De~Gendt, Brems, and Pereira]{Lin2022}
Lin,~P.-C. \latin{et~al.}  Thermal Annealing of Graphene Implanted with Mn at
  Ultralow Energies: From Disordered and Contaminated to Nearly Pristine
  Graphene. \emph{The Journal of Physical Chemistry C} \textbf{2022},
  \emph{126}, 10494--10505\relax
\mciteBstWouldAddEndPuncttrue
\mciteSetBstMidEndSepPunct{\mcitedefaultmidpunct}
{\mcitedefaultendpunct}{\mcitedefaultseppunct}\relax
\EndOfBibitem
\bibitem[Bui \latin{et~al.}(2022)Bui, Rost, Auge, Tu, Zhou, Aguilera,
  Bl{\"u}gel, Ghorbani-Asl, Krasheninnikov, Hashemi, Komsa, Jin, Kibkalo,
  O'Connell, Ramasse, Bangert, Hofs{\"a}ss, Gr{\"u}tzmacher, and
  Kardynal]{Bui2022}
Bui,~M.~N. \latin{et~al.}  Low-energy Se ion implantation in MoS$_2$
  monolayers. \emph{npj 2D Materials and Applications} \textbf{2022}, \emph{6},
  42\relax
\mciteBstWouldAddEndPuncttrue
\mciteSetBstMidEndSepPunct{\mcitedefaultmidpunct}
{\mcitedefaultendpunct}{\mcitedefaultseppunct}\relax
\EndOfBibitem
\bibitem[Bangert \latin{et~al.}(2017)Bangert, Stewart, O’Connell, Courtney,
  Ramasse, Kepaptsoglou, Hofsäss, Amani, Tu, and Kardynal]{Bangert2017}
Bangert,~U.; Stewart,~A.; O’Connell,~E.; Courtney,~E.; Ramasse,~Q.;
  Kepaptsoglou,~D.; Hofsäss,~H.; Amani,~J.; Tu,~J.-S.; Kardynal,~B. Ion-beam
  modification of 2-D materials - single implant atom analysis via annular
  dark-field electron microscopy. \emph{Ultramicroscopy} \textbf{2017},
  \emph{176}, 31--36, 70th Birthday of Robert Sinclair and 65th Birthday of
  Nestor J. Zaluzec PICO 2017 – Fourth Conference on Frontiers of Aberration
  Corrected Electron Microscopy\relax
\mciteBstWouldAddEndPuncttrue
\mciteSetBstMidEndSepPunct{\mcitedefaultmidpunct}
{\mcitedefaultendpunct}{\mcitedefaultseppunct}\relax
\EndOfBibitem
\bibitem[Castellanos-Gomez \latin{et~al.}(2014)Castellanos-Gomez, Buscema,
  Molenaar, Singh, Janssen, van~der Zant, and Steele]{Castellanos_Gomez2014}
Castellanos-Gomez,~A.; Buscema,~M.; Molenaar,~R.; Singh,~V.; Janssen,~L.;
  van~der Zant,~H. S.~J.; Steele,~G.~A. Deterministic transfer of
  two-dimensional materials by all-dry viscoelastic stamping. \emph{2D
  Materials} \textbf{2014}, \emph{1}, 011002\relax
\mciteBstWouldAddEndPuncttrue
\mciteSetBstMidEndSepPunct{\mcitedefaultmidpunct}
{\mcitedefaultendpunct}{\mcitedefaultseppunct}\relax
\EndOfBibitem
\bibitem[Kang \latin{et~al.}(2013)Kang, Tongay, Zhou, Li, and Wu]{Kang2013}
Kang,~J.; Tongay,~S.; Zhou,~J.; Li,~J.; Wu,~J. Band offsets and
  heterostructures of two-dimensional semiconductors. \emph{Applied Physics
  Letters} \textbf{2013}, \emph{102}, 012111\relax
\mciteBstWouldAddEndPuncttrue
\mciteSetBstMidEndSepPunct{\mcitedefaultmidpunct}
{\mcitedefaultendpunct}{\mcitedefaultseppunct}\relax
\EndOfBibitem
\bibitem[Ho and Lai(2019)Ho, and Lai]{Ho2019}
Ho,~C.-H.; Lai,~X.-R. Effect of Cr on the Structure and Property of
  Mo$_{1–x}$Cr$_x$Se$_2$ (0 $\le$ $x$ $\le$ 0.2) and Cr$_2$Se$_3$. \emph{ACS
  Applied Electronic Materials} \textbf{2019}, \emph{1}, 370--378\relax
\mciteBstWouldAddEndPuncttrue
\mciteSetBstMidEndSepPunct{\mcitedefaultmidpunct}
{\mcitedefaultendpunct}{\mcitedefaultseppunct}\relax
\EndOfBibitem
\bibitem[Wang \latin{et~al.}(2015)Wang, Palleau, Amand, Tongay, Marie, and
  Urbaszek]{Wang2015}
Wang,~G.; Palleau,~E.; Amand,~T.; Tongay,~S.; Marie,~X.; Urbaszek,~B.
  Polarization and time-resolved photoluminescence spectroscopy of excitons in
  MoSe$_2$ monolayers. \emph{Applied Physics Letters} \textbf{2015},
  \emph{106}, 112101\relax
\mciteBstWouldAddEndPuncttrue
\mciteSetBstMidEndSepPunct{\mcitedefaultmidpunct}
{\mcitedefaultendpunct}{\mcitedefaultseppunct}\relax
\EndOfBibitem
\bibitem[Fang \latin{et~al.}(2019)Fang, Han, Robert, Semina, Lagarde, Courtade,
  Taniguchi, Watanabe, Amand, Urbaszek, Glazov, and Marie]{Fang2019}
Fang,~H.~H.; Han,~B.; Robert,~C.; Semina,~M.~A.; Lagarde,~D.; Courtade,~E.;
  Taniguchi,~T.; Watanabe,~K.; Amand,~T.; Urbaszek,~B.; Glazov,~M.~M.;
  Marie,~X. Control of the Exciton Radiative Lifetime in van der Waals
  Heterostructures. \emph{Phys. Rev. Lett.} \textbf{2019}, \emph{123},
  067401\relax
\mciteBstWouldAddEndPuncttrue
\mciteSetBstMidEndSepPunct{\mcitedefaultmidpunct}
{\mcitedefaultendpunct}{\mcitedefaultseppunct}\relax
\EndOfBibitem
\bibitem[Selig \latin{et~al.}(2016)Selig, Bergh{\"a}user, Raja, Nagler,
  Sch{\"u}ller, Heinz, Korn, Chernikov, Malic, and Knorr]{Selig2016}
Selig,~M.; Bergh{\"a}user,~G.; Raja,~A.; Nagler,~P.; Sch{\"u}ller,~C.;
  Heinz,~T.~F.; Korn,~T.; Chernikov,~A.; Malic,~E.; Knorr,~A. Excitonic
  linewidth and coherence lifetime in monolayer transition metal
  dichalcogenides. \emph{Nature Communications} \textbf{2016}, \emph{7},
  13279\relax
\mciteBstWouldAddEndPuncttrue
\mciteSetBstMidEndSepPunct{\mcitedefaultmidpunct}
{\mcitedefaultendpunct}{\mcitedefaultseppunct}\relax
\EndOfBibitem
\bibitem[Robert \latin{et~al.}(2016)Robert, Lagarde, Cadiz, Wang, Lassagne,
  Amand, Balocchi, Renucci, Tongay, Urbaszek, and Marie]{Robert2016}
Robert,~C.; Lagarde,~D.; Cadiz,~F.; Wang,~G.; Lassagne,~B.; Amand,~T.;
  Balocchi,~A.; Renucci,~P.; Tongay,~S.; Urbaszek,~B.; Marie,~X. Exciton
  radiative lifetime in transition metal dichalcogenide monolayers. \emph{Phys.
  Rev. B} \textbf{2016}, \emph{93}, 205423\relax
\mciteBstWouldAddEndPuncttrue
\mciteSetBstMidEndSepPunct{\mcitedefaultmidpunct}
{\mcitedefaultendpunct}{\mcitedefaultseppunct}\relax
\EndOfBibitem
\bibitem[Yu \latin{et~al.}(2021)Yu, Deng, Zhang, Borghardt, Kardynal,
  Vučković, and Heinz]{Yu2021}
Yu,~L.; Deng,~M.; Zhang,~J.~L.; Borghardt,~S.; Kardynal,~B.; Vučković,~J.;
  Heinz,~T.~F. Site-Controlled Quantum Emitters in Monolayer MoSe$_2$.
  \emph{Nano Letters} \textbf{2021}, \emph{21}, 2376--2381\relax
\mciteBstWouldAddEndPuncttrue
\mciteSetBstMidEndSepPunct{\mcitedefaultmidpunct}
{\mcitedefaultendpunct}{\mcitedefaultseppunct}\relax
\EndOfBibitem
\bibitem[Reshchikov \latin{et~al.}(2018)Reshchikov, Albarakati, Monavarian,
  Avrutin, and Morkoç]{Reshchikov2018}
Reshchikov,~M.~A.; Albarakati,~N.~M.; Monavarian,~M.; Avrutin,~V.; Morkoç,~H.
  Thermal quenching of the yellow luminescence in GaN. \emph{Journal of Applied
  Physics} \textbf{2018}, \emph{123}, 161520\relax
\mciteBstWouldAddEndPuncttrue
\mciteSetBstMidEndSepPunct{\mcitedefaultmidpunct}
{\mcitedefaultendpunct}{\mcitedefaultseppunct}\relax
\EndOfBibitem
\bibitem[Reshchikov(2021)]{Reshchikov2021}
Reshchikov,~M.~A. Mechanisms of Thermal Quenching of Defect-Related
  Luminescence in Semiconductors. \emph{physica status solidi (a)}
  \textbf{2021}, \emph{218}, 2000101\relax
\mciteBstWouldAddEndPuncttrue
\mciteSetBstMidEndSepPunct{\mcitedefaultmidpunct}
{\mcitedefaultendpunct}{\mcitedefaultseppunct}\relax
\EndOfBibitem
\bibitem[Huang \latin{et~al.}(2016)Huang, Hoang, and Mikkelsen]{Huang2016}
Huang,~J.; Hoang,~T.~B.; Mikkelsen,~M.~H. Probing the origin of excitonic
  states in monolayer WSe$_2$. \emph{Scientific Reports} \textbf{2016},
  \emph{6}, 22414\relax
\mciteBstWouldAddEndPuncttrue
\mciteSetBstMidEndSepPunct{\mcitedefaultmidpunct}
{\mcitedefaultendpunct}{\mcitedefaultseppunct}\relax
\EndOfBibitem
\bibitem[Szyniszewski \latin{et~al.}(2017)Szyniszewski, Mostaani, Drummond, and
  Fal'ko]{Szyniszewski2017}
Szyniszewski,~M.; Mostaani,~E.; Drummond,~N.~D.; Fal'ko,~V.~I. Binding energies
  of trions and biexcitons in two-dimensional semiconductors from diffusion
  quantum Monte Carlo calculations. \emph{Phys. Rev. B} \textbf{2017},
  \emph{95}, 081301\relax
\mciteBstWouldAddEndPuncttrue
\mciteSetBstMidEndSepPunct{\mcitedefaultmidpunct}
{\mcitedefaultendpunct}{\mcitedefaultseppunct}\relax
\EndOfBibitem
\bibitem[Shibata(1998)]{Shibata1998}
Shibata,~H. Negative Thermal Quenching Curves in Photoluminescence of Solids.
  \emph{Japanese Journal of Applied Physics} \textbf{1998}, \emph{37},
  550\relax
\mciteBstWouldAddEndPuncttrue
\mciteSetBstMidEndSepPunct{\mcitedefaultmidpunct}
{\mcitedefaultendpunct}{\mcitedefaultseppunct}\relax
\EndOfBibitem
\bibitem[Watanabe \latin{et~al.}(2006)Watanabe, Sakai, Shibata, Satou, Satou,
  Shibayama, Tampo, Yamada, Matsubara, Sakurai, Ishizuka, Niki, Maeda, and
  Niikura]{Watanabe2006}
Watanabe,~M.; Sakai,~M.; Shibata,~H.; Satou,~C.; Satou,~S.; Shibayama,~T.;
  Tampo,~H.; Yamada,~A.; Matsubara,~K.; Sakurai,~K.; Ishizuka,~S.; Niki,~S.;
  Maeda,~K.; Niikura,~I. Negative thermal quenching of photoluminescence in
  ZnO. \emph{Physica B: Condensed Matter} \textbf{2006}, \emph{376-377},
  711--714, Proceedings of the 23rd International Conference on Defects in
  Semiconductors\relax
\mciteBstWouldAddEndPuncttrue
\mciteSetBstMidEndSepPunct{\mcitedefaultmidpunct}
{\mcitedefaultendpunct}{\mcitedefaultseppunct}\relax
\EndOfBibitem
\bibitem[Lin \latin{et~al.}(2012)Lin, Chen, Xiong, Yang, He, and Luo]{Lin2012}
Lin,~S.~S.; Chen,~B.~G.; Xiong,~W.; Yang,~Y.; He,~H.~P.; Luo,~J. Negative
  thermal quenching of photoluminescence in zinc oxide
  nanowire-core/graphene-shell complexes. \emph{Opt. Express} \textbf{2012},
  \emph{20}, A706--A712\relax
\mciteBstWouldAddEndPuncttrue
\mciteSetBstMidEndSepPunct{\mcitedefaultmidpunct}
{\mcitedefaultendpunct}{\mcitedefaultseppunct}\relax
\EndOfBibitem
\bibitem[Tangi \latin{et~al.}(2017)Tangi, Shakfa, Mishra, Li, Chiu, Ng, Li, and
  Ooi]{Tangi2017}
Tangi,~M.; Shakfa,~M.~K.; Mishra,~P.; Li,~M.-Y.; Chiu,~M.-H.; Ng,~T.~K.;
  Li,~L.-J.; Ooi,~B.~S. Anomalous photoluminescence thermal quenching of
  sandwiched single layer MoS$_2$. \emph{Opt. Mater. Express} \textbf{2017},
  \emph{7}, 3697--3705\relax
\mciteBstWouldAddEndPuncttrue
\mciteSetBstMidEndSepPunct{\mcitedefaultmidpunct}
{\mcitedefaultendpunct}{\mcitedefaultseppunct}\relax
\EndOfBibitem
\bibitem[O’Donnell and Chen(1991)O’Donnell, and Chen]{ODonnell1991}
O’Donnell,~K.~P.; Chen,~X. Temperature dependence of semiconductor band gaps.
  \emph{Applied Physics Letters} \textbf{1991}, \emph{58}, 2924--2926\relax
\mciteBstWouldAddEndPuncttrue
\mciteSetBstMidEndSepPunct{\mcitedefaultmidpunct}
{\mcitedefaultendpunct}{\mcitedefaultseppunct}\relax
\EndOfBibitem
\bibitem[Li \latin{et~al.}(2017)Li, Puretzky, Sang, KC, Tian, Ceballos,
  Mahjouri-Samani, Wang, Unocic, Zhao, Duscher, Cooper, Rouleau, Geohegan, and
  Xiao]{Li2017}
Li,~X.; Puretzky,~A.~A.; Sang,~X.; KC,~S.; Tian,~M.; Ceballos,~F.;
  Mahjouri-Samani,~M.; Wang,~K.; Unocic,~R.~R.; Zhao,~H.; Duscher,~G.;
  Cooper,~V.~R.; Rouleau,~C.~M.; Geohegan,~D.~B.; Xiao,~K. Suppression of
  Defects and Deep Levels Using Isoelectronic Tungsten Substitution in
  Monolayer MoSe$_2$. \emph{Advanced Functional Materials} \textbf{2017},
  \emph{27}, 1603850\relax
\mciteBstWouldAddEndPuncttrue
\mciteSetBstMidEndSepPunct{\mcitedefaultmidpunct}
{\mcitedefaultendpunct}{\mcitedefaultseppunct}\relax
\EndOfBibitem
\bibitem[Tongay \latin{et~al.}(2012)Tongay, Zhou, Ataca, Lo, Matthews, Li,
  Grossman, and Wu]{Tongay2012}
Tongay,~S.; Zhou,~J.; Ataca,~C.; Lo,~K.; Matthews,~T.~S.; Li,~J.;
  Grossman,~J.~C.; Wu,~J. Thermally Driven Crossover from Indirect toward
  Direct Bandgap in 2D Semiconductors: MoSe$_2$ versus MoS$_2$. \emph{Nano
  Letters} \textbf{2012}, \emph{12}, 5576--5580, PMID: 23098085\relax
\mciteBstWouldAddEndPuncttrue
\mciteSetBstMidEndSepPunct{\mcitedefaultmidpunct}
{\mcitedefaultendpunct}{\mcitedefaultseppunct}\relax
\EndOfBibitem
\bibitem[Ross \latin{et~al.}(2013)Ross, Wu, Yu, Ghimire, Jones, Aivazian, Yan,
  Mandrus, Xiao, Yao, and Xu]{Ross2013}
Ross,~J.~S.; Wu,~S.; Yu,~H.; Ghimire,~N.~J.; Jones,~A.~M.; Aivazian,~G.;
  Yan,~J.; Mandrus,~D.~G.; Xiao,~D.; Yao,~W.; Xu,~X. Electrical control of
  neutral and charged excitons in a monolayer semiconductor. \emph{Nature
  Communications} \textbf{2013}, \emph{4}, 1474\relax
\mciteBstWouldAddEndPuncttrue
\mciteSetBstMidEndSepPunct{\mcitedefaultmidpunct}
{\mcitedefaultendpunct}{\mcitedefaultseppunct}\relax
\EndOfBibitem
\bibitem[Choi \latin{et~al.}(2017)Choi, Kim, Jung, Kim, Yu, and
  Chang]{Choi2017}
Choi,~B.~K.; Kim,~M.; Jung,~K.-H.; Kim,~J.; Yu,~K.-S.; Chang,~Y.~J. Temperature
  dependence of band gap in MoSe$_2$ grown by molecular beam epitaxy.
  \emph{Nanoscale Research Letters} \textbf{2017}, \emph{12}, 492\relax
\mciteBstWouldAddEndPuncttrue
\mciteSetBstMidEndSepPunct{\mcitedefaultmidpunct}
{\mcitedefaultendpunct}{\mcitedefaultseppunct}\relax
\EndOfBibitem
\bibitem[Kioseoglou \latin{et~al.}(2016)Kioseoglou, Hanbicki, Currie, Friedman,
  and Jonker]{Kioseoglou2016}
Kioseoglou,~G.; Hanbicki,~A.~T.; Currie,~M.; Friedman,~A.~L.; Jonker,~B.~T.
  Optical polarization and intervalley scattering in single layers of MoS$_2$
  and MoSe$_2$. \emph{Scientific Reports} \textbf{2016}, \emph{6}, 25041\relax
\mciteBstWouldAddEndPuncttrue
\mciteSetBstMidEndSepPunct{\mcitedefaultmidpunct}
{\mcitedefaultendpunct}{\mcitedefaultseppunct}\relax
\EndOfBibitem
\bibitem[Parto \latin{et~al.}(2021)Parto, Azzam, Banerjee, and
  Moody]{Parto2021}
Parto,~K.; Azzam,~S.~I.; Banerjee,~K.; Moody,~G. Defect and strain engineering
  of monolayer WSe$_2$ enables site-controlled single-photon emission up to
  150{\thinspace}K. \emph{Nature Communications} \textbf{2021}, \emph{12},
  3585\relax
\mciteBstWouldAddEndPuncttrue
\mciteSetBstMidEndSepPunct{\mcitedefaultmidpunct}
{\mcitedefaultendpunct}{\mcitedefaultseppunct}\relax
\EndOfBibitem
\bibitem[Srivastava \latin{et~al.}(2015)Srivastava, Sidler, Allain, Lembke,
  Kis, and Imamo{\u{g}}lu]{Srivastava2015}
Srivastava,~A.; Sidler,~M.; Allain,~A.~V.; Lembke,~D.~S.; Kis,~A.;
  Imamo{\u{g}}lu,~A. Valley Zeeman effect in elementary optical excitations of
  monolayer WSe$_2$. \emph{Nature Physics} \textbf{2015}, \emph{11},
  141--147\relax
\mciteBstWouldAddEndPuncttrue
\mciteSetBstMidEndSepPunct{\mcitedefaultmidpunct}
{\mcitedefaultendpunct}{\mcitedefaultseppunct}\relax
\EndOfBibitem
\bibitem[Wang \latin{et~al.}(2015)Wang, Bouet, Glazov, Amand, Ivchenko,
  Palleau, Marie, and Urbaszek]{Wang2015b}
Wang,~G.; Bouet,~L.; Glazov,~M.~M.; Amand,~T.; Ivchenko,~E.~L.; Palleau,~E.;
  Marie,~X.; Urbaszek,~B. Magneto-optics in transition metal diselenide
  monolayers. \emph{2D Materials} \textbf{2015}, \emph{2}, 034002\relax
\mciteBstWouldAddEndPuncttrue
\mciteSetBstMidEndSepPunct{\mcitedefaultmidpunct}
{\mcitedefaultendpunct}{\mcitedefaultseppunct}\relax
\EndOfBibitem
\bibitem[Koperski \latin{et~al.}(2018)Koperski, Molas, Arora, Nogajewski,
  Bartos, Wyzula, Vaclavkova, Kossacki, and Potemski]{Koperski2019}
Koperski,~M.; Molas,~M.~R.; Arora,~A.; Nogajewski,~K.; Bartos,~M.; Wyzula,~J.;
  Vaclavkova,~D.; Kossacki,~P.; Potemski,~M. Orbital, spin and valley
  contributions to Zeeman splitting of excitonic resonances in MoSe$_2$,
  WSe$_2$ and WS$_2$ Monolayers. \emph{2D Materials} \textbf{2018}, \emph{6},
  015001\relax
\mciteBstWouldAddEndPuncttrue
\mciteSetBstMidEndSepPunct{\mcitedefaultmidpunct}
{\mcitedefaultendpunct}{\mcitedefaultseppunct}\relax
\EndOfBibitem
\bibitem[Li \latin{et~al.}(2014)Li, Ludwig, Low, Chernikov, Cui, Arefe, Kim,
  van~der Zande, Rigosi, Hill, Kim, Hone, Li, Smirnov, and Heinz]{Li2014}
Li,~Y.; Ludwig,~J.; Low,~T.; Chernikov,~A.; Cui,~X.; Arefe,~G.; Kim,~Y.~D.;
  van~der Zande,~A.~M.; Rigosi,~A.; Hill,~H.~M.; Kim,~S.~H.; Hone,~J.; Li,~Z.;
  Smirnov,~D.; Heinz,~T.~F. Valley Splitting and Polarization by the Zeeman
  Effect in Monolayer ${\mathrm{MoSe}}_{2}$. \emph{Phys. Rev. Lett.}
  \textbf{2014}, \emph{113}, 266804\relax
\mciteBstWouldAddEndPuncttrue
\mciteSetBstMidEndSepPunct{\mcitedefaultmidpunct}
{\mcitedefaultendpunct}{\mcitedefaultseppunct}\relax
\EndOfBibitem
\bibitem[MacNeill \latin{et~al.}(2015)MacNeill, Heikes, Mak, Anderson,
  Korm\'anyos, Z\'olyomi, Park, and Ralph]{MacNeill2015}
MacNeill,~D.; Heikes,~C.; Mak,~K.~F.; Anderson,~Z.; Korm\'anyos,~A.;
  Z\'olyomi,~V.; Park,~J.; Ralph,~D.~C. Breaking of Valley Degeneracy by
  Magnetic Field in Monolayer ${\mathrm{MoSe}}_{2}$. \emph{Phys. Rev. Lett.}
  \textbf{2015}, \emph{114}, 037401\relax
\mciteBstWouldAddEndPuncttrue
\mciteSetBstMidEndSepPunct{\mcitedefaultmidpunct}
{\mcitedefaultendpunct}{\mcitedefaultseppunct}\relax
\EndOfBibitem
\bibitem[Back \latin{et~al.}(2017)Back, Sidler, Cotlet, Srivastava, Takemura,
  Kroner, and Imamo\ifmmode~\breve{g}\else \u{g}\fi{}lu]{Back2017}
Back,~P.; Sidler,~M.; Cotlet,~O.; Srivastava,~A.; Takemura,~N.; Kroner,~M.;
  Imamo\ifmmode~\breve{g}\else \u{g}\fi{}lu,~A. Giant Paramagnetism-Induced
  Valley Polarization of Electrons in Charge-Tunable Monolayer
  ${\mathrm{MoSe}}_{2}$. \emph{Phys. Rev. Lett.} \textbf{2017}, \emph{118},
  237404\relax
\mciteBstWouldAddEndPuncttrue
\mciteSetBstMidEndSepPunct{\mcitedefaultmidpunct}
{\mcitedefaultendpunct}{\mcitedefaultseppunct}\relax
\EndOfBibitem
\bibitem[Goryca \latin{et~al.}(2019)Goryca, Li, Stier, Taniguchi, Watanabe,
  Courtade, Shree, Robert, Urbaszek, Marie, and Crooker]{Goryca2019}
Goryca,~M.; Li,~J.; Stier,~A.~V.; Taniguchi,~T.; Watanabe,~K.; Courtade,~E.;
  Shree,~S.; Robert,~C.; Urbaszek,~B.; Marie,~X.; Crooker,~S.~A. Revealing
  exciton masses and dielectric properties of monolayer semiconductors with
  high magnetic fields. \emph{Nature Communications} \textbf{2019}, \emph{10},
  4172\relax
\mciteBstWouldAddEndPuncttrue
\mciteSetBstMidEndSepPunct{\mcitedefaultmidpunct}
{\mcitedefaultendpunct}{\mcitedefaultseppunct}\relax
\EndOfBibitem
\bibitem[Cadiz \latin{et~al.}(2017)Cadiz, Courtade, Robert, Wang, Shen, Cai,
  Taniguchi, Watanabe, Carrere, Lagarde, Manca, Amand, Renucci, Tongay, Marie,
  and Urbaszek]{Cadiz2017}
Cadiz,~F. \latin{et~al.}  Excitonic Linewidth Approaching the Homogeneous Limit
  in ${\mathrm{MoS}}_{2}$-Based van der Waals Heterostructures. \emph{Phys.
  Rev. X} \textbf{2017}, \emph{7}, 021026\relax
\mciteBstWouldAddEndPuncttrue
\mciteSetBstMidEndSepPunct{\mcitedefaultmidpunct}
{\mcitedefaultendpunct}{\mcitedefaultseppunct}\relax
\EndOfBibitem
\bibitem[Deilmann \latin{et~al.}(2020)Deilmann, Kr\"uger, and
  Rohlfing]{Deilmann2020}
Deilmann,~T.; Kr\"uger,~P.; Rohlfing,~M. Ab Initio Studies of Exciton $g$
  Factors: Monolayer Transition Metal Dichalcogenides in Magnetic Fields.
  \emph{Phys. Rev. Lett.} \textbf{2020}, \emph{124}, 226402\relax
\mciteBstWouldAddEndPuncttrue
\mciteSetBstMidEndSepPunct{\mcitedefaultmidpunct}
{\mcitedefaultendpunct}{\mcitedefaultseppunct}\relax
\EndOfBibitem
\bibitem[Wo\ifmmode~\acute{z}\else \'{z}\fi{}niak
  \latin{et~al.}(2020)Wo\ifmmode~\acute{z}\else \'{z}\fi{}niak, Faria~Junior,
  Seifert, Chaves, and Kunstmann]{Wozniak2020}
Wo\ifmmode~\acute{z}\else \'{z}\fi{}niak,~T.; Faria~Junior,~P.~E.; Seifert,~G.;
  Chaves,~A.; Kunstmann,~J. Exciton $g$ factors of van der Waals
  heterostructures from first-principles calculations. \emph{Phys. Rev. B}
  \textbf{2020}, \emph{101}, 235408\relax
\mciteBstWouldAddEndPuncttrue
\mciteSetBstMidEndSepPunct{\mcitedefaultmidpunct}
{\mcitedefaultendpunct}{\mcitedefaultseppunct}\relax
\EndOfBibitem
\bibitem[Gruber \latin{et~al.}(2016)Gruber, Wilhelm, P{\'{e}}tuya, Smejkal,
  Kozubek, Hierzenberger, Bayer, Aldazabal, Kazansky, Libisch, Krasheninnikov,
  Schleberger, Facsko, Borisov, Arnau, and Aumayr]{Gruber-2016}
Gruber,~E. \latin{et~al.}  {Ultrafast electronic response of graphene to a
  strong and localized electric field}. \emph{Nature Communications}
  \textbf{2016}, \emph{7}, 13948\relax
\mciteBstWouldAddEndPuncttrue
\mciteSetBstMidEndSepPunct{\mcitedefaultmidpunct}
{\mcitedefaultendpunct}{\mcitedefaultseppunct}\relax
\EndOfBibitem
\bibitem[Ojanpera \latin{et~al.}(2014)Ojanpera, Krasheninnikov, and
  Puska]{Ojanpera-2014}
Ojanpera,~A.; Krasheninnikov,~A.~V.; Puska,~M. Electronic stopping power from
  first-principles calculations with account for core electron excitations and
  projectile ionization. \emph{Physical Review B} \textbf{2014}, \emph{89},
  035120\relax
\mciteBstWouldAddEndPuncttrue
\mciteSetBstMidEndSepPunct{\mcitedefaultmidpunct}
{\mcitedefaultendpunct}{\mcitedefaultseppunct}\relax
\EndOfBibitem
\bibitem[Kalbac \latin{et~al.}(2013)Kalbac, Lehtinen, Krasheninnikov, and
  Keinonen]{Kalbac-2013}
Kalbac,~M.; Lehtinen,~O.; Krasheninnikov,~A.~V.; Keinonen,~J.
  Ion-Irradiation-Induced Defects in Isotopically-Labeled Two Layered Graphene:
  Enhanced In-Situ Annealing of the Damage. \emph{Advanced Materials}
  \textbf{2013}, \emph{25}, 1004--1009\relax
\mciteBstWouldAddEndPuncttrue
\mciteSetBstMidEndSepPunct{\mcitedefaultmidpunct}
{\mcitedefaultendpunct}{\mcitedefaultseppunct}\relax
\EndOfBibitem
\bibitem[Kretschmer \latin{et~al.}(2018)Kretschmer, Maslov, Ghaderzadeh,
  Ghorbani-Asl, Hlawacek, and Krasheninnikov]{Kretschmer-2018}
Kretschmer,~S.; Maslov,~M.; Ghaderzadeh,~S.; Ghorbani-Asl,~M.; Hlawacek,~G.;
  Krasheninnikov,~A.~V. {Supported Two-Dimensional Materials under Ion
  Irradiation: The Substrate Governs Defect Production}. \emph{ACS Applied
  Materials {\&} Interfaces} \textbf{2018}, \emph{10}, 30827--30836\relax
\mciteBstWouldAddEndPuncttrue
\mciteSetBstMidEndSepPunct{\mcitedefaultmidpunct}
{\mcitedefaultendpunct}{\mcitedefaultseppunct}\relax
\EndOfBibitem
\bibitem[Standop \latin{et~al.}(2013)Standop, Lehtinen, Herbig,
  Lewes-Malandrakis, Craes, Kotakoski, Michely, Krasheninnikov, and
  Busse]{Standop13}
Standop,~S.; Lehtinen,~O.; Herbig,~C.; Lewes-Malandrakis,~G.; Craes,~F.;
  Kotakoski,~J.; Michely,~T.; Krasheninnikov,~A.~V.; Busse,~C. Ion Impacts on
  Graphene/Ir(111): Interface Channeling, Vacancy Funnels, and a Nanomesh.
  \emph{Nano Letters} \textbf{2013}, \emph{13}, 1948--1955\relax
\mciteBstWouldAddEndPuncttrue
\mciteSetBstMidEndSepPunct{\mcitedefaultmidpunct}
{\mcitedefaultendpunct}{\mcitedefaultseppunct}\relax
\EndOfBibitem
\bibitem[Karthikeyan \latin{et~al.}(2019)Karthikeyan, Komsa, Batzill, and
  Krasheninnikov]{Karthikeyan19}
Karthikeyan,~J.; Komsa,~H.-P.; Batzill,~M.; Krasheninnikov,~A.~V. Which
  Transition Metal Atoms Can Be Embedded into Two-Dimensional Molybdenum
  Dichalcogenides and Add Magnetism? \emph{Nano Letters} \textbf{2019},
  \emph{19}, 4581--4587\relax
\mciteBstWouldAddEndPuncttrue
\mciteSetBstMidEndSepPunct{\mcitedefaultmidpunct}
{\mcitedefaultendpunct}{\mcitedefaultseppunct}\relax
\EndOfBibitem
\bibitem[Liu \latin{et~al.}(2015)Liu, Xiao, Yao, Xu, and Yao]{Liu2015}
Liu,~G.-B.; Xiao,~D.; Yao,~Y.; Xu,~X.; Yao,~W. Electronic structures and
  theoretical modelling of two-dimensional group-VIB transition metal
  dichalcogenides. \emph{Chem. Soc. Rev.} \textbf{2015}, \emph{44},
  2643--2663\relax
\mciteBstWouldAddEndPuncttrue
\mciteSetBstMidEndSepPunct{\mcitedefaultmidpunct}
{\mcitedefaultendpunct}{\mcitedefaultseppunct}\relax
\EndOfBibitem
\bibitem[Rost(2023)]{Rost2023}
Rost,~S.~H. Computational study of structural and optical properties of
  two-dimensional transition-metal dichalcogenides with implanted defects.
  Dissertation, RWTH Aachen University, Jülich, 2023\relax
\mciteBstWouldAddEndPuncttrue
\mciteSetBstMidEndSepPunct{\mcitedefaultmidpunct}
{\mcitedefaultendpunct}{\mcitedefaultseppunct}\relax
\EndOfBibitem
\bibitem[Iberi \latin{et~al.}(2016)Iberi, Liang, Ievlev, Stanford, Lin, Li,
  Mahjouri-Samani, Jesse, Sumpter, Kalinin, Joy, Xiao, Belianinov, and
  Ovchinnikova]{Iberi2016}
Iberi,~V.; Liang,~L.; Ievlev,~A.~V.; Stanford,~M.~G.; Lin,~M.-W.; Li,~X.;
  Mahjouri-Samani,~M.; Jesse,~S.; Sumpter,~B.~G.; Kalinin,~S.~V.; Joy,~D.~C.;
  Xiao,~K.; Belianinov,~A.; Ovchinnikova,~O.~S. Nanoforging Single Layer
  MoSe$_2$ Through Defect Engineering with Focused Helium Ion Beams.
  \emph{Scientific Reports} \textbf{2016}, \emph{6}, 30481\relax
\mciteBstWouldAddEndPuncttrue
\mciteSetBstMidEndSepPunct{\mcitedefaultmidpunct}
{\mcitedefaultendpunct}{\mcitedefaultseppunct}\relax
\EndOfBibitem
\bibitem[Mahjouri-Samani \latin{et~al.}(2016)Mahjouri-Samani, Liang, Oyedele,
  Kim, Tian, Cross, Wang, Lin, Boulesbaa, Rouleau, Puretzky, Xiao, Yoon, Eres,
  Duscher, Sumpter, and Geohegan]{Samani2016}
Mahjouri-Samani,~M. \latin{et~al.}  Tailoring Vacancies Far Beyond Intrinsic
  Levels Changes the Carrier Type and Optical Response in Monolayer
  MoSe$_{2-x}$ Crystals. \emph{Nano Letters} \textbf{2016}, \emph{16},
  5213--5220, PMID: 27416103\relax
\mciteBstWouldAddEndPuncttrue
\mciteSetBstMidEndSepPunct{\mcitedefaultmidpunct}
{\mcitedefaultendpunct}{\mcitedefaultseppunct}\relax
\EndOfBibitem
\bibitem[Shafqat \latin{et~al.}(2017)Shafqat, Iqbal, and Majid]{Shafqat2017}
Shafqat,~A.; Iqbal,~T.; Majid,~A. A DFT study of intrinsic point defects in
  monolayer MoSe$_2$. \emph{AIP Advances} \textbf{2017}, \emph{7}, 105306\relax
\mciteBstWouldAddEndPuncttrue
\mciteSetBstMidEndSepPunct{\mcitedefaultmidpunct}
{\mcitedefaultendpunct}{\mcitedefaultseppunct}\relax
\EndOfBibitem
\bibitem[Tang \latin{et~al.}(2019)Tang, Mak, and Shan]{Tang19}
Tang,~Y.; Mak,~K.~F.; Shan,~J. Long valley lifetime of dark excitons in
  single-layer WSe$_2$. \emph{Nature Communications} \textbf{2019}, \emph{10},
  4047\relax
\mciteBstWouldAddEndPuncttrue
\mciteSetBstMidEndSepPunct{\mcitedefaultmidpunct}
{\mcitedefaultendpunct}{\mcitedefaultseppunct}\relax
\EndOfBibitem
\bibitem[Kresse and Hafner(1993)Kresse, and Hafner]{Kresse1993}
Kresse,~G.; Hafner,~J. {Ab initio molecular dynamics for liquid metals}.
  \emph{Phys. Rev. B} \textbf{1993}, \emph{47}, 558--561\relax
\mciteBstWouldAddEndPuncttrue
\mciteSetBstMidEndSepPunct{\mcitedefaultmidpunct}
{\mcitedefaultendpunct}{\mcitedefaultseppunct}\relax
\EndOfBibitem
\bibitem[Kresse and Furthm{\"{u}}ller(1996)Kresse, and
  Furthm{\"{u}}ller]{Kresse1996}
Kresse,~G.; Furthm{\"{u}}ller,~J. {Efficiency of ab-initio total energy
  calculations for metals and semiconductors using a plane-wave basis set}.
  \emph{Comput. Mater. Sci.} \textbf{1996}, \emph{6}, 15--50\relax
\mciteBstWouldAddEndPuncttrue
\mciteSetBstMidEndSepPunct{\mcitedefaultmidpunct}
{\mcitedefaultendpunct}{\mcitedefaultseppunct}\relax
\EndOfBibitem
\bibitem[Perdew \latin{et~al.}(1996)Perdew, Burke, and
  Ernzerhof]{Perdew1996_PRL}
Perdew,~J.~P.; Burke,~K.; Ernzerhof,~M. {Generalized Gradient Approximation
  Made Simple}. \emph{Phys. Rev. Lett.} \textbf{1996}, \emph{77},
  3865--3868\relax
\mciteBstWouldAddEndPuncttrue
\mciteSetBstMidEndSepPunct{\mcitedefaultmidpunct}
{\mcitedefaultendpunct}{\mcitedefaultseppunct}\relax
\EndOfBibitem
\bibitem[fla(2022)]{flapw}
Jülich FLAPW code family. \url{https://www.flapw.de/}, 2022; Accessed:
  2022-11-15\relax
\mciteBstWouldAddEndPuncttrue
\mciteSetBstMidEndSepPunct{\mcitedefaultmidpunct}
{\mcitedefaultendpunct}{\mcitedefaultseppunct}\relax
\EndOfBibitem
\bibitem[Friedrich \latin{et~al.}(2010)Friedrich, Bl{\"{u}}gel, and
  Schindlmayr]{Friedrich2010}
Friedrich,~C.; Bl{\"{u}}gel,~S.; Schindlmayr,~A. {Efficient implementation of
  the GW approximation within the all-electron FLAPW method}. \emph{Phys. Rev.
  B} \textbf{2010}, \emph{81}, 125102\relax
\mciteBstWouldAddEndPuncttrue
\mciteSetBstMidEndSepPunct{\mcitedefaultmidpunct}
{\mcitedefaultendpunct}{\mcitedefaultseppunct}\relax
\EndOfBibitem
\bibitem[Friedrich \latin{et~al.}(2021)Friedrich, Bl{\"{u}}gel, and
  Schindlmayr]{Friedrich2021}
Friedrich,~C.; Bl{\"{u}}gel,~S.; Schindlmayr,~A. {Erratum: Efficient
  implementation of the GW approximation within the all-electron FLAPW method
  (Physical Review B (2010) 81 (125102) DOI: 10.1103/PhysRevB.81.125102)}.
  \emph{ibid.} \textbf{2021}, \emph{104}, 039901\relax
\mciteBstWouldAddEndPuncttrue
\mciteSetBstMidEndSepPunct{\mcitedefaultmidpunct}
{\mcitedefaultendpunct}{\mcitedefaultseppunct}\relax
\EndOfBibitem
\bibitem[Adler(1962)]{Adler1962}
Adler,~S.~L. Quantum theory of the dielectric constant in real solids.
  \emph{Phys. Rev.} \textbf{1962}, \emph{126}, 413--420\relax
\mciteBstWouldAddEndPuncttrue
\mciteSetBstMidEndSepPunct{\mcitedefaultmidpunct}
{\mcitedefaultendpunct}{\mcitedefaultseppunct}\relax
\EndOfBibitem
\bibitem[Wiser(1963)]{Wiser63}
Wiser,~N. Dielectric Constant with Local Field Effects Included. \emph{Phys.
  Rev.} \textbf{1963}, \emph{129}, 62--69\relax
\mciteBstWouldAddEndPuncttrue
\mciteSetBstMidEndSepPunct{\mcitedefaultmidpunct}
{\mcitedefaultendpunct}{\mcitedefaultseppunct}\relax
\EndOfBibitem
\bibitem[Rubel \latin{et~al.}(2014)Rubel, Bokhanchuk, Ahmed, and
  Assmann]{Rubel2014}
Rubel,~O.; Bokhanchuk,~A.; Ahmed,~S.~J.; Assmann,~E. Unfolding the band
  structure of disordered solids: From bound states to high-mobility Kane
  fermions. \emph{Phys. Rev. B} \textbf{2014}, \emph{90}, 115202\relax
\mciteBstWouldAddEndPuncttrue
\mciteSetBstMidEndSepPunct{\mcitedefaultmidpunct}
{\mcitedefaultendpunct}{\mcitedefaultseppunct}\relax
\EndOfBibitem
\bibitem[Andersen(1975)]{Anderson1975}
Andersen,~O.~K. Linear methods in band theory. \emph{Phys. Rev. B}
  \textbf{1975}, \emph{12}, 3060--3083\relax
\mciteBstWouldAddEndPuncttrue
\mciteSetBstMidEndSepPunct{\mcitedefaultmidpunct}
{\mcitedefaultendpunct}{\mcitedefaultseppunct}\relax
\EndOfBibitem
\bibitem[Tonndorf \latin{et~al.}(2013)Tonndorf, Schmidt, B\"{o}ttger, Zhang,
  B\"{o}rner, Liebig, Albrecht, Kloc, Gordan, Zahn, de~Vasconcellos, and
  Bratschitsch]{Tonndorf2013}
Tonndorf,~P.; Schmidt,~R.; B\"{o}ttger,~P.; Zhang,~X.; B\"{o}rner,~J.;
  Liebig,~A.; Albrecht,~M.; Kloc,~C.; Gordan,~O.; Zahn,~D. R.~T.;
  de~Vasconcellos,~S.~M.; Bratschitsch,~R. Photoluminescence emission and Raman
  response of monolayer MoS$_2$, MoSe$_2$, and WSe$_2$. \emph{Opt. Express}
  \textbf{2013}, \emph{21}, 4908--4916\relax
\mciteBstWouldAddEndPuncttrue
\mciteSetBstMidEndSepPunct{\mcitedefaultmidpunct}
{\mcitedefaultendpunct}{\mcitedefaultseppunct}\relax
\EndOfBibitem
\bibitem[{Forschungszentrum J\"{u}lich GmbH}(2017)]{HNF}
{Forschungszentrum J\"{u}lich GmbH} HNF - Helmholtz Nano Facility.
  \emph{Journal of large scale research facilities} \textbf{2017}, \emph{3},
  A112\relax
\mciteBstWouldAddEndPuncttrue
\mciteSetBstMidEndSepPunct{\mcitedefaultmidpunct}
{\mcitedefaultendpunct}{\mcitedefaultseppunct}\relax
\EndOfBibitem
\end{mcitethebibliography}


\begin{thebibliography}{32}%
\makeatletter
\providecommand \@ifxundefined [1]{%
 \@ifx{#1\undefined}
}%
\providecommand \@ifnum [1]{%
 \ifnum #1\expandafter \@firstoftwo
 \else \expandafter \@secondoftwo
 \fi
}%
\providecommand \@ifx [1]{%
 \ifx #1\expandafter \@firstoftwo
 \else \expandafter \@secondoftwo
 \fi
}%
\providecommand \natexlab [1]{#1}%
\providecommand \enquote  [1]{``#1''}%
\providecommand \bibnamefont  [1]{#1}%
\providecommand \bibfnamefont [1]{#1}%
\providecommand \citenamefont [1]{#1}%
\providecommand \href@noop [0]{\@secondoftwo}%
\providecommand \href [0]{\begingroup \@sanitize@url \@href}%
\providecommand \@href[1]{\@@startlink{#1}\@@href}%
\providecommand \@@href[1]{\endgroup#1\@@endlink}%
\providecommand \@sanitize@url [0]{\catcode `\\12\catcode `\$12\catcode
  `\&12\catcode `\#12\catcode `\^12\catcode `\_12\catcode `\%12\relax}%
\providecommand \@@startlink[1]{}%
\providecommand \@@endlink[0]{}%
\providecommand \url  [0]{\begingroup\@sanitize@url \@url }%
\providecommand \@url [1]{\endgroup\@href {#1}{\urlprefix }}%
\providecommand \urlprefix  [0]{URL }%
\providecommand \Eprint [0]{\href }%
\providecommand \doibase [0]{https://doi.org/}%
\providecommand \selectlanguage [0]{\@gobble}%
\providecommand \bibinfo  [0]{\@secondoftwo}%
\providecommand \bibfield  [0]{\@secondoftwo}%
\providecommand \translation [1]{[#1]}%
\providecommand \BibitemOpen [0]{}%
\providecommand \bibitemStop [0]{}%
\providecommand \bibitemNoStop [0]{.\EOS\space}%
\providecommand \EOS [0]{\spacefactor3000\relax}%
\providecommand \BibitemShut  [1]{\csname bibitem#1\endcsname}%
\let\auto@bib@innerbib\@empty
\bibitem [{\citenamefont {Mayer}(1997)}]{Mayer97}%
  \BibitemOpen
  \bibfield  {author} {\bibinfo {author} {\bibfnamefont {M.}~\bibnamefont
  {Mayer}},\ }\href {https://home.mpcdf.mpg.de/~mam/Report%20IPP%209-113.pdf}
  {\bibinfo {title} {Simnra user’s guide}},\ \bibinfo {howpublished} {Report
  IPP 9/113} (\bibinfo {year} {1997}),\ \bibinfo {note} {accessed at 10:33
  12.04.2023}\BibitemShut {NoStop}%
\bibitem [{\citenamefont {MacNeill}\ \emph {et~al.}(2015)\citenamefont
  {MacNeill}, \citenamefont {Heikes}, \citenamefont {Mak}, \citenamefont
  {Anderson}, \citenamefont {Korm\'anyos}, \citenamefont {Z\'olyomi},
  \citenamefont {Park},\ and\ \citenamefont {Ralph}}]{MacNeill2015}%
  \BibitemOpen
  \bibfield  {author} {\bibinfo {author} {\bibfnamefont {D.}~\bibnamefont
  {MacNeill}}, \bibinfo {author} {\bibfnamefont {C.}~\bibnamefont {Heikes}},
  \bibinfo {author} {\bibfnamefont {K.~F.}\ \bibnamefont {Mak}}, \bibinfo
  {author} {\bibfnamefont {Z.}~\bibnamefont {Anderson}}, \bibinfo {author}
  {\bibfnamefont {A.}~\bibnamefont {Korm\'anyos}}, \bibinfo {author}
  {\bibfnamefont {V.}~\bibnamefont {Z\'olyomi}}, \bibinfo {author}
  {\bibfnamefont {J.}~\bibnamefont {Park}},\ and\ \bibinfo {author}
  {\bibfnamefont {D.~C.}\ \bibnamefont {Ralph}},\ }\bibfield  {title} {\bibinfo
  {title} {Breaking of valley degeneracy by magnetic field in monolayer
  ${\mathrm{mose}}_{2}$},\ }\href
  {https://doi.org/10.1103/PhysRevLett.114.037401} {\bibfield  {journal}
  {\bibinfo  {journal} {Phys. Rev. Lett.}\ }\textbf {\bibinfo {volume} {114}},\
  \bibinfo {pages} {037401} (\bibinfo {year} {2015})}\BibitemShut {NoStop}%
\bibitem [{\citenamefont {Kioseoglou}\ \emph {et~al.}(2016)\citenamefont
  {Kioseoglou}, \citenamefont {Hanbicki}, \citenamefont {Currie}, \citenamefont
  {Friedman},\ and\ \citenamefont {Jonker}}]{Kioseoglou2016}%
  \BibitemOpen
  \bibfield  {author} {\bibinfo {author} {\bibfnamefont {G.}~\bibnamefont
  {Kioseoglou}}, \bibinfo {author} {\bibfnamefont {A.~T.}\ \bibnamefont
  {Hanbicki}}, \bibinfo {author} {\bibfnamefont {M.}~\bibnamefont {Currie}},
  \bibinfo {author} {\bibfnamefont {A.~L.}\ \bibnamefont {Friedman}},\ and\
  \bibinfo {author} {\bibfnamefont {B.~T.}\ \bibnamefont {Jonker}},\ }\bibfield
   {title} {\bibinfo {title} {Optical polarization and intervalley scattering
  in single layers of mos$_2$ and mose$_2$},\ }\href
  {https://doi.org/10.1038/srep25041} {\bibfield  {journal} {\bibinfo
  {journal} {Scientific Reports}\ }\textbf {\bibinfo {volume} {6}},\ \bibinfo
  {pages} {25041} (\bibinfo {year} {2016})}\BibitemShut {NoStop}%
\bibitem [{\citenamefont {Wang}\ \emph {et~al.}(2015)\citenamefont {Wang},
  \citenamefont {Palleau}, \citenamefont {Amand}, \citenamefont {Tongay},
  \citenamefont {Marie},\ and\ \citenamefont {Urbaszek}}]{Wang2015}%
  \BibitemOpen
  \bibfield  {author} {\bibinfo {author} {\bibfnamefont {G.}~\bibnamefont
  {Wang}}, \bibinfo {author} {\bibfnamefont {E.}~\bibnamefont {Palleau}},
  \bibinfo {author} {\bibfnamefont {T.}~\bibnamefont {Amand}}, \bibinfo
  {author} {\bibfnamefont {S.}~\bibnamefont {Tongay}}, \bibinfo {author}
  {\bibfnamefont {X.}~\bibnamefont {Marie}},\ and\ \bibinfo {author}
  {\bibfnamefont {B.}~\bibnamefont {Urbaszek}},\ }\bibfield  {title} {\bibinfo
  {title} {Polarization and time-resolved photoluminescence spectroscopy of
  excitons in mose$_2$ monolayers},\ }\href {https://doi.org/10.1063/1.4916089}
  {\bibfield  {journal} {\bibinfo  {journal} {Applied Physics Letters}\
  }\textbf {\bibinfo {volume} {106}},\ \bibinfo {pages} {112101} (\bibinfo
  {year} {2015})},\ \Eprint
  {https://arxiv.org/abs/https://doi.org/10.1063/1.4916089}
  {https://doi.org/10.1063/1.4916089} \BibitemShut {NoStop}%
\bibitem [{\citenamefont {Rost}(2023)}]{Rost2023}%
  \BibitemOpen
  \bibfield  {author} {\bibinfo {author} {\bibfnamefont {S.~H.}\ \bibnamefont
  {Rost}},\ }\emph {\bibinfo {title} {Computational study of structural and
  optical properties of two-dimensional transition-metal dichalcogenides with
  implanted defects}},\ \href@noop {} {\bibinfo {type} {Dissertation}},\
  \bibinfo  {school} {RWTH Aachen University}, \bibinfo {address} {Jülich}
  (\bibinfo {year} {2023})\BibitemShut {NoStop}%
\bibitem [{\citenamefont {Iberi}\ \emph {et~al.}(2016)\citenamefont {Iberi},
  \citenamefont {Liang}, \citenamefont {Ievlev}, \citenamefont {Stanford},
  \citenamefont {Lin}, \citenamefont {Li}, \citenamefont {Mahjouri-Samani},
  \citenamefont {Jesse}, \citenamefont {Sumpter}, \citenamefont {Kalinin},
  \citenamefont {Joy}, \citenamefont {Xiao}, \citenamefont {Belianinov},\ and\
  \citenamefont {Ovchinnikova}}]{Iberi2016}%
  \BibitemOpen
  \bibfield  {author} {\bibinfo {author} {\bibfnamefont {V.}~\bibnamefont
  {Iberi}}, \bibinfo {author} {\bibfnamefont {L.}~\bibnamefont {Liang}},
  \bibinfo {author} {\bibfnamefont {A.~V.}\ \bibnamefont {Ievlev}}, \bibinfo
  {author} {\bibfnamefont {M.~G.}\ \bibnamefont {Stanford}}, \bibinfo {author}
  {\bibfnamefont {M.-W.}\ \bibnamefont {Lin}}, \bibinfo {author} {\bibfnamefont
  {X.}~\bibnamefont {Li}}, \bibinfo {author} {\bibfnamefont {M.}~\bibnamefont
  {Mahjouri-Samani}}, \bibinfo {author} {\bibfnamefont {S.}~\bibnamefont
  {Jesse}}, \bibinfo {author} {\bibfnamefont {B.~G.}\ \bibnamefont {Sumpter}},
  \bibinfo {author} {\bibfnamefont {S.~V.}\ \bibnamefont {Kalinin}}, \bibinfo
  {author} {\bibfnamefont {D.~C.}\ \bibnamefont {Joy}}, \bibinfo {author}
  {\bibfnamefont {K.}~\bibnamefont {Xiao}}, \bibinfo {author} {\bibfnamefont
  {A.}~\bibnamefont {Belianinov}},\ and\ \bibinfo {author} {\bibfnamefont
  {O.~S.}\ \bibnamefont {Ovchinnikova}},\ }\bibfield  {title} {\bibinfo {title}
  {Nanoforging single layer mose$_2$ through defect engineering with focused
  helium ion beams},\ }\href {https://doi.org/10.1038/srep30481} {\bibfield
  {journal} {\bibinfo  {journal} {Scientific Reports}\ }\textbf {\bibinfo
  {volume} {6}},\ \bibinfo {pages} {30481} (\bibinfo {year}
  {2016})}\BibitemShut {NoStop}%
\bibitem [{\citenamefont {Mahjouri-Samani}\ \emph {et~al.}(2016)\citenamefont
  {Mahjouri-Samani}, \citenamefont {Liang}, \citenamefont {Oyedele},
  \citenamefont {Kim}, \citenamefont {Tian}, \citenamefont {Cross},
  \citenamefont {Wang}, \citenamefont {Lin}, \citenamefont {Boulesbaa},
  \citenamefont {Rouleau}, \citenamefont {Puretzky}, \citenamefont {Xiao},
  \citenamefont {Yoon}, \citenamefont {Eres}, \citenamefont {Duscher},
  \citenamefont {Sumpter},\ and\ \citenamefont {Geohegan}}]{Samani2016}%
  \BibitemOpen
  \bibfield  {author} {\bibinfo {author} {\bibfnamefont {M.}~\bibnamefont
  {Mahjouri-Samani}}, \bibinfo {author} {\bibfnamefont {L.}~\bibnamefont
  {Liang}}, \bibinfo {author} {\bibfnamefont {A.}~\bibnamefont {Oyedele}},
  \bibinfo {author} {\bibfnamefont {Y.-S.}\ \bibnamefont {Kim}}, \bibinfo
  {author} {\bibfnamefont {M.}~\bibnamefont {Tian}}, \bibinfo {author}
  {\bibfnamefont {N.}~\bibnamefont {Cross}}, \bibinfo {author} {\bibfnamefont
  {K.}~\bibnamefont {Wang}}, \bibinfo {author} {\bibfnamefont {M.-W.}\
  \bibnamefont {Lin}}, \bibinfo {author} {\bibfnamefont {A.}~\bibnamefont
  {Boulesbaa}}, \bibinfo {author} {\bibfnamefont {C.~M.}\ \bibnamefont
  {Rouleau}}, \bibinfo {author} {\bibfnamefont {A.~A.}\ \bibnamefont
  {Puretzky}}, \bibinfo {author} {\bibfnamefont {K.}~\bibnamefont {Xiao}},
  \bibinfo {author} {\bibfnamefont {M.}~\bibnamefont {Yoon}}, \bibinfo {author}
  {\bibfnamefont {G.}~\bibnamefont {Eres}}, \bibinfo {author} {\bibfnamefont
  {G.}~\bibnamefont {Duscher}}, \bibinfo {author} {\bibfnamefont {B.~G.}\
  \bibnamefont {Sumpter}},\ and\ \bibinfo {author} {\bibfnamefont {D.~B.}\
  \bibnamefont {Geohegan}},\ }\bibfield  {title} {\bibinfo {title} {Tailoring
  vacancies far beyond intrinsic levels changes the carrier type and optical
  response in monolayer mose$_{2-x}$ crystals},\ }\href
  {https://doi.org/10.1021/acs.nanolett.6b02263} {\bibfield  {journal}
  {\bibinfo  {journal} {Nano Letters}\ }\textbf {\bibinfo {volume} {16}},\
  \bibinfo {pages} {5213} (\bibinfo {year} {2016})},\ \bibinfo {note} {pMID:
  27416103},\ \Eprint
  {https://arxiv.org/abs/https://doi.org/10.1021/acs.nanolett.6b02263}
  {https://doi.org/10.1021/acs.nanolett.6b02263} \BibitemShut {NoStop}%
\bibitem [{\citenamefont {Shafqat}\ \emph {et~al.}(2017)\citenamefont
  {Shafqat}, \citenamefont {Iqbal},\ and\ \citenamefont {Majid}}]{Shafqat2017}%
  \BibitemOpen
  \bibfield  {author} {\bibinfo {author} {\bibfnamefont {A.}~\bibnamefont
  {Shafqat}}, \bibinfo {author} {\bibfnamefont {T.}~\bibnamefont {Iqbal}},\
  and\ \bibinfo {author} {\bibfnamefont {A.}~\bibnamefont {Majid}},\ }\bibfield
   {title} {\bibinfo {title} {A dft study of intrinsic point defects in
  monolayer mose$_2$},\ }\href {https://doi.org/10.1063/1.4999524} {\bibfield
  {journal} {\bibinfo  {journal} {AIP Advances}\ }\textbf {\bibinfo {volume}
  {7}},\ \bibinfo {pages} {105306} (\bibinfo {year} {2017})},\ \Eprint
  {https://arxiv.org/abs/https://doi.org/10.1063/1.4999524}
  {https://doi.org/10.1063/1.4999524} \BibitemShut {NoStop}%
\bibitem [{\citenamefont {Tanoh}\ \emph {et~al.}(2021)\citenamefont {Tanoh},
  \citenamefont {Alexander-Webber}, \citenamefont {Fan}, \citenamefont
  {Gauriot}, \citenamefont {Xiao}, \citenamefont {Pandya}, \citenamefont {Li},
  \citenamefont {Hofmann},\ and\ \citenamefont {Rao}}]{Tanoh2021}%
  \BibitemOpen
  \bibfield  {author} {\bibinfo {author} {\bibfnamefont {A.~O.~A.}\
  \bibnamefont {Tanoh}}, \bibinfo {author} {\bibfnamefont {J.}~\bibnamefont
  {Alexander-Webber}}, \bibinfo {author} {\bibfnamefont {Y.}~\bibnamefont
  {Fan}}, \bibinfo {author} {\bibfnamefont {N.}~\bibnamefont {Gauriot}},
  \bibinfo {author} {\bibfnamefont {J.}~\bibnamefont {Xiao}}, \bibinfo {author}
  {\bibfnamefont {R.}~\bibnamefont {Pandya}}, \bibinfo {author} {\bibfnamefont
  {Z.}~\bibnamefont {Li}}, \bibinfo {author} {\bibfnamefont {S.}~\bibnamefont
  {Hofmann}},\ and\ \bibinfo {author} {\bibfnamefont {A.}~\bibnamefont {Rao}},\
  }\bibfield  {title} {\bibinfo {title} {Giant photoluminescence enhancement in
  mose$_2$ monolayers treated with oleic acid ligands},\ }\href
  {https://doi.org/10.1039/D0NA01014F} {\bibfield  {journal} {\bibinfo
  {journal} {Nanoscale Adv.}\ }\textbf {\bibinfo {volume} {3}},\ \bibinfo
  {pages} {4216} (\bibinfo {year} {2021})}\BibitemShut {NoStop}%
\bibitem [{\citenamefont {Amani}\ \emph {et~al.}(2015)\citenamefont {Amani},
  \citenamefont {Lien}, \citenamefont {Kiriya}, \citenamefont {Xiao},
  \citenamefont {Azcatl}, \citenamefont {Noh}, \citenamefont {Madhvapathy},
  \citenamefont {Addou}, \citenamefont {KC}, \citenamefont {Dubey},
  \citenamefont {Cho}, \citenamefont {Wallace}, \citenamefont {Lee},
  \citenamefont {He}, \citenamefont {Ager}, \citenamefont {Zhang},
  \citenamefont {Yablonovitch},\ and\ \citenamefont {Javey}}]{Amani2015}%
  \BibitemOpen
  \bibfield  {author} {\bibinfo {author} {\bibfnamefont {M.}~\bibnamefont
  {Amani}}, \bibinfo {author} {\bibfnamefont {D.-H.}\ \bibnamefont {Lien}},
  \bibinfo {author} {\bibfnamefont {D.}~\bibnamefont {Kiriya}}, \bibinfo
  {author} {\bibfnamefont {J.}~\bibnamefont {Xiao}}, \bibinfo {author}
  {\bibfnamefont {A.}~\bibnamefont {Azcatl}}, \bibinfo {author} {\bibfnamefont
  {J.}~\bibnamefont {Noh}}, \bibinfo {author} {\bibfnamefont {S.~R.}\
  \bibnamefont {Madhvapathy}}, \bibinfo {author} {\bibfnamefont
  {R.}~\bibnamefont {Addou}}, \bibinfo {author} {\bibfnamefont
  {S.}~\bibnamefont {KC}}, \bibinfo {author} {\bibfnamefont {M.}~\bibnamefont
  {Dubey}}, \bibinfo {author} {\bibfnamefont {K.}~\bibnamefont {Cho}}, \bibinfo
  {author} {\bibfnamefont {R.~M.}\ \bibnamefont {Wallace}}, \bibinfo {author}
  {\bibfnamefont {S.-C.}\ \bibnamefont {Lee}}, \bibinfo {author} {\bibfnamefont
  {J.-H.}\ \bibnamefont {He}}, \bibinfo {author} {\bibfnamefont {J.~W.}\
  \bibnamefont {Ager}}, \bibinfo {author} {\bibfnamefont {X.}~\bibnamefont
  {Zhang}}, \bibinfo {author} {\bibfnamefont {E.}~\bibnamefont
  {Yablonovitch}},\ and\ \bibinfo {author} {\bibfnamefont {A.}~\bibnamefont
  {Javey}},\ }\bibfield  {title} {\bibinfo {title} {Near-unity
  photoluminescence quantum yield in mos$_2$},\ }\href
  {https://doi.org/10.1126/science.aad2114} {\bibfield  {journal} {\bibinfo
  {journal} {Science}\ }\textbf {\bibinfo {volume} {350}},\ \bibinfo {pages}
  {1065} (\bibinfo {year} {2015})},\ \Eprint
  {https://arxiv.org/abs/https://www.science.org/doi/pdf/10.1126/science.aad2114}
  {https://www.science.org/doi/pdf/10.1126/science.aad2114} \BibitemShut
  {NoStop}%
\bibitem [{\citenamefont {Li}\ \emph {et~al.}(2017)\citenamefont {Li},
  \citenamefont {Puretzky}, \citenamefont {Sang}, \citenamefont {KC},
  \citenamefont {Tian}, \citenamefont {Ceballos}, \citenamefont
  {Mahjouri-Samani}, \citenamefont {Wang}, \citenamefont {Unocic},
  \citenamefont {Zhao}, \citenamefont {Duscher}, \citenamefont {Cooper},
  \citenamefont {Rouleau}, \citenamefont {Geohegan},\ and\ \citenamefont
  {Xiao}}]{Li2017}%
  \BibitemOpen
  \bibfield  {author} {\bibinfo {author} {\bibfnamefont {X.}~\bibnamefont
  {Li}}, \bibinfo {author} {\bibfnamefont {A.~A.}\ \bibnamefont {Puretzky}},
  \bibinfo {author} {\bibfnamefont {X.}~\bibnamefont {Sang}}, \bibinfo {author}
  {\bibfnamefont {S.}~\bibnamefont {KC}}, \bibinfo {author} {\bibfnamefont
  {M.}~\bibnamefont {Tian}}, \bibinfo {author} {\bibfnamefont {F.}~\bibnamefont
  {Ceballos}}, \bibinfo {author} {\bibfnamefont {M.}~\bibnamefont
  {Mahjouri-Samani}}, \bibinfo {author} {\bibfnamefont {K.}~\bibnamefont
  {Wang}}, \bibinfo {author} {\bibfnamefont {R.~R.}\ \bibnamefont {Unocic}},
  \bibinfo {author} {\bibfnamefont {H.}~\bibnamefont {Zhao}}, \bibinfo {author}
  {\bibfnamefont {G.}~\bibnamefont {Duscher}}, \bibinfo {author} {\bibfnamefont
  {V.~R.}\ \bibnamefont {Cooper}}, \bibinfo {author} {\bibfnamefont {C.~M.}\
  \bibnamefont {Rouleau}}, \bibinfo {author} {\bibfnamefont {D.~B.}\
  \bibnamefont {Geohegan}},\ and\ \bibinfo {author} {\bibfnamefont
  {K.}~\bibnamefont {Xiao}},\ }\bibfield  {title} {\bibinfo {title}
  {Suppression of defects and deep levels using isoelectronic tungsten
  substitution in monolayer mose$_2$},\ }\href
  {https://doi.org/https://doi.org/10.1002/adfm.201603850} {\bibfield
  {journal} {\bibinfo  {journal} {Advanced Functional Materials}\ }\textbf
  {\bibinfo {volume} {27}},\ \bibinfo {pages} {1603850} (\bibinfo {year}
  {2017})}\BibitemShut {NoStop}%
\bibitem [{\citenamefont {Klein}\ \emph {et~al.}(2019)\citenamefont {Klein},
  \citenamefont {Lorke}, \citenamefont {Florian}, \citenamefont {Sigger},
  \citenamefont {Sigl}, \citenamefont {Rey}, \citenamefont {Wierzbowski},
  \citenamefont {Cerne}, \citenamefont {Müller}, \citenamefont {Mitterreiter},
  \citenamefont {Zimmermann}, \citenamefont {Taniguchi}, \citenamefont
  {Watanabe}, \citenamefont {Wurstbauer}, \citenamefont {Kaniber},
  \citenamefont {Knap}, \citenamefont {Schmidt}, \citenamefont {Finley},\ and\
  \citenamefont {Holleitner}}]{Klein2019}%
  \BibitemOpen
  \bibfield  {author} {\bibinfo {author} {\bibfnamefont {J.}~\bibnamefont
  {Klein}}, \bibinfo {author} {\bibfnamefont {M.}~\bibnamefont {Lorke}},
  \bibinfo {author} {\bibfnamefont {M.}~\bibnamefont {Florian}}, \bibinfo
  {author} {\bibfnamefont {F.}~\bibnamefont {Sigger}}, \bibinfo {author}
  {\bibfnamefont {L.}~\bibnamefont {Sigl}}, \bibinfo {author} {\bibfnamefont
  {S.}~\bibnamefont {Rey}}, \bibinfo {author} {\bibfnamefont {J.}~\bibnamefont
  {Wierzbowski}}, \bibinfo {author} {\bibfnamefont {J.}~\bibnamefont {Cerne}},
  \bibinfo {author} {\bibfnamefont {K.}~\bibnamefont {Müller}}, \bibinfo
  {author} {\bibfnamefont {E.}~\bibnamefont {Mitterreiter}}, \bibinfo {author}
  {\bibfnamefont {P.}~\bibnamefont {Zimmermann}}, \bibinfo {author}
  {\bibfnamefont {T.}~\bibnamefont {Taniguchi}}, \bibinfo {author}
  {\bibfnamefont {K.}~\bibnamefont {Watanabe}}, \bibinfo {author}
  {\bibfnamefont {U.}~\bibnamefont {Wurstbauer}}, \bibinfo {author}
  {\bibfnamefont {M.}~\bibnamefont {Kaniber}}, \bibinfo {author} {\bibfnamefont
  {M.}~\bibnamefont {Knap}}, \bibinfo {author} {\bibfnamefont {R.}~\bibnamefont
  {Schmidt}}, \bibinfo {author} {\bibfnamefont {J.~J.}\ \bibnamefont
  {Finley}},\ and\ \bibinfo {author} {\bibfnamefont {A.~W.}\ \bibnamefont
  {Holleitner}},\ }\bibfield  {title} {\bibinfo {title} {Site-selectively
  generated photon emitters in monolayer ${\mathrm{mos}}_{2}$ via local helium
  ion irradiation},\ }\href@noop {} {\bibfield  {journal} {\bibinfo  {journal}
  {Nat. Commun.}\ }\textbf {\bibinfo {volume} {10}},\ \bibinfo {pages} {2755}
  (\bibinfo {year} {2019})}\BibitemShut {NoStop}%
\bibitem [{\citenamefont {Mitterreiter}\ \emph {et~al.}(2021)\citenamefont
  {Mitterreiter}, \citenamefont {Schuler}, \citenamefont {Micevic},
  \citenamefont {Hernangómez-Pérez}, \citenamefont {Barthelmi}, \citenamefont
  {Cochrane}, \citenamefont {Kiemle}, \citenamefont {Sigger}, \citenamefont
  {Klein}, \citenamefont {Wong}, \citenamefont {Watanabe}, \citenamefont
  {Taniguchi}, \citenamefont {Lorke}, \citenamefont {Jahnke}, \citenamefont
  {Finley}, \citenamefont {Schwartzberg}, \citenamefont {Qiu}, \citenamefont
  {Refaely-Abramson}, \citenamefont {Holleitner}, \citenamefont
  {Weber-Bargioni},\ and\ \citenamefont {Kastl}}]{Mitterreiter2021}%
  \BibitemOpen
  \bibfield  {author} {\bibinfo {author} {\bibfnamefont {E.}~\bibnamefont
  {Mitterreiter}}, \bibinfo {author} {\bibfnamefont {B.}~\bibnamefont
  {Schuler}}, \bibinfo {author} {\bibfnamefont {A.}~\bibnamefont {Micevic}},
  \bibinfo {author} {\bibfnamefont {D.}~\bibnamefont {Hernangómez-Pérez}},
  \bibinfo {author} {\bibfnamefont {K.}~\bibnamefont {Barthelmi}}, \bibinfo
  {author} {\bibfnamefont {K.~A.}\ \bibnamefont {Cochrane}}, \bibinfo {author}
  {\bibfnamefont {J.}~\bibnamefont {Kiemle}}, \bibinfo {author} {\bibfnamefont
  {F.}~\bibnamefont {Sigger}}, \bibinfo {author} {\bibfnamefont
  {J.}~\bibnamefont {Klein}}, \bibinfo {author} {\bibfnamefont {E.~S.}\
  \bibnamefont {Wong}, \bibfnamefont {Edward~Barnard}}, \bibinfo {author}
  {\bibfnamefont {K.}~\bibnamefont {Watanabe}}, \bibinfo {author}
  {\bibfnamefont {T.}~\bibnamefont {Taniguchi}}, \bibinfo {author}
  {\bibfnamefont {M.}~\bibnamefont {Lorke}}, \bibinfo {author} {\bibfnamefont
  {F.}~\bibnamefont {Jahnke}}, \bibinfo {author} {\bibfnamefont {J.~J.}\
  \bibnamefont {Finley}}, \bibinfo {author} {\bibfnamefont {A.~M.}\
  \bibnamefont {Schwartzberg}}, \bibinfo {author} {\bibfnamefont {D.~Y.}\
  \bibnamefont {Qiu}}, \bibinfo {author} {\bibfnamefont {S.}~\bibnamefont
  {Refaely-Abramson}}, \bibinfo {author} {\bibfnamefont {A.~W.}\ \bibnamefont
  {Holleitner}}, \bibinfo {author} {\bibfnamefont {A.}~\bibnamefont
  {Weber-Bargioni}},\ and\ \bibinfo {author} {\bibfnamefont {C.}~\bibnamefont
  {Kastl}},\ }\bibfield  {title} {\bibinfo {title} {The role of chalcogen
  vacancies for atomic defect emission in ${\mathrm{mos}}_{2}$},\ }\href@noop
  {} {\bibfield  {journal} {\bibinfo  {journal} {Nat. Commun.}\ }\textbf
  {\bibinfo {volume} {12}},\ \bibinfo {pages} {3822} (\bibinfo {year}
  {2021})}\BibitemShut {NoStop}%
\bibitem [{\citenamefont {O’Donnell}\ and\ \citenamefont
  {Chen}(1991)}]{ODonnell1991}%
  \BibitemOpen
  \bibfield  {author} {\bibinfo {author} {\bibfnamefont {K.~P.}\ \bibnamefont
  {O’Donnell}}\ and\ \bibinfo {author} {\bibfnamefont {X.}~\bibnamefont
  {Chen}},\ }\bibfield  {title} {\bibinfo {title} {Temperature dependence of
  semiconductor band gaps},\ }\href {https://doi.org/10.1063/1.104723}
  {\bibfield  {journal} {\bibinfo  {journal} {Applied Physics Letters}\
  }\textbf {\bibinfo {volume} {58}},\ \bibinfo {pages} {2924} (\bibinfo {year}
  {1991})},\ \Eprint {https://arxiv.org/abs/https://doi.org/10.1063/1.104723}
  {https://doi.org/10.1063/1.104723} \BibitemShut {NoStop}%
\bibitem [{\citenamefont {Lucchese}\ \emph {et~al.}(2010)\citenamefont
  {Lucchese}, \citenamefont {Stavale}, \citenamefont {Ferreira}, \citenamefont
  {Vilani}, \citenamefont {Moutinho}, \citenamefont {Capaz}, \citenamefont
  {Achete},\ and\ \citenamefont {Jorio}}]{Lucchese2010}%
  \BibitemOpen
  \bibfield  {author} {\bibinfo {author} {\bibfnamefont {M.}~\bibnamefont
  {Lucchese}}, \bibinfo {author} {\bibfnamefont {F.}~\bibnamefont {Stavale}},
  \bibinfo {author} {\bibfnamefont {E.~M.}\ \bibnamefont {Ferreira}}, \bibinfo
  {author} {\bibfnamefont {C.}~\bibnamefont {Vilani}}, \bibinfo {author}
  {\bibfnamefont {M.}~\bibnamefont {Moutinho}}, \bibinfo {author}
  {\bibfnamefont {R.~B.}\ \bibnamefont {Capaz}}, \bibinfo {author}
  {\bibfnamefont {C.}~\bibnamefont {Achete}},\ and\ \bibinfo {author}
  {\bibfnamefont {A.}~\bibnamefont {Jorio}},\ }\bibfield  {title} {\bibinfo
  {title} {Quantifying ion-induced defects and raman relaxation length in
  graphene},\ }\href
  {https://doi.org/https://doi.org/10.1016/j.carbon.2009.12.057} {\bibfield
  {journal} {\bibinfo  {journal} {Carbon}\ }\textbf {\bibinfo {volume} {48}},\
  \bibinfo {pages} {1592} (\bibinfo {year} {2010})}\BibitemShut {NoStop}%
\bibitem [{\citenamefont {Lehtinen}\ \emph {et~al.}(2010)\citenamefont
  {Lehtinen}, \citenamefont {Kotakoski}, \citenamefont {Krasheninnikov},
  \citenamefont {Tolvanen}, \citenamefont {Nordlund},\ and\ \citenamefont
  {Keinonen}}]{Lehtinen2010}%
  \BibitemOpen
  \bibfield  {author} {\bibinfo {author} {\bibfnamefont {O.}~\bibnamefont
  {Lehtinen}}, \bibinfo {author} {\bibfnamefont {J.}~\bibnamefont {Kotakoski}},
  \bibinfo {author} {\bibfnamefont {A.~V.}\ \bibnamefont {Krasheninnikov}},
  \bibinfo {author} {\bibfnamefont {A.}~\bibnamefont {Tolvanen}}, \bibinfo
  {author} {\bibfnamefont {K.}~\bibnamefont {Nordlund}},\ and\ \bibinfo
  {author} {\bibfnamefont {J.}~\bibnamefont {Keinonen}},\ }\bibfield  {title}
  {\bibinfo {title} {Effects of ion bombardment on a two-dimensional target:
  Atomistic simulations of graphene irradiation},\ }\href
  {https://doi.org/10.1103/PhysRevB.81.153401} {\bibfield  {journal} {\bibinfo
  {journal} {Phys. Rev. B}\ }\textbf {\bibinfo {volume} {81}},\ \bibinfo
  {pages} {153401} (\bibinfo {year} {2010})}\BibitemShut {NoStop}%
\bibitem [{\citenamefont {Junge}\ \emph {et~al.}(2023)\citenamefont {Junge},
  \citenamefont {Auge}, \citenamefont {Zarkua},\ and\ \citenamefont
  {Hofsäss}}]{Junge2023}%
  \BibitemOpen
  \bibfield  {author} {\bibinfo {author} {\bibfnamefont {F.}~\bibnamefont
  {Junge}}, \bibinfo {author} {\bibfnamefont {M.}~\bibnamefont {Auge}},
  \bibinfo {author} {\bibfnamefont {Z.}~\bibnamefont {Zarkua}},\ and\ \bibinfo
  {author} {\bibfnamefont {H.}~\bibnamefont {Hofsäss}},\ }\bibfield  {title}
  {\bibinfo {title} {Lateral controlled doping and defect engineering of
  graphene by ultra-low-energy ion implantation},\ }\bibfield  {journal}
  {\bibinfo  {journal} {Nanomaterials}\ }\textbf {\bibinfo {volume} {13}},\
  \href {https://doi.org/10.3390/nano13040658} {10.3390/nano13040658} (\bibinfo
  {year} {2023})\BibitemShut {NoStop}%
\bibitem [{\citenamefont {Kianinia}\ \emph {et~al.}(2020)\citenamefont
  {Kianinia}, \citenamefont {White}, \citenamefont {Fröch}, \citenamefont
  {Bradac},\ and\ \citenamefont {Aharonovich}}]{Kianinia2020}%
  \BibitemOpen
  \bibfield  {author} {\bibinfo {author} {\bibfnamefont {M.}~\bibnamefont
  {Kianinia}}, \bibinfo {author} {\bibfnamefont {S.}~\bibnamefont {White}},
  \bibinfo {author} {\bibfnamefont {J.~E.}\ \bibnamefont {Fröch}}, \bibinfo
  {author} {\bibfnamefont {C.}~\bibnamefont {Bradac}},\ and\ \bibinfo {author}
  {\bibfnamefont {I.}~\bibnamefont {Aharonovich}},\ }\bibfield  {title}
  {\bibinfo {title} {Generation of spin defects in hexagonal boron nitride},\
  }\href {https://doi.org/10.1021/acsphotonics.0c00614} {\bibfield  {journal}
  {\bibinfo  {journal} {ACS Photonics}\ }\textbf {\bibinfo {volume} {7}},\
  \bibinfo {pages} {2147} (\bibinfo {year} {2020})},\ \Eprint
  {https://arxiv.org/abs/https://doi.org/10.1021/acsphotonics.0c00614}
  {https://doi.org/10.1021/acsphotonics.0c00614} \BibitemShut {NoStop}%
\bibitem [{\citenamefont {Shi}\ \emph {et~al.}(2016)\citenamefont {Shi},
  \citenamefont {Lin}, \citenamefont {Tan}, \citenamefont {Qiao}, \citenamefont
  {Zhang},\ and\ \citenamefont {Tan}}]{Shi2016}%
  \BibitemOpen
  \bibfield  {author} {\bibinfo {author} {\bibfnamefont {W.}~\bibnamefont
  {Shi}}, \bibinfo {author} {\bibfnamefont {M.-L.}\ \bibnamefont {Lin}},
  \bibinfo {author} {\bibfnamefont {Q.-H.}\ \bibnamefont {Tan}}, \bibinfo
  {author} {\bibfnamefont {X.-F.}\ \bibnamefont {Qiao}}, \bibinfo {author}
  {\bibfnamefont {J.}~\bibnamefont {Zhang}},\ and\ \bibinfo {author}
  {\bibfnamefont {P.-H.}\ \bibnamefont {Tan}},\ }\bibfield  {title} {\bibinfo
  {title} {Raman and photoluminescence spectra of two-dimensional
  nanocrystallites of monolayer ws$_2$ and wse$_2$},\ }\href
  {https://doi.org/10.1088/2053-1583/3/2/025016} {\bibfield  {journal}
  {\bibinfo  {journal} {2D Materials}\ }\textbf {\bibinfo {volume} {3}},\
  \bibinfo {pages} {025016} (\bibinfo {year} {2016})}\BibitemShut {NoStop}%
\bibitem [{\citenamefont {Murray}\ \emph {et~al.}(2016)\citenamefont {Murray},
  \citenamefont {Haynes}, \citenamefont {Zhao}, \citenamefont {Perry},
  \citenamefont {Hatem},\ and\ \citenamefont {Jones}}]{Murray2016}%
  \BibitemOpen
  \bibfield  {author} {\bibinfo {author} {\bibfnamefont {R.}~\bibnamefont
  {Murray}}, \bibinfo {author} {\bibfnamefont {K.}~\bibnamefont {Haynes}},
  \bibinfo {author} {\bibfnamefont {X.}~\bibnamefont {Zhao}}, \bibinfo {author}
  {\bibfnamefont {S.}~\bibnamefont {Perry}}, \bibinfo {author} {\bibfnamefont
  {C.}~\bibnamefont {Hatem}},\ and\ \bibinfo {author} {\bibfnamefont
  {K.}~\bibnamefont {Jones}},\ }\bibfield  {title} {\bibinfo {title} {The
  effect of low energy ion implantation on mos$_2$},\ }\href
  {https://doi.org/10.1149/2.0111611jss} {\bibfield  {journal} {\bibinfo
  {journal} {ECS Journal of Solid State Science and Technology}\ }\textbf
  {\bibinfo {volume} {5}},\ \bibinfo {pages} {Q3050} (\bibinfo {year}
  {2016})}\BibitemShut {NoStop}%
\bibitem [{\citenamefont {Maguire}\ \emph {et~al.}(2018)\citenamefont
  {Maguire}, \citenamefont {Fox}, \citenamefont {Zhou}, \citenamefont {Wang},
  \citenamefont {O'Brien}, \citenamefont {Jadwiszczak}, \citenamefont {Cullen},
  \citenamefont {McManus}, \citenamefont {Bateman}, \citenamefont {McEvoy},
  \citenamefont {Duesberg},\ and\ \citenamefont {Zhang}}]{Maguire2018}%
  \BibitemOpen
  \bibfield  {author} {\bibinfo {author} {\bibfnamefont {P.}~\bibnamefont
  {Maguire}}, \bibinfo {author} {\bibfnamefont {D.~S.}\ \bibnamefont {Fox}},
  \bibinfo {author} {\bibfnamefont {Y.}~\bibnamefont {Zhou}}, \bibinfo {author}
  {\bibfnamefont {Q.}~\bibnamefont {Wang}}, \bibinfo {author} {\bibfnamefont
  {M.}~\bibnamefont {O'Brien}}, \bibinfo {author} {\bibfnamefont
  {J.}~\bibnamefont {Jadwiszczak}}, \bibinfo {author} {\bibfnamefont {C.~P.}\
  \bibnamefont {Cullen}}, \bibinfo {author} {\bibfnamefont {J.}~\bibnamefont
  {McManus}}, \bibinfo {author} {\bibfnamefont {S.}~\bibnamefont {Bateman}},
  \bibinfo {author} {\bibfnamefont {N.}~\bibnamefont {McEvoy}}, \bibinfo
  {author} {\bibfnamefont {G.~S.}\ \bibnamefont {Duesberg}},\ and\ \bibinfo
  {author} {\bibfnamefont {H.}~\bibnamefont {Zhang}},\ }\bibfield  {title}
  {\bibinfo {title} {Defect sizing, separation, and substrate effects in
  ion-irradiated monolayer two-dimensional materials},\ }\href
  {https://doi.org/10.1103/PhysRevB.98.134109} {\bibfield  {journal} {\bibinfo
  {journal} {Phys. Rev. B}\ }\textbf {\bibinfo {volume} {98}},\ \bibinfo
  {pages} {134109} (\bibinfo {year} {2018})}\BibitemShut {NoStop}%
\bibitem [{\citenamefont {Klein}\ \emph {et~al.}(2021)\citenamefont {Klein},
  \citenamefont {Sigl}, \citenamefont {Gyger}, \citenamefont {Barthelmi},
  \citenamefont {Florian}, \citenamefont {Rey}, \citenamefont {Taniguchi},
  \citenamefont {Watanabe}, \citenamefont {Jahnke}, \citenamefont {Kastl},
  \citenamefont {Zwiller}, \citenamefont {Jöns}, \citenamefont {Müller},
  \citenamefont {Wurstbauer}, \citenamefont {Finley},\ and\ \citenamefont
  {Holleitner}}]{Klein2021}%
  \BibitemOpen
  \bibfield  {author} {\bibinfo {author} {\bibfnamefont {J.}~\bibnamefont
  {Klein}}, \bibinfo {author} {\bibfnamefont {L.}~\bibnamefont {Sigl}},
  \bibinfo {author} {\bibfnamefont {S.}~\bibnamefont {Gyger}}, \bibinfo
  {author} {\bibfnamefont {K.}~\bibnamefont {Barthelmi}}, \bibinfo {author}
  {\bibfnamefont {M.}~\bibnamefont {Florian}}, \bibinfo {author} {\bibfnamefont
  {S.}~\bibnamefont {Rey}}, \bibinfo {author} {\bibfnamefont {T.}~\bibnamefont
  {Taniguchi}}, \bibinfo {author} {\bibfnamefont {K.}~\bibnamefont {Watanabe}},
  \bibinfo {author} {\bibfnamefont {F.}~\bibnamefont {Jahnke}}, \bibinfo
  {author} {\bibfnamefont {C.}~\bibnamefont {Kastl}}, \bibinfo {author}
  {\bibfnamefont {V.}~\bibnamefont {Zwiller}}, \bibinfo {author} {\bibfnamefont
  {K.~D.}\ \bibnamefont {Jöns}}, \bibinfo {author} {\bibfnamefont
  {K.}~\bibnamefont {Müller}}, \bibinfo {author} {\bibfnamefont
  {U.}~\bibnamefont {Wurstbauer}}, \bibinfo {author} {\bibfnamefont {J.~J.}\
  \bibnamefont {Finley}},\ and\ \bibinfo {author} {\bibfnamefont {A.~W.}\
  \bibnamefont {Holleitner}},\ }\bibfield  {title} {\bibinfo {title}
  {Engineering the luminescence and generation of individual defect emitters in
  atomically thin mos$_2$},\ }\href
  {https://doi.org/10.1021/acsphotonics.0c01907} {\bibfield  {journal}
  {\bibinfo  {journal} {ACS Photonics}\ }\textbf {\bibinfo {volume} {8}},\
  \bibinfo {pages} {669} (\bibinfo {year} {2021})},\ \Eprint
  {https://arxiv.org/abs/https://doi.org/10.1021/acsphotonics.0c01907}
  {https://doi.org/10.1021/acsphotonics.0c01907} \BibitemShut {NoStop}%
\bibitem [{\citenamefont {Ghorbani-Asl}\ \emph {et~al.}(2017)\citenamefont
  {Ghorbani-Asl}, \citenamefont {Kretschmer}, \citenamefont {Spearot},\ and\
  \citenamefont {Krasheninnikov}}]{Ghorbani-Asl2017}%
  \BibitemOpen
  \bibfield  {author} {\bibinfo {author} {\bibfnamefont {M.}~\bibnamefont
  {Ghorbani-Asl}}, \bibinfo {author} {\bibfnamefont {S.}~\bibnamefont
  {Kretschmer}}, \bibinfo {author} {\bibfnamefont {D.~E.}\ \bibnamefont
  {Spearot}},\ and\ \bibinfo {author} {\bibfnamefont {A.~V.}\ \bibnamefont
  {Krasheninnikov}},\ }\bibfield  {title} {\bibinfo {title} {Two-dimensional
  mos$_2$ under ion irradiation: from controlled defect production to
  electronic structure engineering},\ }\href
  {https://doi.org/10.1088/2053-1583/aa6b17} {\bibfield  {journal} {\bibinfo
  {journal} {2D Materials}\ }\textbf {\bibinfo {volume} {4}},\ \bibinfo {pages}
  {025078} (\bibinfo {year} {2017})}\BibitemShut {NoStop}%
\bibitem [{\citenamefont {Ghaderzadeh}\ \emph {et~al.}(2020)\citenamefont
  {Ghaderzadeh}, \citenamefont {Ladygin}, \citenamefont {Ghorbani-Asl},
  \citenamefont {Hlawacek}, \citenamefont {Schleberger},\ and\ \citenamefont
  {Krasheninnikov}}]{Ghaderzadeh2020}%
  \BibitemOpen
  \bibfield  {author} {\bibinfo {author} {\bibfnamefont {S.}~\bibnamefont
  {Ghaderzadeh}}, \bibinfo {author} {\bibfnamefont {V.}~\bibnamefont
  {Ladygin}}, \bibinfo {author} {\bibfnamefont {M.}~\bibnamefont
  {Ghorbani-Asl}}, \bibinfo {author} {\bibfnamefont {G.}~\bibnamefont
  {Hlawacek}}, \bibinfo {author} {\bibfnamefont {M.}~\bibnamefont
  {Schleberger}},\ and\ \bibinfo {author} {\bibfnamefont {A.~V.}\ \bibnamefont
  {Krasheninnikov}},\ }\bibfield  {title} {\bibinfo {title} {Freestanding and
  supported mos$_2$ monolayers under cluster irradiation: Insights from
  molecular dynamics simulations},\ }\href
  {https://doi.org/10.1021/acsami.0c09255} {\bibfield  {journal} {\bibinfo
  {journal} {ACS Applied Materials \& Interfaces}\ }\textbf {\bibinfo {volume}
  {12}},\ \bibinfo {pages} {37454} (\bibinfo {year} {2020})},\ \bibinfo {note}
  {pMID: 32814400},\ \Eprint
  {https://arxiv.org/abs/https://doi.org/10.1021/acsami.0c09255}
  {https://doi.org/10.1021/acsami.0c09255} \BibitemShut {NoStop}%
\bibitem [{\citenamefont {Azam}\ \emph {et~al.}(2022)\citenamefont {Azam},
  \citenamefont {Boebinger}, \citenamefont {Jaiswal}, \citenamefont {Unocic},
  \citenamefont {Fathi-Hafshejani},\ and\ \citenamefont
  {Mahjouri-Samani}}]{Azam2022}%
  \BibitemOpen
  \bibfield  {author} {\bibinfo {author} {\bibfnamefont {N.}~\bibnamefont
  {Azam}}, \bibinfo {author} {\bibfnamefont {M.~G.}\ \bibnamefont {Boebinger}},
  \bibinfo {author} {\bibfnamefont {S.}~\bibnamefont {Jaiswal}}, \bibinfo
  {author} {\bibfnamefont {R.~R.}\ \bibnamefont {Unocic}}, \bibinfo {author}
  {\bibfnamefont {P.}~\bibnamefont {Fathi-Hafshejani}},\ and\ \bibinfo {author}
  {\bibfnamefont {M.}~\bibnamefont {Mahjouri-Samani}},\ }\bibfield  {title}
  {\bibinfo {title} {Laser-assisted synthesis of monolayer 2d mose$_2$ crystals
  with tunable vacancy concentrations: Implications for gas and biosensing},\
  }\href {https://doi.org/10.1021/acsanm.2c01458} {\bibfield  {journal}
  {\bibinfo  {journal} {ACS Applied Nano Materials}\ }\textbf {\bibinfo
  {volume} {5}},\ \bibinfo {pages} {9129} (\bibinfo {year} {2022})},\ \Eprint
  {https://arxiv.org/abs/https://doi.org/10.1021/acsanm.2c0145}
  {https://doi.org/10.1021/acsanm.2c0145} \BibitemShut {NoStop}%
\bibitem [{\citenamefont {Kim}\ \emph {et~al.}(2022)\citenamefont {Kim},
  \citenamefont {Van~Quang}, \citenamefont {Nguyen}, \citenamefont {Kim},
  \citenamefont {Lee}, \citenamefont {Lee}, \citenamefont {Cho}, \citenamefont
  {Seong}, \citenamefont {Kim},\ and\ \citenamefont {Chang}}]{Kim2022}%
  \BibitemOpen
  \bibfield  {author} {\bibinfo {author} {\bibfnamefont {H.~J.}\ \bibnamefont
  {Kim}}, \bibinfo {author} {\bibfnamefont {N.}~\bibnamefont {Van~Quang}},
  \bibinfo {author} {\bibfnamefont {T.~H.}\ \bibnamefont {Nguyen}}, \bibinfo
  {author} {\bibfnamefont {S.}~\bibnamefont {Kim}}, \bibinfo {author}
  {\bibfnamefont {Y.}~\bibnamefont {Lee}}, \bibinfo {author} {\bibfnamefont
  {I.~H.}\ \bibnamefont {Lee}}, \bibinfo {author} {\bibfnamefont
  {S.}~\bibnamefont {Cho}}, \bibinfo {author} {\bibfnamefont {M.-J.}\
  \bibnamefont {Seong}}, \bibinfo {author} {\bibfnamefont {K.}~\bibnamefont
  {Kim}},\ and\ \bibinfo {author} {\bibfnamefont {Y.~J.}\ \bibnamefont
  {Chang}},\ }\bibfield  {title} {\bibinfo {title} {Tuning of thermoelectric
  properties of mose$_2$ thin films under helium ion irradiation},\ }\href
  {https://doi.org/10.1186/s11671-022-03665-9} {\bibfield  {journal} {\bibinfo
  {journal} {Nanoscale Research Letters}\ }\textbf {\bibinfo {volume} {17}},\
  \bibinfo {pages} {26} (\bibinfo {year} {2022})}\BibitemShut {NoStop}%
\bibitem [{\citenamefont {Xiao}\ \emph {et~al.}(2020)\citenamefont {Xiao},
  \citenamefont {Ruan}, \citenamefont {Bao}, \citenamefont {Luo}, \citenamefont
  {Huang}, \citenamefont {Tang}, \citenamefont {Shen}, \citenamefont {Cheng},\
  and\ \citenamefont {Chu}}]{Xiao2020}%
  \BibitemOpen
  \bibfield  {author} {\bibinfo {author} {\bibfnamefont {D.}~\bibnamefont
  {Xiao}}, \bibinfo {author} {\bibfnamefont {Q.}~\bibnamefont {Ruan}}, \bibinfo
  {author} {\bibfnamefont {D.-L.}\ \bibnamefont {Bao}}, \bibinfo {author}
  {\bibfnamefont {Y.}~\bibnamefont {Luo}}, \bibinfo {author} {\bibfnamefont
  {C.}~\bibnamefont {Huang}}, \bibinfo {author} {\bibfnamefont
  {S.}~\bibnamefont {Tang}}, \bibinfo {author} {\bibfnamefont {J.}~\bibnamefont
  {Shen}}, \bibinfo {author} {\bibfnamefont {C.}~\bibnamefont {Cheng}},\ and\
  \bibinfo {author} {\bibfnamefont {P.~K.}\ \bibnamefont {Chu}},\ }\bibfield
  {title} {\bibinfo {title} {Effects of ion energy and density on the plasma
  etching-induced surface area, edge electrical field, and multivacancies in
  mose$_2$ nanosheets for enhancement of the hydrogen evolution reaction},\
  }\href {https://doi.org/https://doi.org/10.1002/smll.202001470} {\bibfield
  {journal} {\bibinfo  {journal} {Small}\ }\textbf {\bibinfo {volume} {16}},\
  \bibinfo {pages} {2001470} (\bibinfo {year} {2020})},\ \Eprint
  {https://arxiv.org/abs/https://onlinelibrary.wiley.com/doi/pdf/10.1002/smll.202001470}
  {https://onlinelibrary.wiley.com/doi/pdf/10.1002/smll.202001470} \BibitemShut
  {NoStop}%
\bibitem [{\citenamefont {Hennessy}\ \emph {et~al.}(2022)\citenamefont
  {Hennessy}, \citenamefont {O'Connell}, \citenamefont {Auge}, \citenamefont
  {Moynihan}, \citenamefont {Hofsäss},\ and\ \citenamefont
  {Bangert}}]{Hennessy2022}%
  \BibitemOpen
  \bibfield  {author} {\bibinfo {author} {\bibfnamefont {M.}~\bibnamefont
  {Hennessy}}, \bibinfo {author} {\bibfnamefont {E.~N.}\ \bibnamefont
  {O'Connell}}, \bibinfo {author} {\bibfnamefont {M.}~\bibnamefont {Auge}},
  \bibinfo {author} {\bibfnamefont {E.}~\bibnamefont {Moynihan}}, \bibinfo
  {author} {\bibfnamefont {H.}~\bibnamefont {Hofsäss}},\ and\ \bibinfo
  {author} {\bibfnamefont {U.}~\bibnamefont {Bangert}},\ }\bibfield  {title}
  {\bibinfo {title} {Quantification of ion-implanted single-atom dopants in
  monolayer mos$_2$ via haadf stem using the temul toolkit},\ }\href
  {https://doi.org/10.1017/S1431927622000757} {\bibfield  {journal} {\bibinfo
  {journal} {Microscopy and Microanalysis}\ }\textbf {\bibinfo {volume} {28}},\
  \bibinfo {pages} {1407–1416} (\bibinfo {year} {2022})}\BibitemShut
  {NoStop}%
\bibitem [{\citenamefont {Robert}\ \emph {et~al.}(2021)\citenamefont {Robert},
  \citenamefont {Park}, \citenamefont {Cadiz}, \citenamefont {Lombez},
  \citenamefont {Ren}, \citenamefont {Tornatzky}, \citenamefont {Rowe},
  \citenamefont {Paget}, \citenamefont {Sirotti}, \citenamefont {Yang},
  \citenamefont {Van~Tuan}, \citenamefont {Taniguchi}, \citenamefont
  {Urbaszek}, \citenamefont {Watanabe}, \citenamefont {Amand}, \citenamefont
  {Dery},\ and\ \citenamefont {Marie}}]{Robert2021}%
  \BibitemOpen
  \bibfield  {author} {\bibinfo {author} {\bibfnamefont {C.}~\bibnamefont
  {Robert}}, \bibinfo {author} {\bibfnamefont {S.}~\bibnamefont {Park}},
  \bibinfo {author} {\bibfnamefont {F.}~\bibnamefont {Cadiz}}, \bibinfo
  {author} {\bibfnamefont {L.}~\bibnamefont {Lombez}}, \bibinfo {author}
  {\bibfnamefont {L.}~\bibnamefont {Ren}}, \bibinfo {author} {\bibfnamefont
  {H.}~\bibnamefont {Tornatzky}}, \bibinfo {author} {\bibfnamefont
  {A.}~\bibnamefont {Rowe}}, \bibinfo {author} {\bibfnamefont {D.}~\bibnamefont
  {Paget}}, \bibinfo {author} {\bibfnamefont {F.}~\bibnamefont {Sirotti}},
  \bibinfo {author} {\bibfnamefont {M.}~\bibnamefont {Yang}}, \bibinfo {author}
  {\bibfnamefont {D.}~\bibnamefont {Van~Tuan}}, \bibinfo {author}
  {\bibfnamefont {T.}~\bibnamefont {Taniguchi}}, \bibinfo {author}
  {\bibfnamefont {B.}~\bibnamefont {Urbaszek}}, \bibinfo {author}
  {\bibfnamefont {K.}~\bibnamefont {Watanabe}}, \bibinfo {author}
  {\bibfnamefont {T.}~\bibnamefont {Amand}}, \bibinfo {author} {\bibfnamefont
  {H.}~\bibnamefont {Dery}},\ and\ \bibinfo {author} {\bibfnamefont
  {X.}~\bibnamefont {Marie}},\ }\bibfield  {title} {\bibinfo {title}
  {Spin/valley pumping of resident electrons in wse$_2$ and ws$_2$
  monolayers},\ }\href {https://doi.org/10.1038/s41467-021-25747-5} {\bibfield
  {journal} {\bibinfo  {journal} {Nature Communications}\ }\textbf {\bibinfo
  {volume} {12}},\ \bibinfo {pages} {5455} (\bibinfo {year}
  {2021})}\BibitemShut {NoStop}%
\bibitem [{\citenamefont {Li}\ \emph {et~al.}(2021)\citenamefont {Li},
  \citenamefont {Biswas}, \citenamefont {Hail},\ and\ \citenamefont
  {Atwater}}]{Li2021b}%
  \BibitemOpen
  \bibfield  {author} {\bibinfo {author} {\bibfnamefont {M.}~\bibnamefont
  {Li}}, \bibinfo {author} {\bibfnamefont {S.}~\bibnamefont {Biswas}}, \bibinfo
  {author} {\bibfnamefont {C.~U.}\ \bibnamefont {Hail}},\ and\ \bibinfo
  {author} {\bibfnamefont {H.~A.}\ \bibnamefont {Atwater}},\ }\bibfield
  {title} {\bibinfo {title} {Refractive index modulation in monolayer
  molybdenum diselenide},\ }\href
  {https://doi.org/10.1021/acs.nanolett.1c02199} {\bibfield  {journal}
  {\bibinfo  {journal} {Nano Letters}\ }\textbf {\bibinfo {volume} {21}},\
  \bibinfo {pages} {7602} (\bibinfo {year} {2021})},\ \bibinfo {note} {pMID:
  34468150},\ \Eprint
  {https://arxiv.org/abs/https://doi.org/10.1021/acs.nanolett.1c02199}
  {https://doi.org/10.1021/acs.nanolett.1c02199} \BibitemShut {NoStop}%
\bibitem [{\citenamefont {Pierret}\ \emph {et~al.}(2022)\citenamefont
  {Pierret}, \citenamefont {Mele}, \citenamefont {Graef}, \citenamefont
  {Palomo}, \citenamefont {Taniguchi}, \citenamefont {Watanabe}, \citenamefont
  {Li}, \citenamefont {Toury}, \citenamefont {Journet}, \citenamefont {Steyer},
  \citenamefont {Garnier}, \citenamefont {Loiseau}, \citenamefont {Berroir},
  \citenamefont {Bocquillon}, \citenamefont {Fève}, \citenamefont {Voisin},
  \citenamefont {Baudin}, \citenamefont {Rosticher},\ and\ \citenamefont
  {Plaçais}}]{Pierret2022}%
  \BibitemOpen
  \bibfield  {author} {\bibinfo {author} {\bibfnamefont {A.}~\bibnamefont
  {Pierret}}, \bibinfo {author} {\bibfnamefont {D.}~\bibnamefont {Mele}},
  \bibinfo {author} {\bibfnamefont {H.}~\bibnamefont {Graef}}, \bibinfo
  {author} {\bibfnamefont {J.}~\bibnamefont {Palomo}}, \bibinfo {author}
  {\bibfnamefont {T.}~\bibnamefont {Taniguchi}}, \bibinfo {author}
  {\bibfnamefont {K.}~\bibnamefont {Watanabe}}, \bibinfo {author}
  {\bibfnamefont {Y.}~\bibnamefont {Li}}, \bibinfo {author} {\bibfnamefont
  {B.}~\bibnamefont {Toury}}, \bibinfo {author} {\bibfnamefont
  {C.}~\bibnamefont {Journet}}, \bibinfo {author} {\bibfnamefont
  {P.}~\bibnamefont {Steyer}}, \bibinfo {author} {\bibfnamefont
  {V.}~\bibnamefont {Garnier}}, \bibinfo {author} {\bibfnamefont
  {A.}~\bibnamefont {Loiseau}}, \bibinfo {author} {\bibfnamefont {J.-M.}\
  \bibnamefont {Berroir}}, \bibinfo {author} {\bibfnamefont {E.}~\bibnamefont
  {Bocquillon}}, \bibinfo {author} {\bibfnamefont {G.}~\bibnamefont {Fève}},
  \bibinfo {author} {\bibfnamefont {C.}~\bibnamefont {Voisin}}, \bibinfo
  {author} {\bibfnamefont {E.}~\bibnamefont {Baudin}}, \bibinfo {author}
  {\bibfnamefont {M.}~\bibnamefont {Rosticher}},\ and\ \bibinfo {author}
  {\bibfnamefont {B.}~\bibnamefont {Plaçais}},\ }\bibfield  {title} {\bibinfo
  {title} {Dielectric permittivity, conductivity and breakdown field of
  hexagonal boron nitride},\ }\href {https://doi.org/10.1088/2053-1591/ac4fe1}
  {\bibfield  {journal} {\bibinfo  {journal} {Materials Research Express}\
  }\textbf {\bibinfo {volume} {9}},\ \bibinfo {pages} {065901} (\bibinfo {year}
  {2022})}\BibitemShut {NoStop}%
\bibitem [{\citenamefont {Laturia}\ and\ \citenamefont
  {Vandenberghe}(2018)}]{Laturia2018}%
  \BibitemOpen
  \bibfield  {author} {\bibinfo {author} {\bibfnamefont {M.~L.}\ \bibnamefont
  {Laturia}, \bibfnamefont {Akashand Van de~Put}}\ and\ \bibinfo {author}
  {\bibfnamefont {W.~G.}\ \bibnamefont {Vandenberghe}},\ }\bibfield  {title}
  {\bibinfo {title} {Dielectric properties of hexagonal boron nitride and
  transition metal dichalcogenides: from monolayer to bulk},\ }\href
  {https://doi.org/10.1038/s41699-018-0050-x} {\bibfield  {journal} {\bibinfo
  {journal} {npj 2D Materials and Applications}\ }\textbf {\bibinfo {volume}
  {2}},\ \bibinfo {pages} {6} (\bibinfo {year} {2018})}\BibitemShut {NoStop}%
\end{thebibliography}%

\end{document}


\title{Optical properties of MoSe$_2$ monolayer implanted with ultra-low energy Cr ions - Supplementary Information}


\author{Minh N. Bui}
\email{m.bui@fz-juelich.de}
\affiliation{Peter Gr\"{u}nberg Institute 9 (PGI-9), Forschungszentrum J\"{u}lich, 52425 J\"{u}lich, Germany}
\affiliation{Department of Physics, RWTH Aachen University, 
52074 Aachen, Germany}

\author{Stefan Rost}
\affiliation{Peter Gr\"{u}nberg Institute 1 (PGI-1) and Institute for Advanced Simulation 1 (IAS-1), Forschungszentrum J\"{u}lich and JARA, 52425 J\"{u}lich, Germany}
\affiliation{Department of Physics, RWTH Aachen University, 
52074 Aachen, Germany}

\author{Manuel Auge}
\affiliation{II. Institute of Physics, University of G\"{o}ttingen, 37077 G\"{o}ttingen, Germany}

\author{Lanqing Zhou}
\affiliation{Peter Gr\"{u}nberg Institute 9 (PGI-9), Forschungszentrum J\"{u}lich, 52425 J\"{u}lich, Germany}
\affiliation{Department of Physics, RWTH Aachen University, 
52074 Aachen, Germany}

\author{Christoph Friedrich}
\affiliation{Peter Gr\"{u}nberg Institute 1 (PGI-1) and Institute for Advanced Simulation 1 (IAS-1), Forschungszentrum J\"{u}lich and JARA, 52425 J\"{u}lich, Germany}

\author{Stefan Bl\"{u}gel}
\affiliation{Peter Gr\"{u}nberg Institute 1 (PGI-1) and Institute for Advanced Simulation 1 (IAS-1), Forschungszentrum J\"{u}lich and JARA, 52425 J\"{u}lich, Germany}
\affiliation{Department of Physics, RWTH Aachen University, 
52074 Aachen, Germany}


\author{Silvan Kretschmer}
\affiliation{Institute of Ion Beam Physics and Materials Research, Helmholtz‐Zentrum Dresden‐Rossendorf, 01328 Dresden, Germany}

\author{Arkady V. Krasheninnikov}
\affiliation{Institute of Ion Beam Physics and Materials Research, Helmholtz‐Zentrum Dresden‐Rossendorf, 01328 Dresden, Germany}
\affiliation{Department of Applied Physics, Aalto University School of Science, P.O.Box 11100, 00076 Aalto, Finland}



\author{Kenji Watanabe}
\affiliation{Research Center for Functional Materials, National Institute for Materials Science, 1-1 Namiki, Tsukuba 305-0044, Japan\looseness=-1}

\author{Takashi Taniguchi}
\affiliation{International Center for Materials Nanoarchitectonics, National Institute for Materials Science, 1-1 Namiki, Tsukuba 305-0044, Japan\looseness=-1}


\author{Hans C. Hofsäss}
\affiliation{II. Institute of Physics, University of G\"{o}ttingen, 37077 G\"{o}ttingen, Germany}

\author{Detlev Gr\"{u}tzmacher}
\author{Beata E.~Kardyna\l}
\email{b.kardynal@fz-juelich.de}
\affiliation{Peter Gr\"{u}nberg Institute 9 (PGI-9), Forschungszentrum J\"{u}lich, 52425 J\"{u}lich, Germany}
\affiliation{Department of Physics, RWTH Aachen University, 
52074 Aachen, Germany}

\maketitle

\setcounter{table}{0}
\renewcommand{\thetable}{S\arabic{table}}%
\setcounter{figure}{0}
\renewcommand{\thefigure}{S\arabic{figure}}%
\setcounter{section}{0}
\renewcommand{\thesection}{S\arabic{section}}

\tableofcontents

\newpage

\section{Supplementary note 1 - Rutherford backscattering spectrometry}
\label{section:RBS}

\begin{figure}[htbp]
    \centering
    \includegraphics[width=0.5\textwidth]{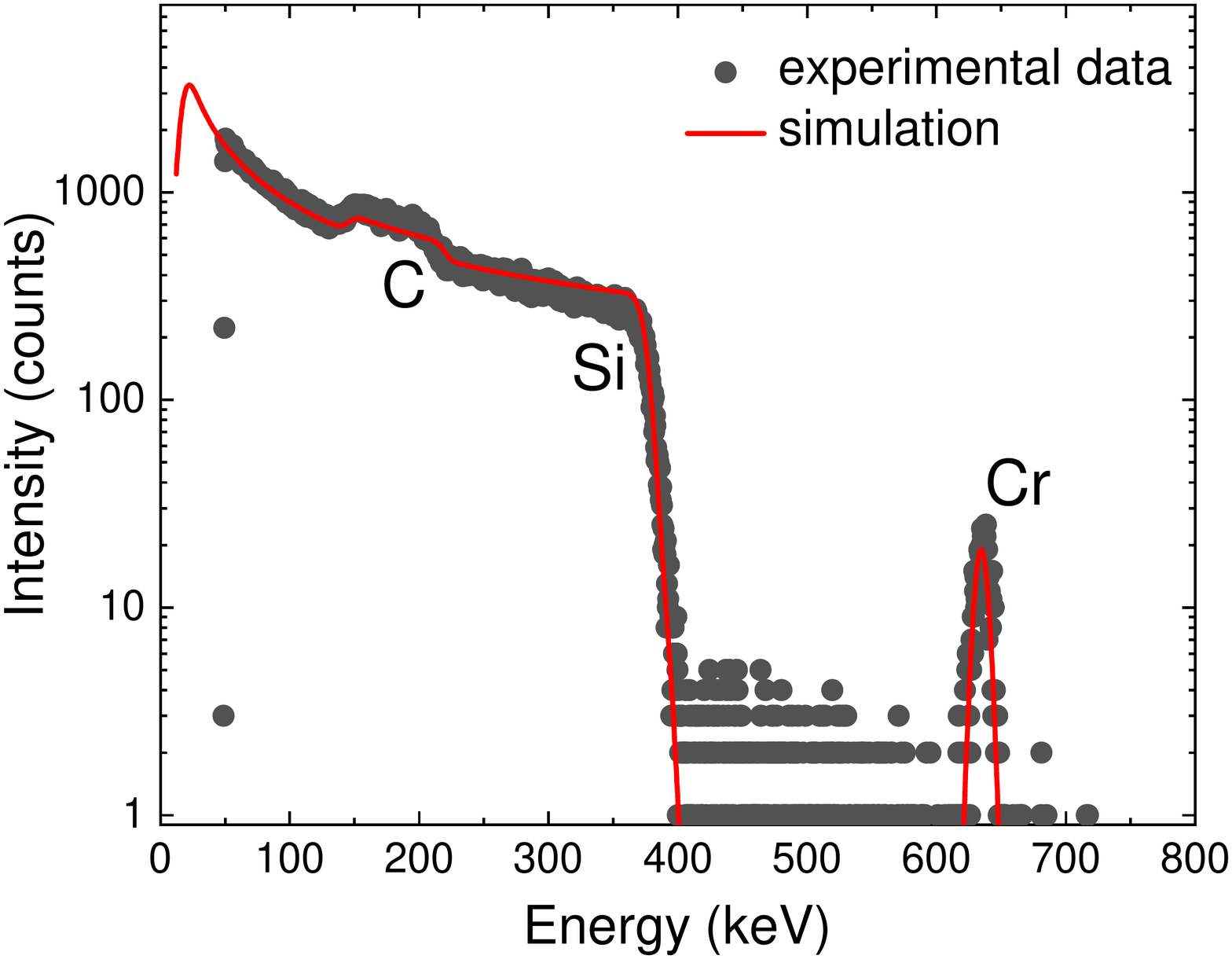}
     \caption{\textbf{RBS spectrum for Se implanted in ta-C on a Si substrate.} The implantation was performed at 25 eV with a fluence of 3\,$\times$\,10$^{15}$ cm$^{-2}$. Fit curve was done using SIMNRA.}
    \label{fig:RBS spectrum}
\end{figure}

Rutherford backscattering spectrometry (RBS) studies were carried out to verify the nominal implantation fluence of our experiments. An 860 keV helium ion beam was generated using a Cockcroft-Walton accelerator and then directed onto the sample surface. The backscattered ions were detected by a silicon charged-particle radiation detector under an angle of 165° to the ion beam direction. The semiconductor detector has an active area of 25 mm$^{2}$ and is located at a distance of 88 mm. Figure \ref{fig:RBS spectrum} shows an RBS spectrum for Cr implantation at 25 eV with a fluence of 3\,$\times$\,10$^{15}$ cm$^{-2}$ on a Si/ta-C substrate. Fitting procedure using SIMNRA \cite{Mayer97} gives Cr density of 3.03\,$\pm$\,0.13\,$\times$\,10$^{15}$ cm$^{-2}$, which agrees well with the given values in the Methods section. The test implantations are performed with a fluence of 3\,$\times$\,10$^{15}$ cm$^{-2}$ to achieve a high count rate in RBS in a reasonable acquisition time and to perform a better fit in SIMNRA.

\newpage
\section{Supplementary note 2 - Spatial distribution of D peak}

\begin{figure}[ht]
\begin{subfigure}[c]{.32\textwidth}
 \caption{}
 \vspace{-10pt}
    \includegraphics[trim={10 30 10 30},clip,width=\textwidth]{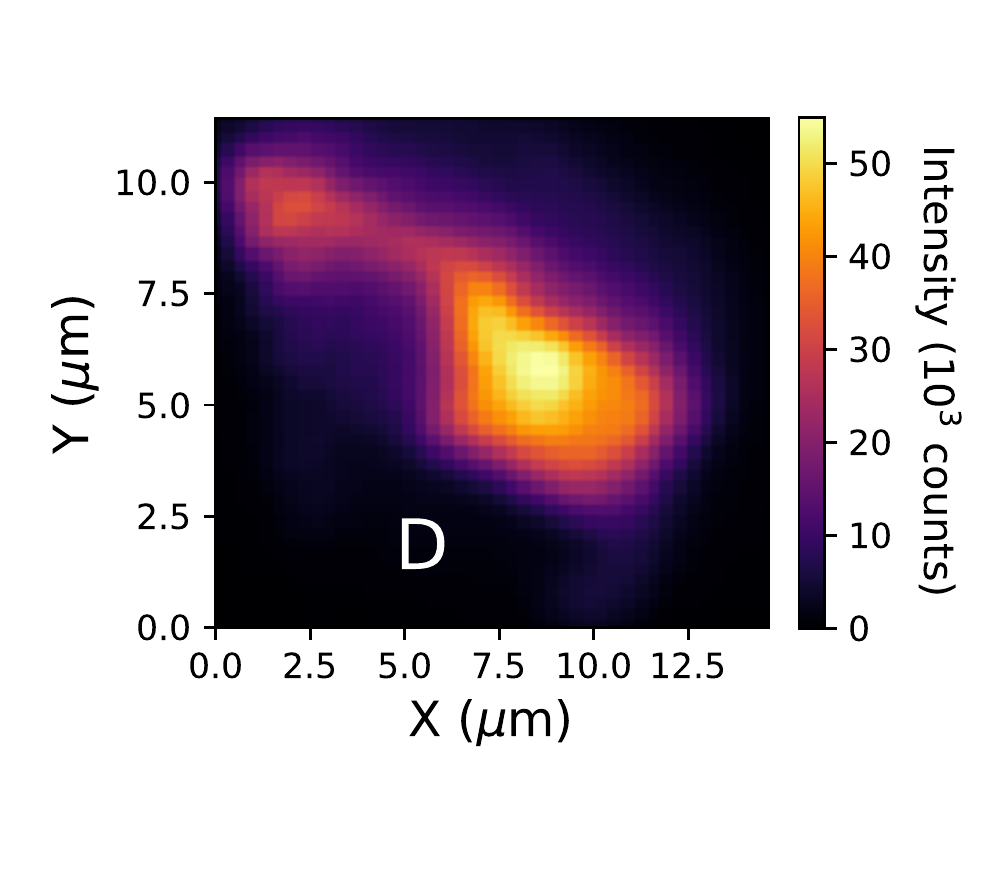}
    \label{fig:2D map D}
    \end{subfigure}
\begin{subfigure}[c]{.32\textwidth}
 \caption{}
 \vspace{-10pt}
    \includegraphics[trim={10 30 10 30},clip,width=\textwidth]{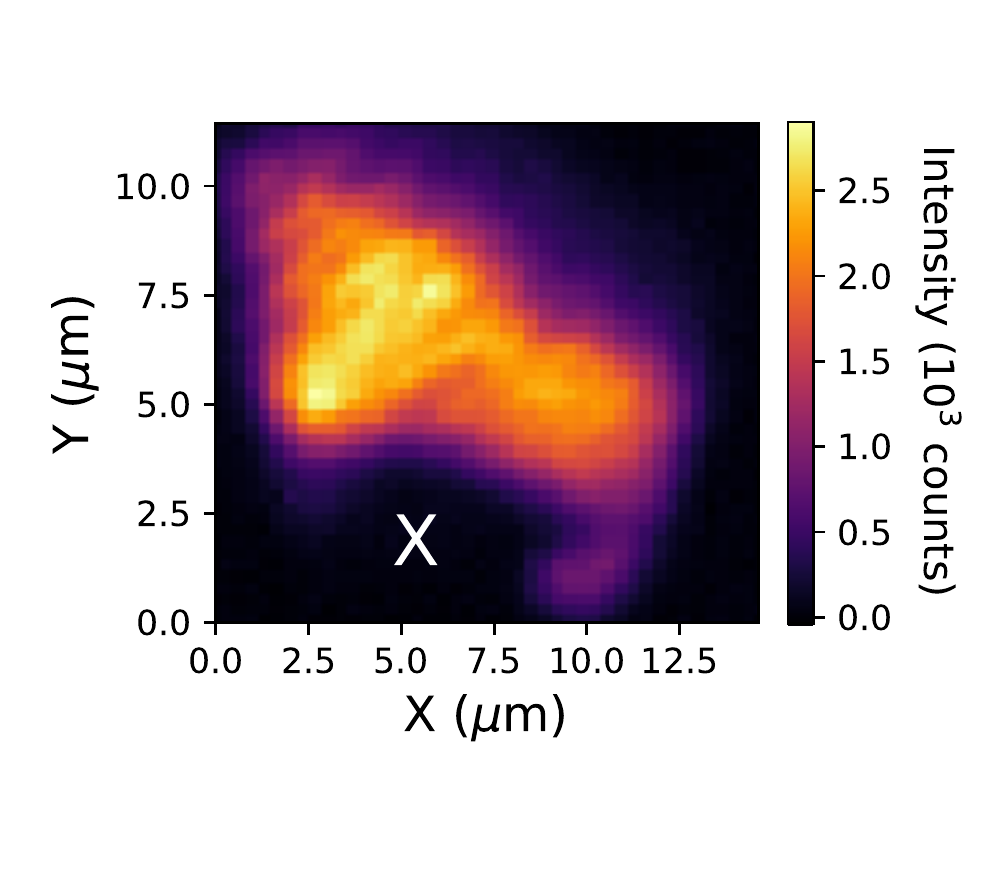}
    \label{fig:2D map X0}
    \end{subfigure}
\begin{subfigure}[c]{.32\textwidth}
 \caption{}
 \vspace{-10pt}
    \includegraphics[trim={10 30 10 30},clip,width=\textwidth]{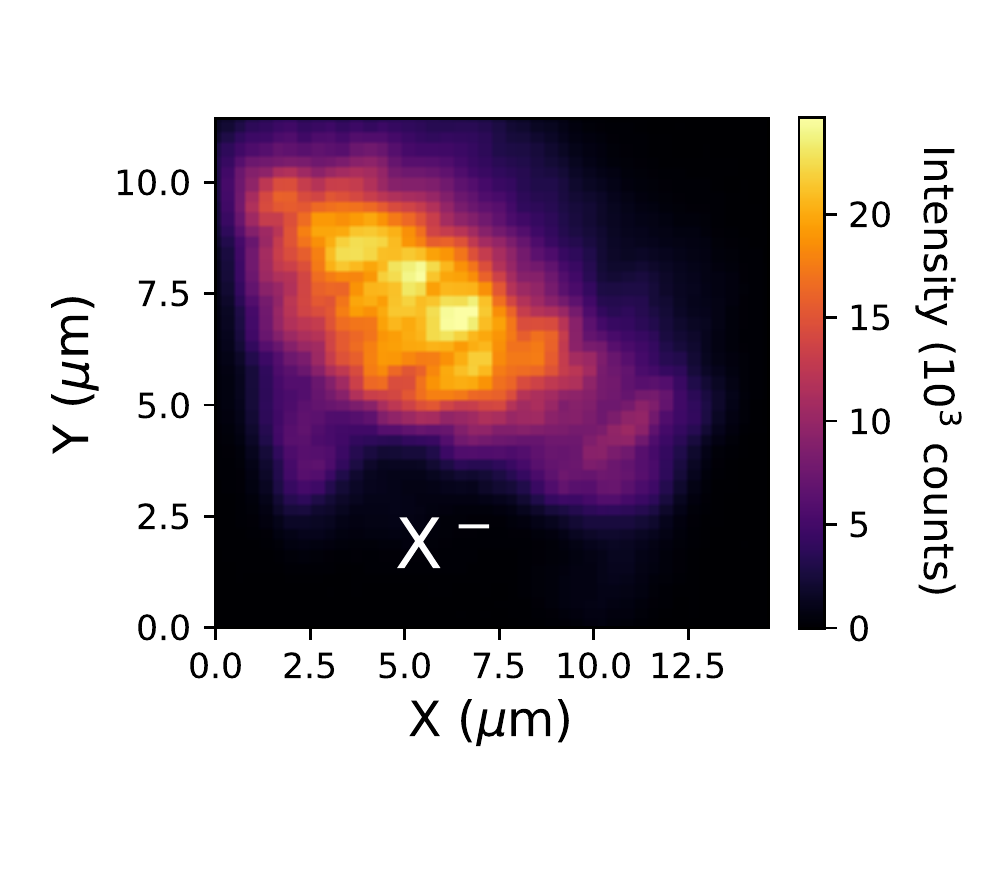}
    \label{fig:2D map X-}
    \end{subfigure}
\begin{subfigure}[c]{.32\textwidth}
 \vspace{-10pt}
 \caption{}
 \vspace{-10pt}
 \hspace{-10pt}
    \includegraphics[width=\textwidth]{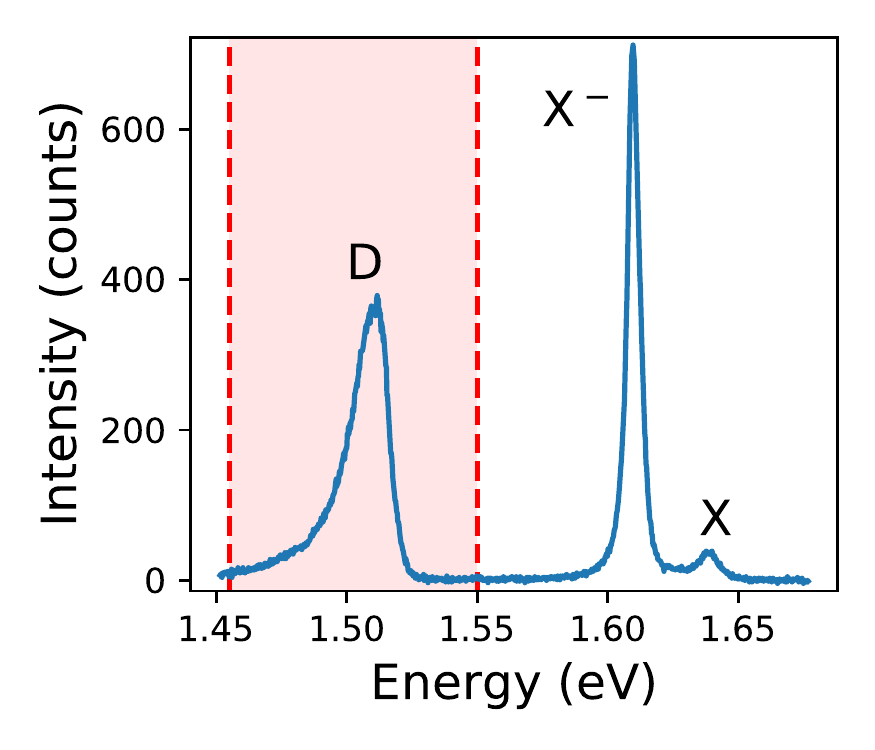}
    \label{fig:2D map D spectrum}
    \end{subfigure}
\begin{subfigure}[c]{.32\textwidth}
 \vspace{-10pt}
 \caption{}
 \vspace{-10pt}
 \hspace{-10pt}
    \includegraphics[width=\textwidth]{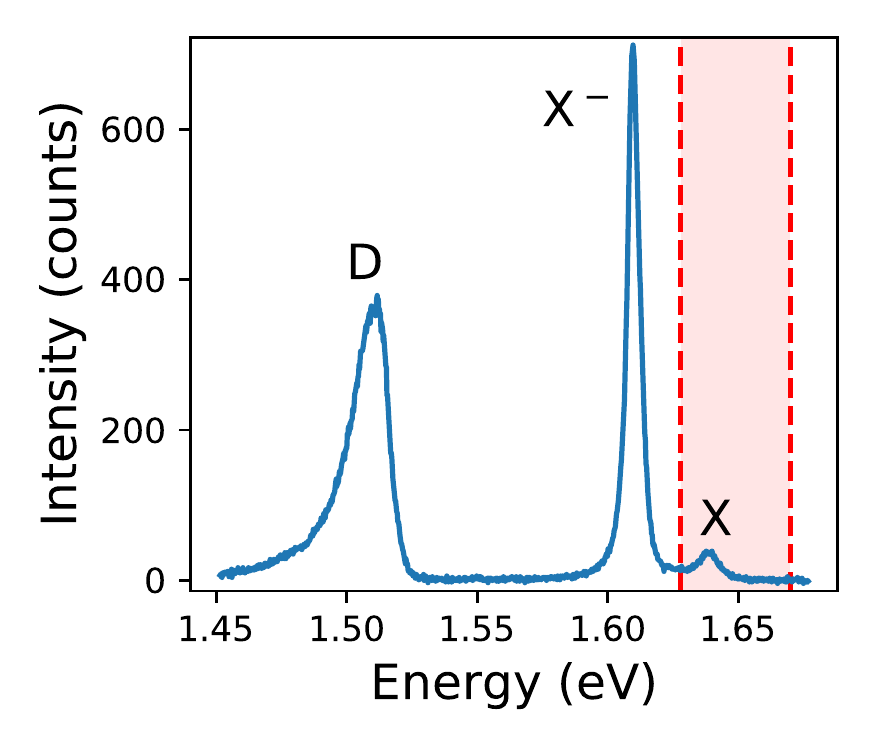}
    \label{fig:2D map X0 spectrum}
    \end{subfigure}
\begin{subfigure}[c]{.32\textwidth}
 \vspace{-10pt}
 \caption{}
 \vspace{-10pt}
 \hspace{-10pt}
    \includegraphics[width=\textwidth]{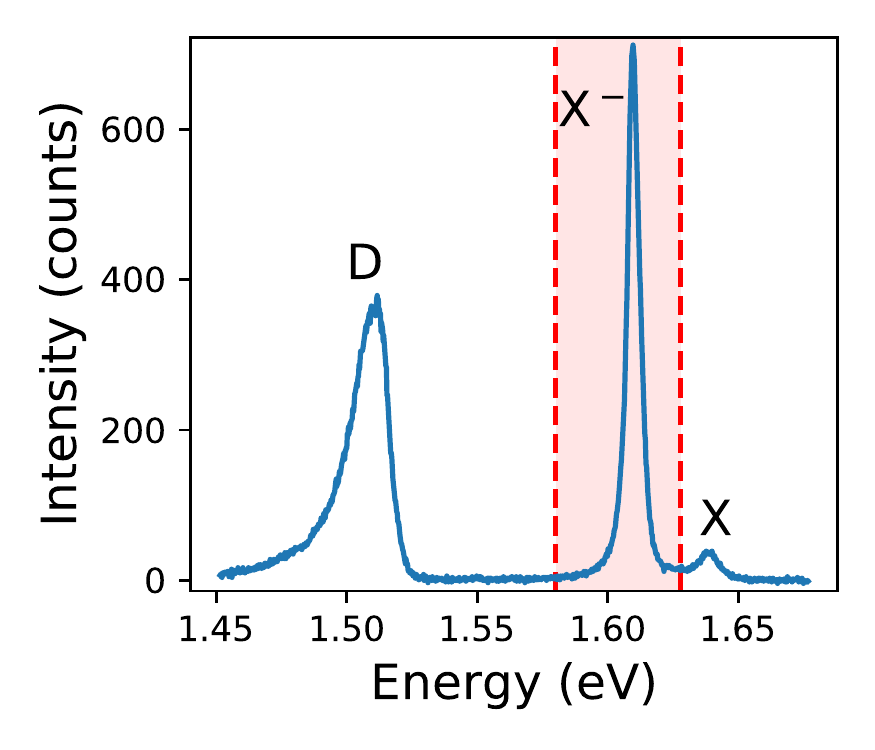}
    \label{fig:2D map X- spectrum}
    \end{subfigure}
\caption{\textbf{$\upmu$PL maps of Cr-implanted MoSe$_2$ ML.} The PL intensity was integrated around (a) D, (b) X, and (c) X$^-$ emission. The integrated PL signal is summed over the energy range indicated in the shaded region in respective single spectra in (d), (e), and (f).}
\label{fig:2D maps}
\end{figure}

Figure \ref{fig:2D maps} shows the 2D integrated micro-PL ($\upmu$-PL) maps for D, X and X$^-$ emissions. Compared to the excitonic emission map of X and X$^-$, D emission is observed only in the implanted MoSe$_2$ ML area and does not come from the hBN or Gr layer underneath. The emission is also not from localised sites, e.g. wrinkles, scratches or bubbles in the ML, but rather the whole ML. The apparent difference between the bright and dark halves in the ML is likely from inhomogeneous doping provided by the bottom Gr gate.

\newpage
\section{Supplementary note 3 - Polarization resolved PL}

\begin{figure}[h]
    \centering
    \includegraphics[width=0.5\textwidth]{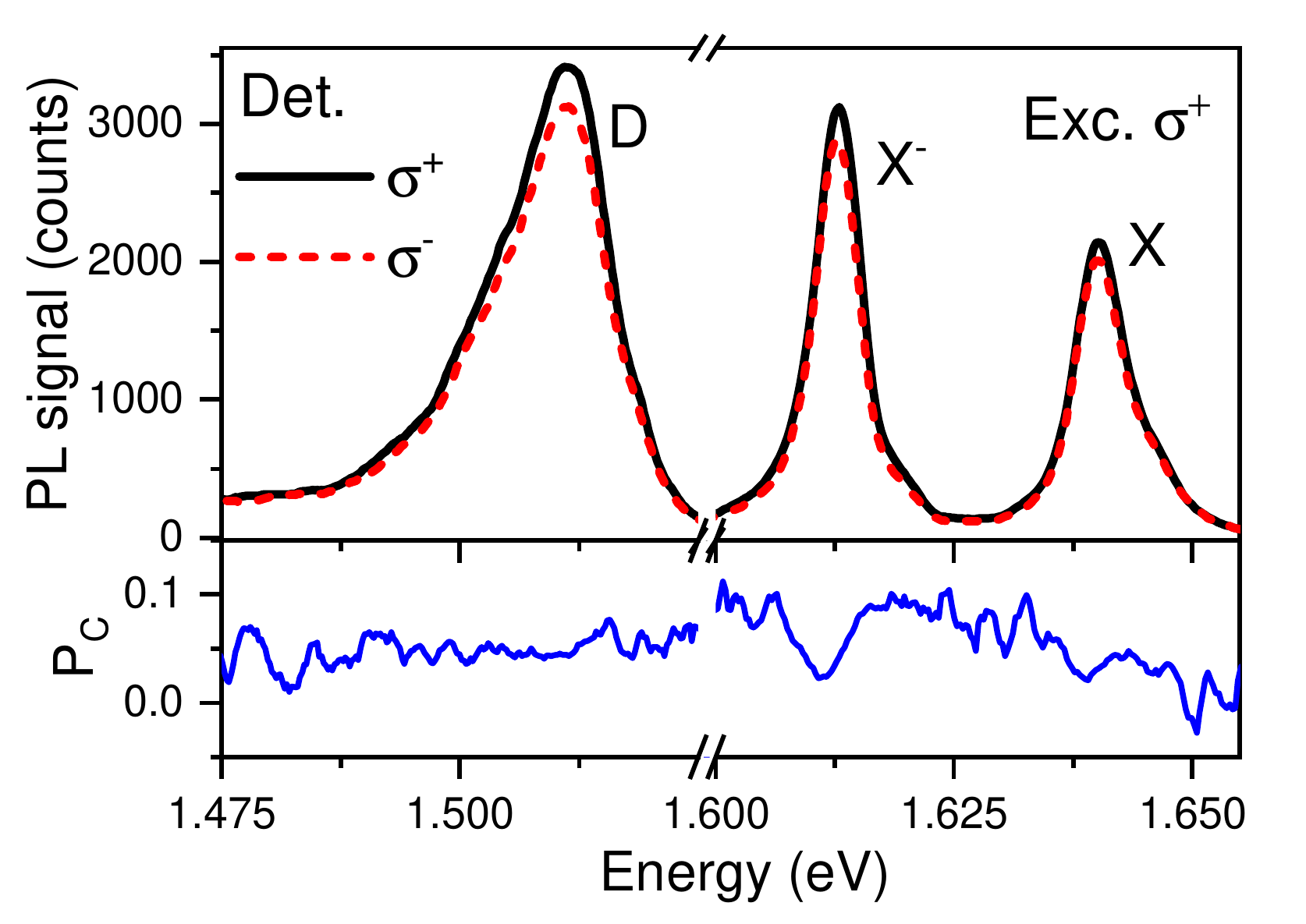}
\caption{\textbf{Polarization-dependent PL of Cr-implanted MoSe$_2$ at 10 K.}
The sample is excited with left ($\sigma^+$) circularly polarised light, and PL is detected in left - red curve - and right ($\sigma^-$) - dotted black curve - states. D, X$^-$ and X exhibit low circular dichroism $P_C$ at below 0.05.}
\label{fig:polarization}
\end{figure}

A linear polariser, a $\lambda$/2-waveplate and a $\lambda$/4-waveplate were inserted in the excitation path (before the beamsplitter) to set the polarisation state of the laser beam from a diode laser (655 nm in wavelength and 4.59 $\upmu$W in power on the sample). On the detection path, a $\lambda$/4-waveplate, a $\lambda$/2-waveplate and a linear polariser were placed to set the detected polarisation state.
D line shares polarisation properties with the X and X$^-$ as shown in figure \ref{fig:polarization}. Excitation with circularly polarised light resulted in a low degree of circular dichroism for X and X$^-$ emissions, typical for MoSe$_2$ \cite{MacNeill2015, Kioseoglou2016, Wang2015}. The degree of dichroism, defined as $P_C = (I_+ - I_-)/(I_+ + I_-)$ where $I_{\pm}$ are PL intensity detected in $\sigma_{\pm}$ states, is similar for D compared with X and X$^-$.

\newpage
\section{Supplementary note 4 - Raman and PL spectroscopy studies of vacancies in MoSe$_2$ ML}

\begin{figure}[h]
\vspace{-10pt}
\begin{subfigure}[c]{.4\textwidth}
 \caption{}
 \vspace{-10pt}
 \hspace{-10pt}
    \centering
    \includegraphics[width=\textwidth]{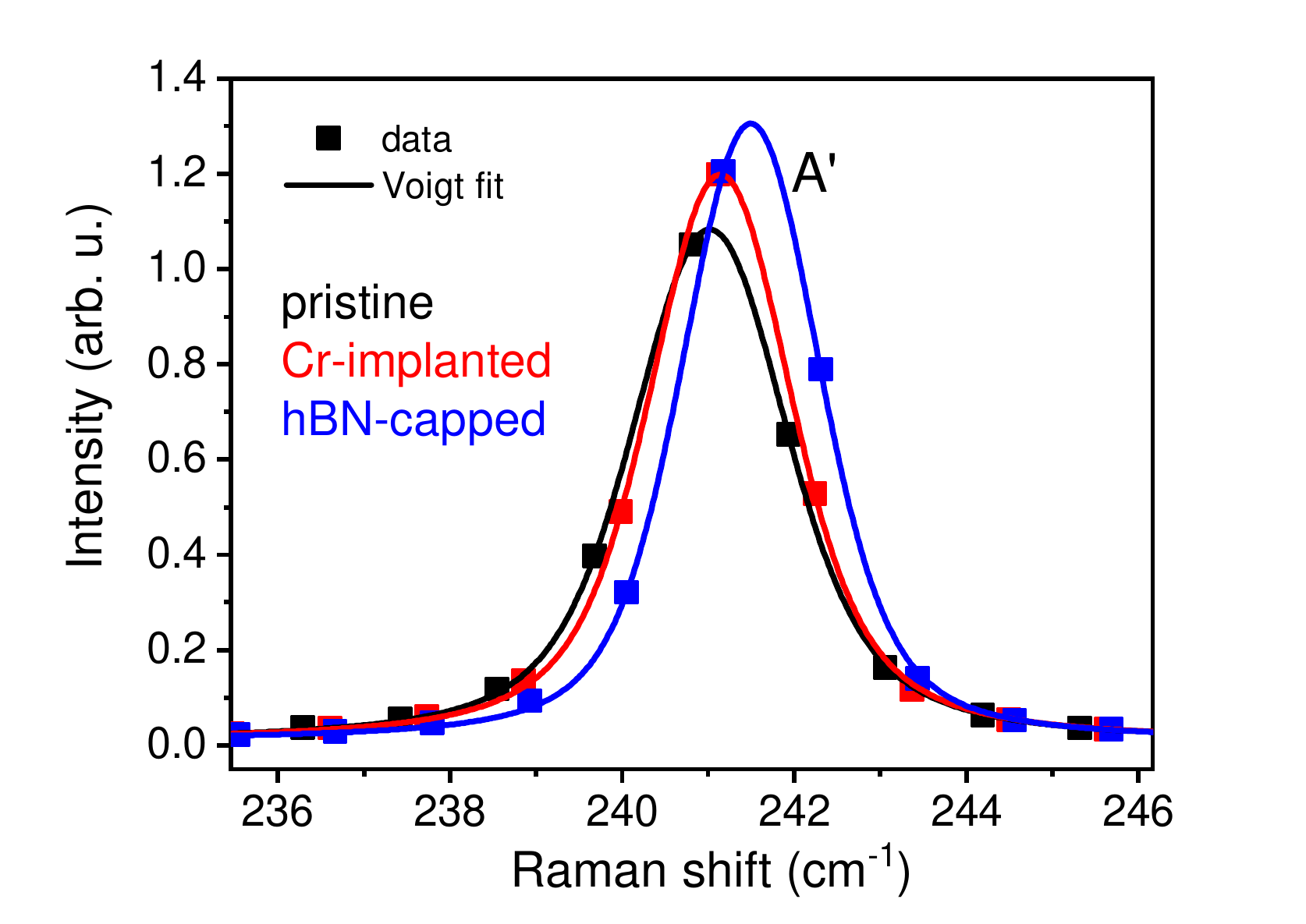}
    \label{fig:RT Raman}
    \end{subfigure}
\begin{subfigure}[c]{.4\textwidth}
 \caption{}
 \vspace{-10pt}
 \hspace{-10pt}
    \centering
    \includegraphics[width=\textwidth]{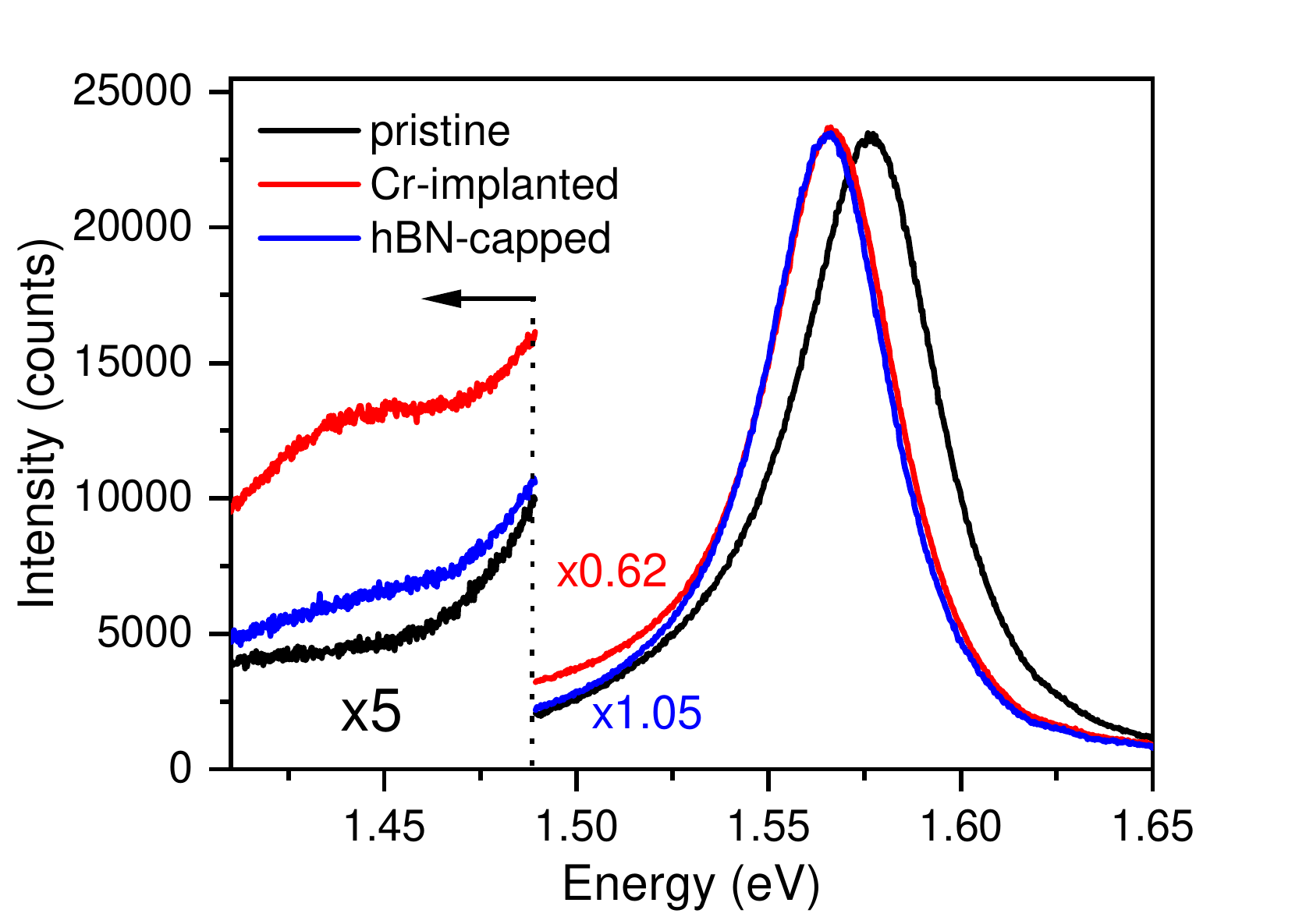}
    \label{fig:RT PL}
    \end{subfigure}
\begin{subfigure}[c]{.4\textwidth}
 \caption{}
 \vspace{-10pt}
 \hspace{-10pt}
    \centering
    \includegraphics[width=\textwidth]{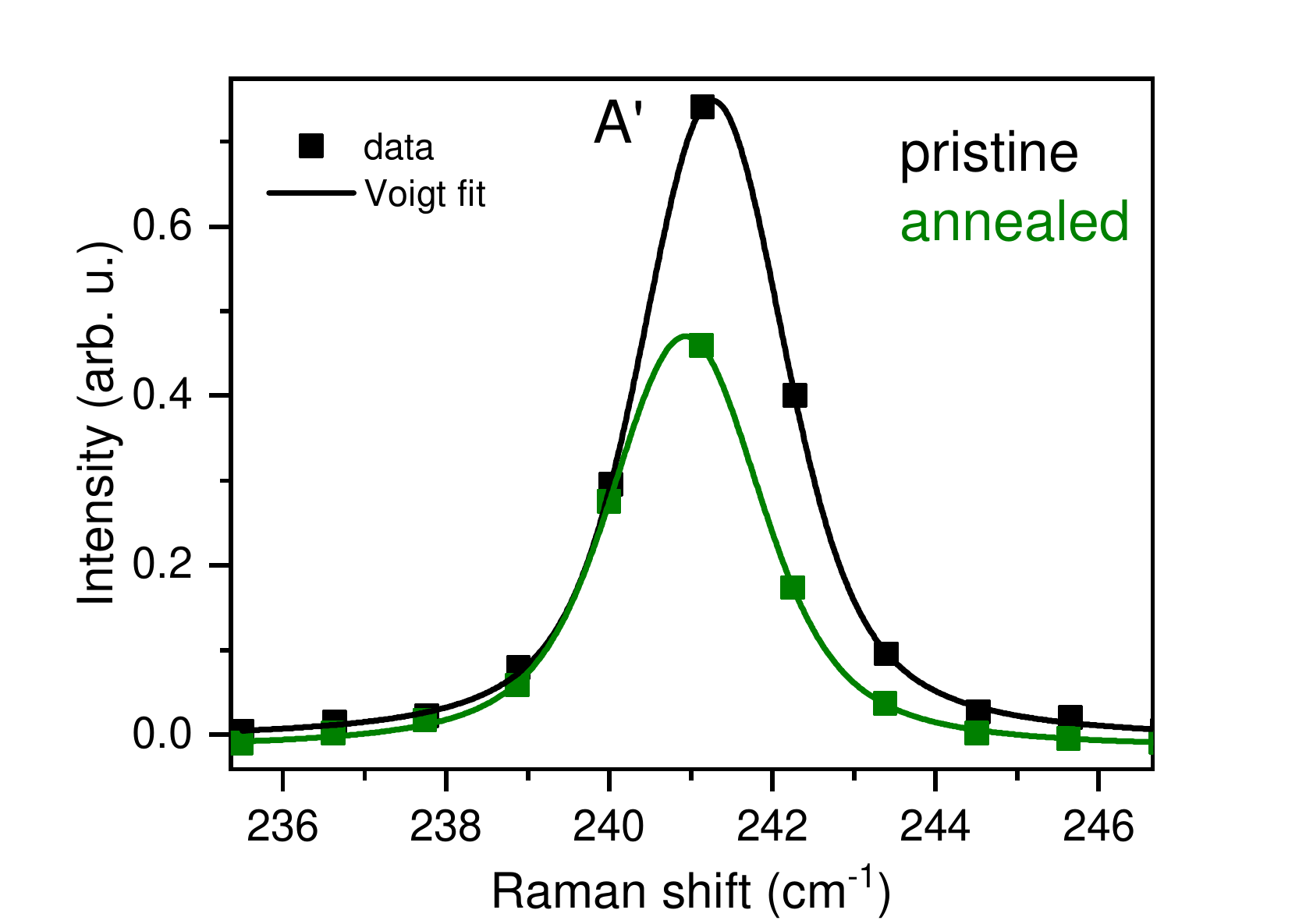}
    \label{fig:vacancies anneal Raman}
    \end{subfigure}
\begin{subfigure}[c]{.4\textwidth}
 \caption{}
 \vspace{-10pt}
 \hspace{-10pt}
    \centering
    \includegraphics[width=\textwidth]{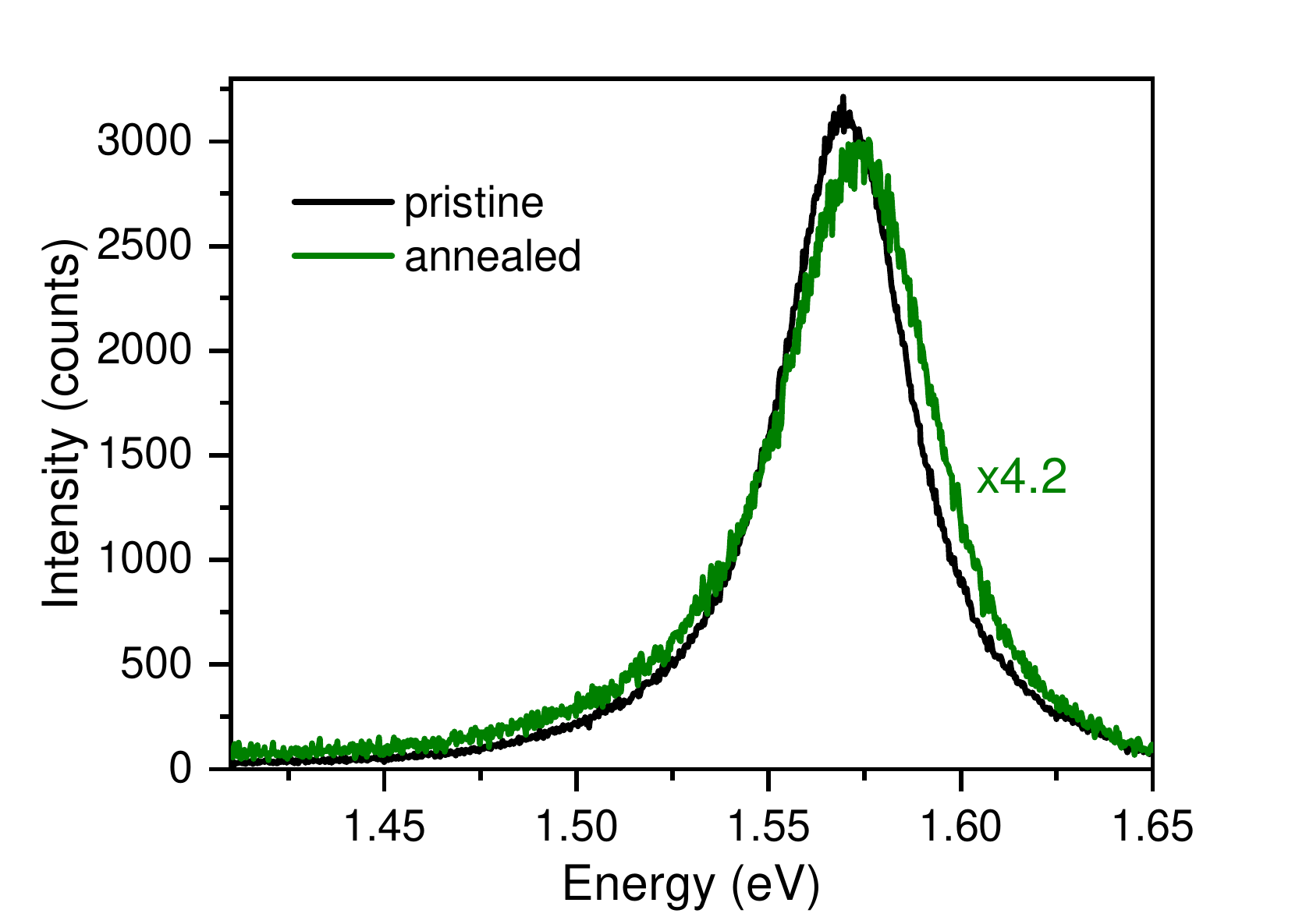}
    \label{fig:vacancies anneal PL}
    \end{subfigure}
\begin{subfigure}[c]{.4\textwidth}
 \caption{}
 \vspace{-10pt}
 \hspace{-10pt}
    \centering
    \includegraphics[width=\textwidth]{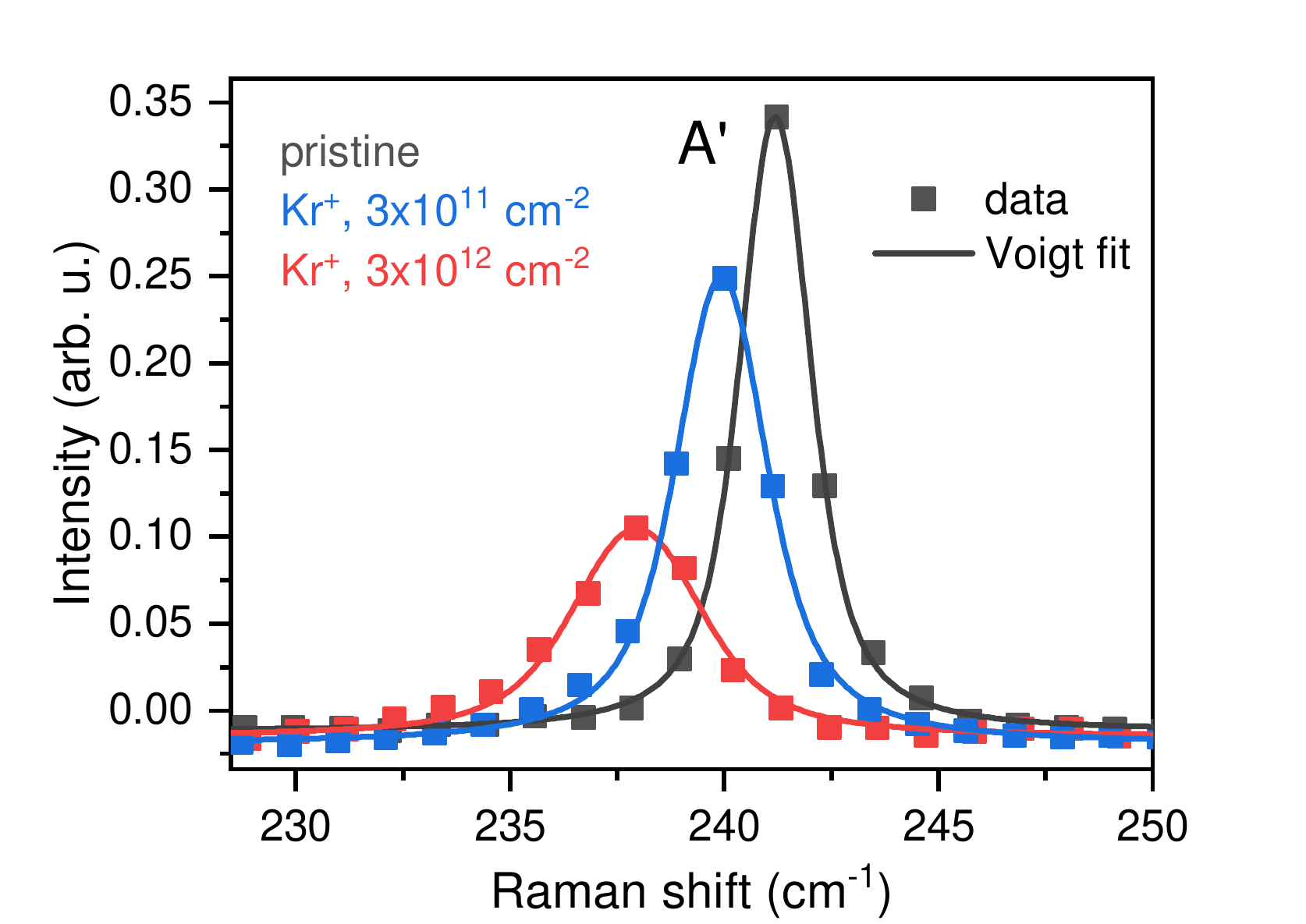}
    \label{fig:vacancies Kr Raman}
    \end{subfigure}
\begin{subfigure}[c]{.4\textwidth}
 \caption{}
 \vspace{-10pt}
 \hspace{-10pt}
    \centering
    \includegraphics[width=\textwidth]{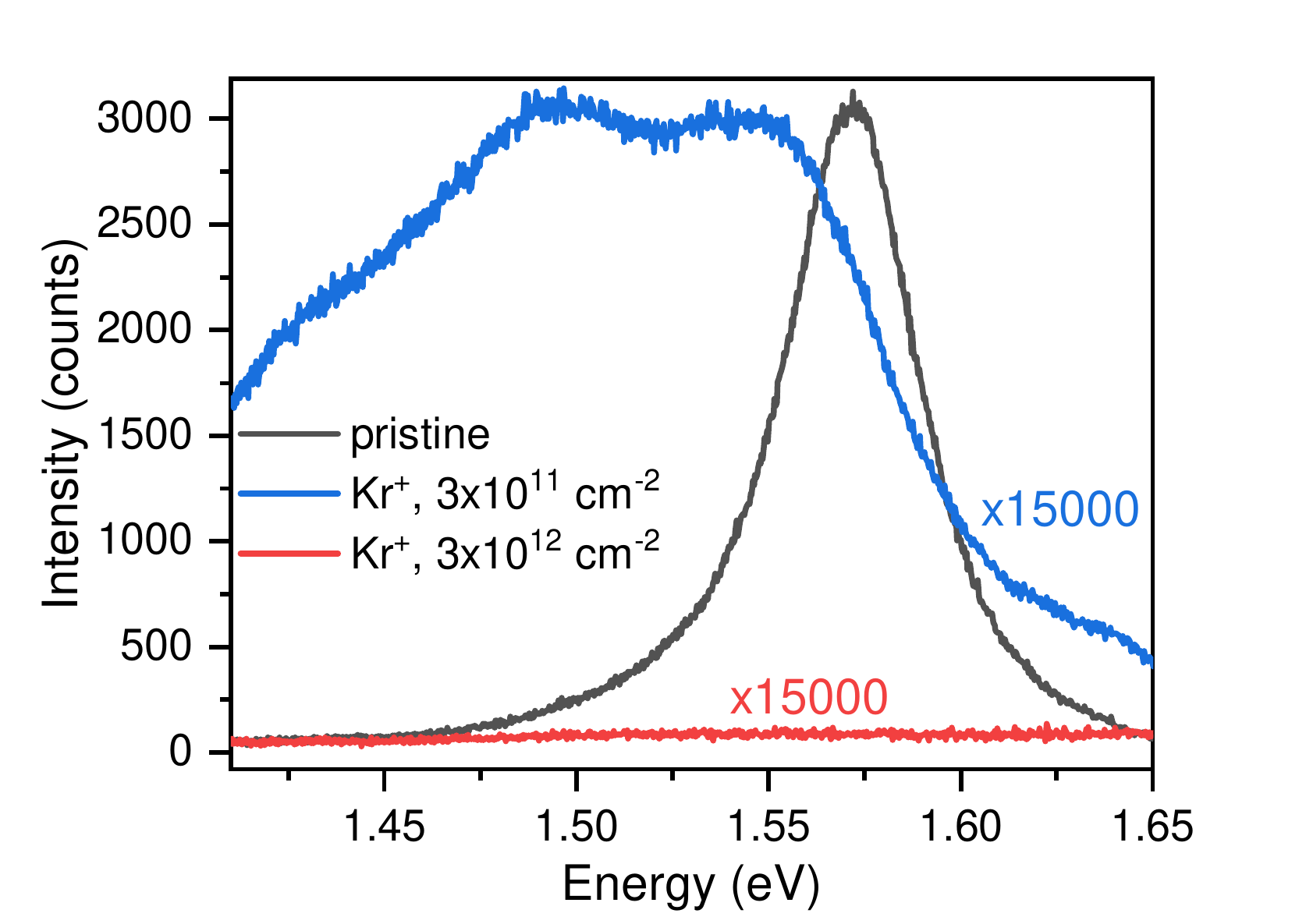}
    \label{fig:vacancies Kr PL}
    \end{subfigure}
\caption{\textbf{Optical spectroscopy of defected MoSe$_2$ ML.} Room temperature Raman (a,c,e) and PL (b,d,f) spectra of Cr-implanted MoSe$_2$ ML (a,b), annealed ML (c,d) and Kr-implanted MLs (e,f), with a  pristine ML’s spectrum for comparison. Raman spectra are all normalised to Si signal at 520.5 cm$^{-1}$.}
\label{fig:vacancies}
\end{figure}

According to MD simulation (table \ref{tab:MDsim}), we should exclude the Se vacancy as the signal's origin. Vacancy introduces a donor state at the Fermi level hybridised with the valence band and an acceptor state deep in the bandgap \cite{Rost2023, Iberi2016, Samani2016, Shafqat2017}. An optical transition between the defect levels is not allowed (figure \ref{fig:mose2_cr_compare_imag_multiple_pristine}). The lowest allowed energy optical transition is between the valence band state at the $\Gamma$ point and the deep acceptor level. The existing literature reports that vacancies form non-radiative recombination sites that quench PL \cite{Tanoh2021, Amani2015} or that they contribute to sub-bandgap PL but only at low temperature \cite{Li2017, Klein2019, Mitterreiter2021}, which is in contrast with our data, (see RT PL from our samples in figure \ref{fig:RT PL}). In addition, the energy shift of the D line with the temperature \cite{ODonnell1991} (figure \ref{fig:bandgap}) is weaker than that of X, unlike the fast change reported for vacancies in MoSe$_2$ \cite{Li2017}. At sufficient density (around 8$\%$), vacancies can blueshift the PL and downshift the Raman lines \cite{Samani2016}. In contrast, we did not see such changes in our Cr-implanted MoSe$_2$ ML (figure \ref{fig:vacancies}).

To ensure that one can exclude the role of vacancies in the D emission, room temperature Raman and PL spectra of MoSe$_2$ MLs with vacancies were compared to Cr-implanted MoSe$_2$. To create vacancies, one sample of MoSe$_2$ was annealed at 300 $\degree$C in low vacuum (5\,$\times$\,10$^{-2}$ mbar) for 2.5 hours \cite{Mitterreiter2021}. Another two samples were implanted with Kr$^+$, which, together with other noble gases, is commonly used for creating vacancies and their complexes in 2D materials \cite{Lucchese2010, Lehtinen2010, Junge2023, Kianinia2020, Shi2016, Murray2016, Maguire2018, Klein2019, Mitterreiter2021, Klein2021}. Ion implantation was done at 25 eV energy for introducing vacancies \cite{Ghorbani-Asl2017, Ghaderzadeh2020}, with fluences of 3\,$\times$\,10$^{11}$ and 3\,$\times$\,10$^{12}$ cm$^{-2}$ at elevated temperature of 220 $\degree$C, after being pre-annealed for 30 minutes at the same temperature in the implant chamber to remove volatile adsorbates. The higher fluence corresponds to the same Cr fluence in the main experiment. The lower fluence is about 10 times higher than the upper limit of potential vacancies density predicted for the Cr implanted sample by the atomistic MD simulation in the main text.

Figure \ref{fig:vacancies} shows the room temperature Raman and PL spectra of MoSe$_2$ MLs before and after Cr implantation, annealing and Kr implantation (as already described above or in the main text). Since slight variations in PL and Raman spectra are possible from sample to sample, the spectra from the same ML before and after processing are shown. 
Cr-implantation introduces a small upshift in the out-of-plane Raman vibrational mode A$'$ (figure \ref{fig:RT Raman}). The upshift is expected for Cr atoms in the lattice because it stiffens the lattice bonds and increases the restoring force and the A$'$ frequency. 
Annealing and Kr implantation downshift this Raman line (figures \ref{fig:vacancies anneal Raman} and \ref{fig:vacancies Kr Raman}). Such a downshift has been reported before for MoSe$_2$ MLs with vacancies \cite{Samani2016, Azam2022, Kim2022, Xiao2020}. It was explained by the lattice bond loosening and lowering the restoring force. PL emission of the Cr-implanted ML is 15\,meV redshifted compared with the pristine ML. It also shows an emission band around 120 meV below the free exciton line (figure \ref{fig:RT PL}). This low-energy emission is also observed at low temperatures. On the other hand, the annealed ML has a very slight blueshift (figure \ref{fig:vacancies anneal PL}), consistent with several other reports \cite{Samani2016, Azam2022}. Kr-implanted MLs' PL is heavily quenched (figure \ref{fig:vacancies Kr PL}). Some signal is visible only when the excitation laser power is raised 15000 times (from 17 nW to 261 $\upmu$W). The disagreement from these experiments doesn't favour vacancies being the origin of the D emission from the Cr-implanted ML.

Raman and PL spectra presented in this section were acquired under ambient conditions (room temperature and pressure) using a confocal Raman microscope (Renishaw inVia) with 532\,nm excitation laser (Coherent Compass 315M 150SL). The laser power was set between 0.017 and 261 $\upmu$W for sufficient signal intensity while preventing the sample from heating during exposure. An objective lens (50$\times$, NA = 0.75, Leica N-plan EPI) collected the Raman signal, which was then dispersed by a 2400 l/mm grating on the CCD camera, giving a spectral resolution of 1 cm$^{-1}$. PL signal was dispersed with a 600 l/mm grating, yielding a resolution of 0.147 nm.

\newpage
\section{Supplementary note 5 - MD simulation and TEM studies of Cr implantation in MoS$_2$}
\label{section:MoS2+Cr}

To compare the types of defects obtained in the simulations to the available experimental data, we also modelled ion implantation into ML MoS$_2$. The results are presented in supplementary table \ref{tab:MDsim_mos2}. We note that
interstitials cannot exist in ML MoS$_2$ due to a smaller unit cell size than that of MoSe$_2$.
Configurations predicted by the simulation were observed in the atomic-resolution TEM characterisation of MoS$_2$ implanted with Cr under similar conditions \cite{Hennessy2022}. It was noted that the relative abundance of defects, determined by the intensity-based atomic site assignment, was heavily affected by the presence of carbon contamination on the ML surface. In addition to the poor resolution in the experimental images, this renders accurate quantification of the elements difficult.

\begin{table}[h]
\caption{\textbf{Results of DFT MD simulations of 25 eV Cr ion irradiation on single layer MoS$_2$.} 
The probabilities $p$ of likely defect configurations to appear along with the formation energies $E_f$ of
these configurations are listed. }
\centering
\begin{tabular}{lrr}
                & $p$ & $E_f$ [eV]  \\
adatom              & 0.12 &  -0.98           \\
X-sub               & 0.12 &  -0.74           \\
interstitial        & 0.00 &  0.93           \\
Cr@Mo               & 0.41 &  3.06            \\
Cr@S                & 0.12 &  2.87            \\
V$_{\text{S}}$        & 0.13 &  6.53            \\
passed through      & 0.10 &  0.00            \\                  
\end{tabular}
\label{tab:MDsim_mos2}
\end{table}

\newpage

\section{Supplementary note 6 - Unfolded band structures including implanted defects}
\label{section:DFT}

The main text shows the dielectric functions of MoSe$_2$ MLs containing several Cr-implantation introduced defects. In the following, their unfolded band structures are shown, which means that the band structures are shown in the first Brillouin zone of the pristine unit cell to visualise better the perturbation that the defect introduces. The methodology is briefly recapitulated in the method section of the main text.

Single atomic defects should show up as flat energy levels if the computational supercell is large enough for the periodic images of the defect atom not to interact. It is not always the case here due to the computational resources that such large supercell would require. Wherever a defect state crosses and interacts with a band of the host lattice, a hybridisation of the bands is visible.

\begin{figure}[h]
\begin{subfigure}[c]{.49\textwidth}
 \caption{Cr@Mo}
  \centering
 \includegraphics[trim={25 20 44 17},clip,width=\textwidth]{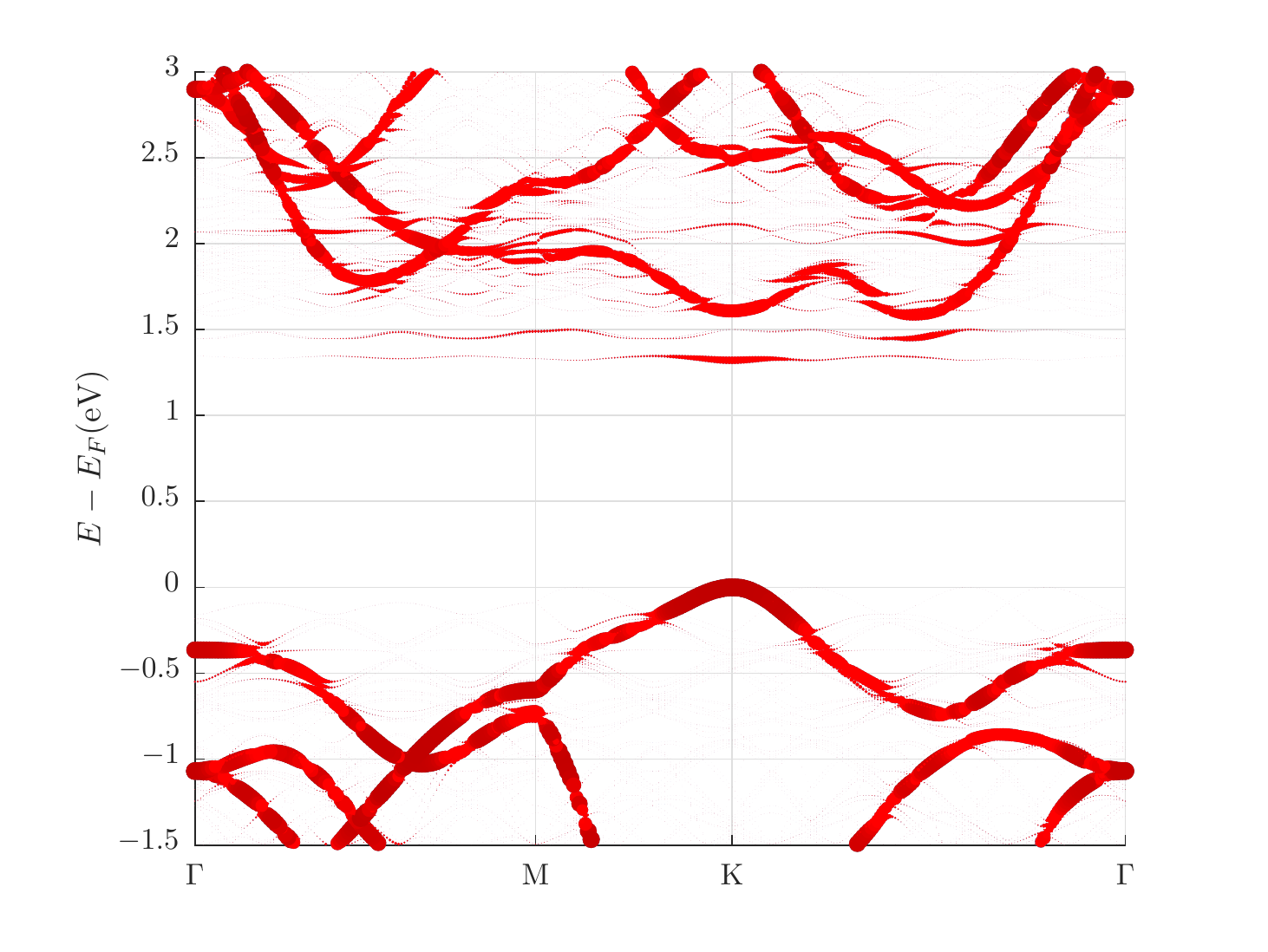}
\end{subfigure}
\begin{subfigure}[c]{.49\textwidth}
 \caption{Cr@Se}
  \centering
 \includegraphics[trim={25 20 44 17},clip,width=\textwidth]{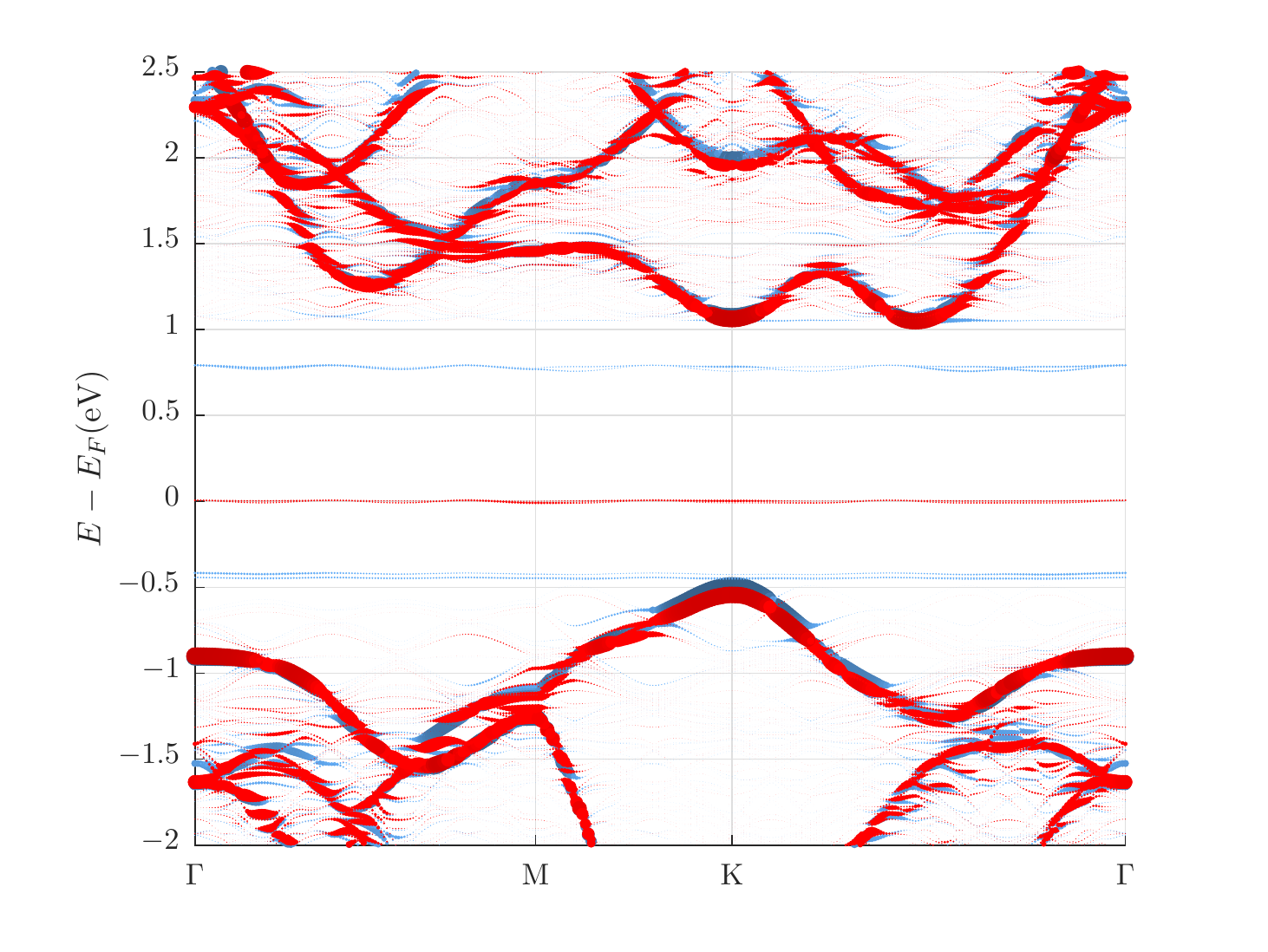}
\end{subfigure}
\begin{subfigure}[c]{.49\textwidth}
 \caption{Cr interstitial}
  \centering
 \includegraphics[trim={25 20 44 17},clip,width=\textwidth]{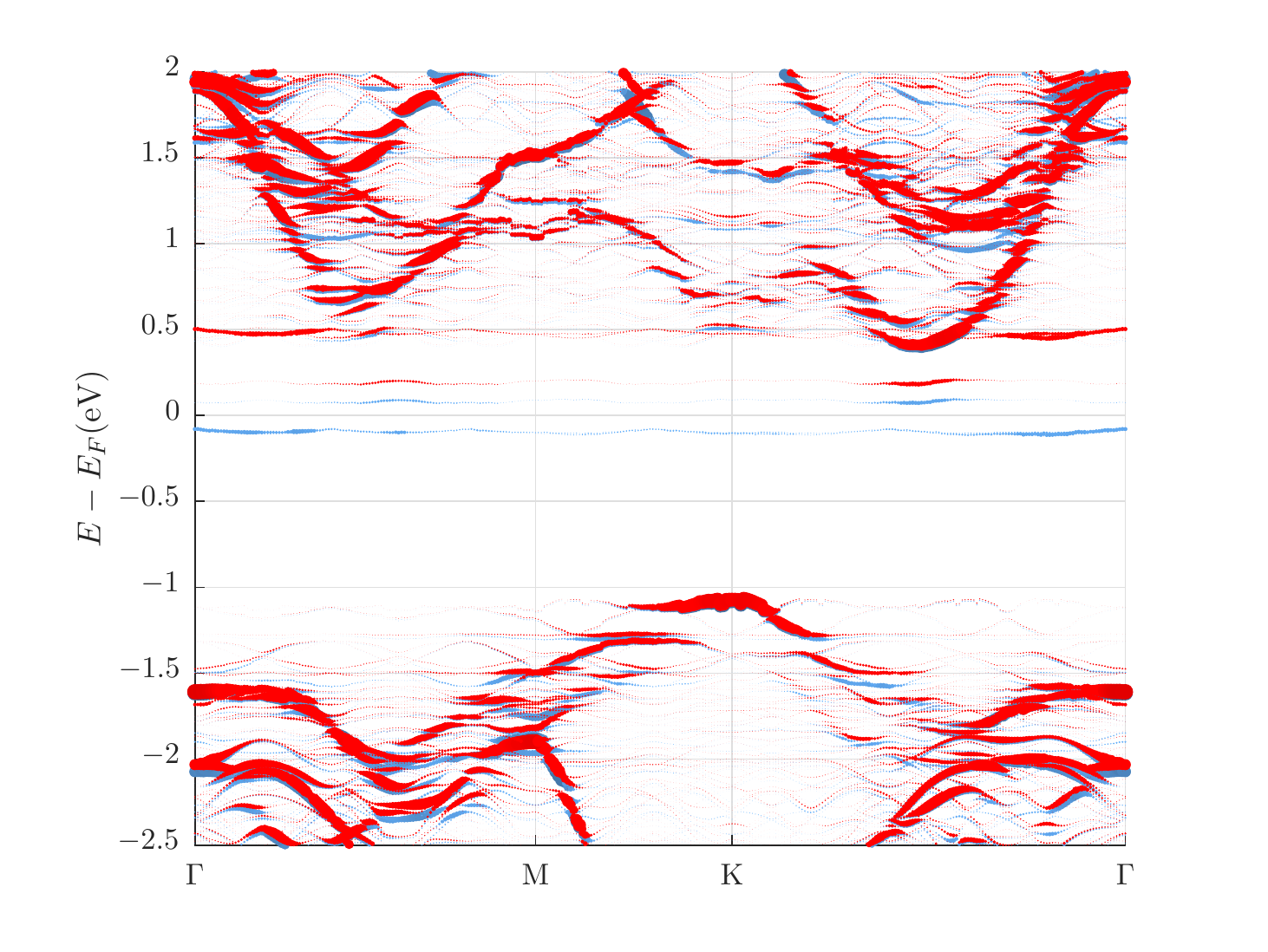}
 \end{subfigure}
\begin{subfigure}[c]{.49\textwidth}
 \caption{X-sub}
  \centering
 \includegraphics[trim={25 20 44 17},clip,width=\textwidth]{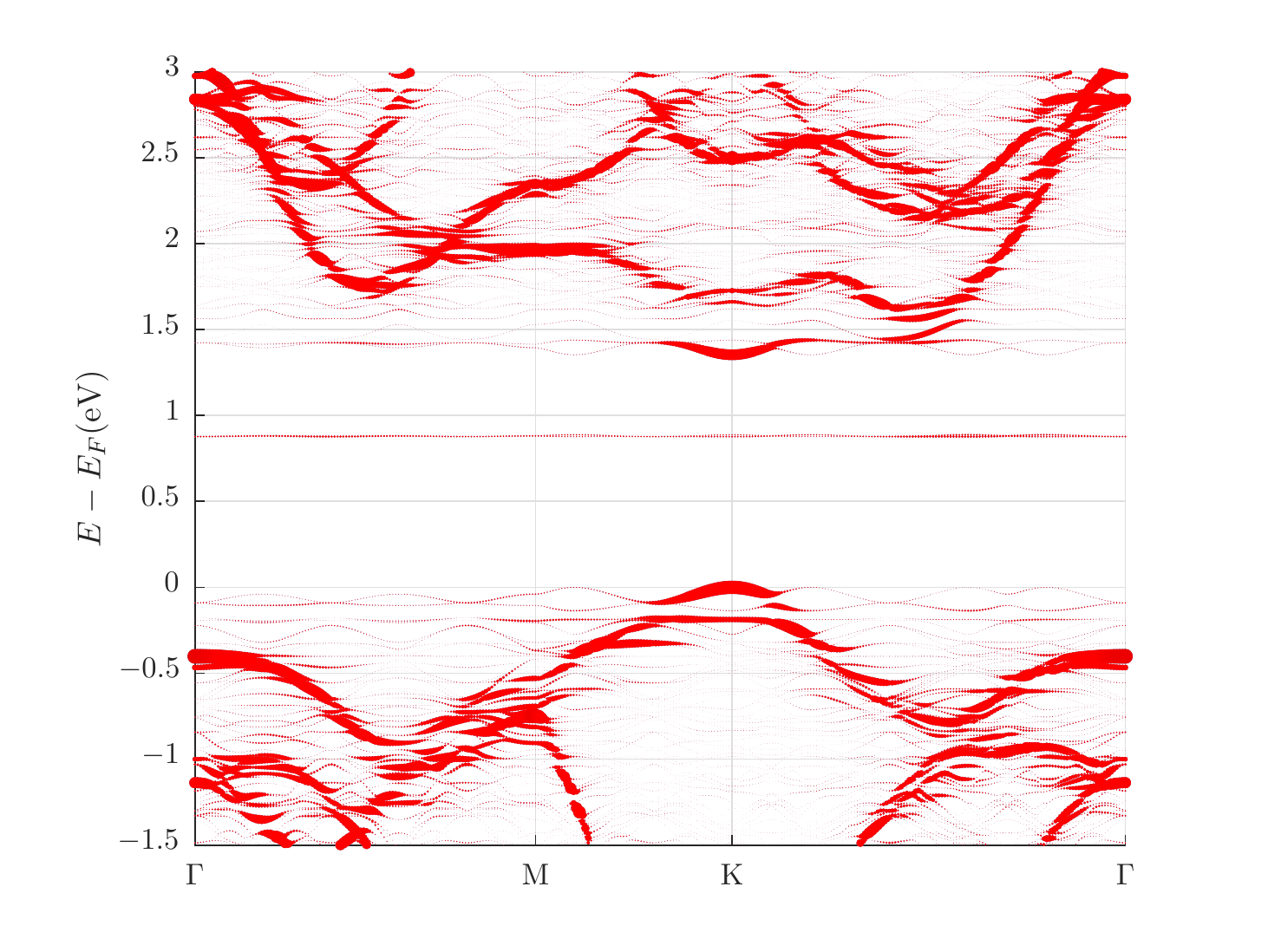}
\end{subfigure}
 \caption{\textbf{DFT calculation for unfolded band structure of Cr-implanted MoSe$_2$ ML with different Cr defect configurations:} (a) Cr@Mo, (b) Cr@Se, (c) interstitial Cr atom, (d) X-sub. Calculations were performed in 5$\times$5 supercells (with a reciprocal cutoff radius of $4.1$~Bohr$^{-1}$ for (d), and $3.6$~Bohr$^{-1}$ for its respective absorption spectrum in the main text)~\cite{Rost2023}. The calculations include two spins, shown here in red and blue, without spin-orbit coupling. Defects (and bands) in (a) and (d) are spin degenerate. Defect-induced states in (b) and (c) can be occupied by one electron. The influence of the implanted defect is visible via the defect-induced states inside the bandgap. A thick symbol size (high unfolding weight) means that the symmetry of this state corresponds to the one of the primitive cell (pristine material).}
 \label{fig:DFT_band}
\end{figure}

\newpage
\section{Supplementary note 7 - Charge carrier density conversion from gate voltage}
\label{section:carrier}

The carrier concentration $n$ is estimated by the simple parallel plate capacitor model \cite{Robert2021, Li2021b}. The bottom hBN flake acts as a dielectric of the capacitor with dielectric constant $\epsilon_{\text{hBN}}$ $\approx$ 3.4 \cite{Li2021b, Pierret2022, Laturia2018}. Upon applying gate voltage $V_g$, we have  
$$C = \frac{Q}{(V_g-V_0)} = \epsilon_{\text{hBN}} \epsilon_0 \frac{A}{t}$$
with $C$ the capacitance of the graphite/hBN/MoSe$_2$ stack,  $Q$ the charge on MoSe$_2$ ML at $V_g$, $V_0$ being the gate voltage at charge neutrality point (taken to be at minimum trions PL intensity), $\epsilon_0$ the vacuum permittivity, $A$ the hBN flake's area, and the flake's thickness $t$ = 20 nm. Rearranging the above equation gives the charge carrier density $n$ on the ML
$$n = \frac{Q}{eA} = \epsilon_{\text{hBN}} \epsilon_0 \frac{(V_g-V_0)}{et}$$
where $e$ = -1.602\,$\times$\,10$^{-19}$ C is the electron charge.

\bibliography{SI_text.bib}